%% file: MortarXFEM_R2.tex
\documentclass[11pt]{article}
\usepackage{verbatim,amsmath,amssymb}
\usepackage{epsfig,float,color}
\usepackage{geometry}
\usepackage{setspace}
\usepackage{natbib}
\usepackage{hyperref}
\usepackage[all]{hypcap}
\usepackage{epstopdf}
\usepackage{mathrsfs}
\usepackage{multirow}
\usepackage{graphicx}
\definecolor{darkblue}{rgb}{0,0,1}
\hypersetup{pdftex=true, colorlinks=true, breaklinks=true, linkcolor=darkblue, menucolor=darkblue, pagecolor=darkblue, citecolor=darkblue, urlcolor=darkblue}

\input{neco_updated.tex}

\newcommand{\DNOTERS}[1]{}

\newcommand{\MYNOTE}[1]{}

\begin{document}

\begin{center}
\Large{\bf{A segmentation-free isogeometric extended mortar contact method}}\\
\end{center}

\begin{center}
\normalsize{Thang X. Duong$^1$,  Laura De Lorenzis$^2$, Roger A. Sauer\footnote{corresponding author, email: sauer@aices.rwth-aachen.de}}
\vspace{1mm}

\textit{$^1$ Aachen Institute for Advanced Study in Computational Engineering Science (AICES), RWTH Aachen University, Germany }

\textit{$^2$ Institute of Applied Mechanics, Technische Universit{\"a}t Braunschweig, Germany} 

\setcounter{footnote}{2}
\vspace{4mm}
Published\footnote{This pdf is the personal version of an article whose final publication is available at \href{http://dx.doi.org/10.1007/s00466-018-1599-0}{http://link.springer.com/}} 
in \textit{Computational Mechanics},
\href{http://dx.doi.org/10.1007/s00466-018-1599-0}{DOI: 10.1007/s00466-018-1599-0} \\
Submitted on 4th December 2017, Revised on 19th June 2018, Accepted on 28th June 2018
\end{center}
\vspace{2mm}



\rule{\linewidth}{.15mm}
{\bf Abstract:}
This paper presents a new isogeometric mortar contact formulation based on an extended finite element interpolation to capture physical pressure discontinuities at the contact boundary. The so called \textit{two-half-pass} algorithm is employed, which leads to an unbiased formulation and, when applied to the mortar setting, has the additional advantage that the mortar coupling term is no longer present in the contact forces. As a result, the computationally expensive segmentation at overlapping master-slave element boundaries, usually required in mortar methods (although often simplified with loss of accuracy), is not needed from the outset.
For the numerical integration of general contact problems, the so-called \textit{refined boundary quadrature} is employed, which is based on adaptive partitioning of contact elements along the contact boundary. 
\textcolor{black}{The contact patch test shows that the proposed formulation passes the test without using either segmentation or refined boundary quadrature.} Several numerical examples are presented to demonstrate the robustness and accuracy of the proposed formulation. 


\vspace{-5mm}

\rule{\linewidth}{.15mm}

\vspace{-8.5mm}
\hspace{5mm}

%


\section{Introduction}\label{s:intro} 
This paper aims at developing a new mortar method that is unbiased and is able to capture physical discontinuities of the contact pressure at contact boundaries. We are concerned with nonlinear contact involving arbitrarily large deformation and sliding in the context of isogeometric analysis (IGA) \citep{hughes05}. 

In mortar methods, the contact constraints are enforced in a weak sense. This approach originates from domain decomposition, which provides a rigorous mathematical background for proving stability and optimal convergence rates, see e.g.~\citet{puso04a,Hartmann2008,Hesch2008,temizer2012,Laura2014} and references therein.

Existing mortar formulations differ mainly on the choice of shape functions for the approximation of the contact pressure. We will refer to these shape functions as {\em mortar shape functions} \citep{duong17a}. For instance, in the formulation of \citet{puso04a}, the mortar shape functions are identical to the standard shape functions of the displacement field,  \textcolor{black}{\citet{yang05} and \citet{Lorenzis2012}} employ weighted standard shape functions, and \citet{Apop2012} use dual shape functions that are constructed from a local biorthogonality condition. As shown in \citet{duong17a}, all above choices can be derived systematically from least-squares conditions. 
Note that all existing mortar formulations are biased, i.e.~the computational results depend on the choice of slave and master surface. 
\\[1.5mm]
The application of IGA to contact problems has been shown to be highly advantageous in terms of robustness \citep{Laura2014}. This is due to the fact that the IGA  discretizations (i.e.~NURBS, T-Splines) can provide smoothness of any order across element boundaries. This feature is generally beneficial for contact computations with large displacements and large sliding. A number of works  have studied mortar methods in the context of IGA (e.g.\DNOTERS{Lu et al. (2011) not in mortar list}~\cite{temizer2011,Lorenzis2012,Kim12,Dittmann14,Buffa15,Apop2016}).\\[1.5mm]
However, within the discretized system, there may exist various discontinuities of the contact pressure that can reduce either  the accuracy or the robustness of the simulation. Here,  we classify discontinuities into two categories: {\em{physical}} and {\em artificial} discontinuities \textcolor{black}{as is shown in Fig.~\ref{f:discont}}. The first one is independent of the contact formulation and discretization,  and it should be recovered as accurately as possible in contact computations.  In contrast,  artificial discontinuities may appear due to the discretization and the underlying contact formulation used to describe the contact problem. Artificial discontinuities are thus further classified \textcolor{black}{into \textit{normal} and \textit{interpolation} discontinuities. \textit{Normal} discontinuities\footnote{\textcolor{black}{Also referred to as {\em geometric} discontinuities in \citet{Richard96,duong-phd}}}} are caused by discontinuities in the normal vectors, while \textcolor{black}{interpolation} discontinuities\footnote{\textcolor{black}{Also referred to as {\em theoretical} discontinuities in \citet{duong-phd}}}  appear due to discontinuities of shape functions and their products. An example of \textcolor{black}{artificial} discontinuities are the discontinuities of the mortar coupling term\footnote{\textcolor{black}{i.e.~the finite element (FE) force term acting on the master surface in the standard full-pass mortar method (see e.g.~Eq.~\eqref{e:fcsm})}} at the overlapping boundaries of slave and master elements in standard mortar formulations.  These discontinuities appear even for high order NURBS \textcolor{black}{discretizations \citep{Apop2016}. 
\begin{figure}[!htp]
\begin{center} \unitlength1cm
\begin{picture}(0,17.5)
\put(-7.5,-.1){\includegraphics[width=0.95\textwidth]{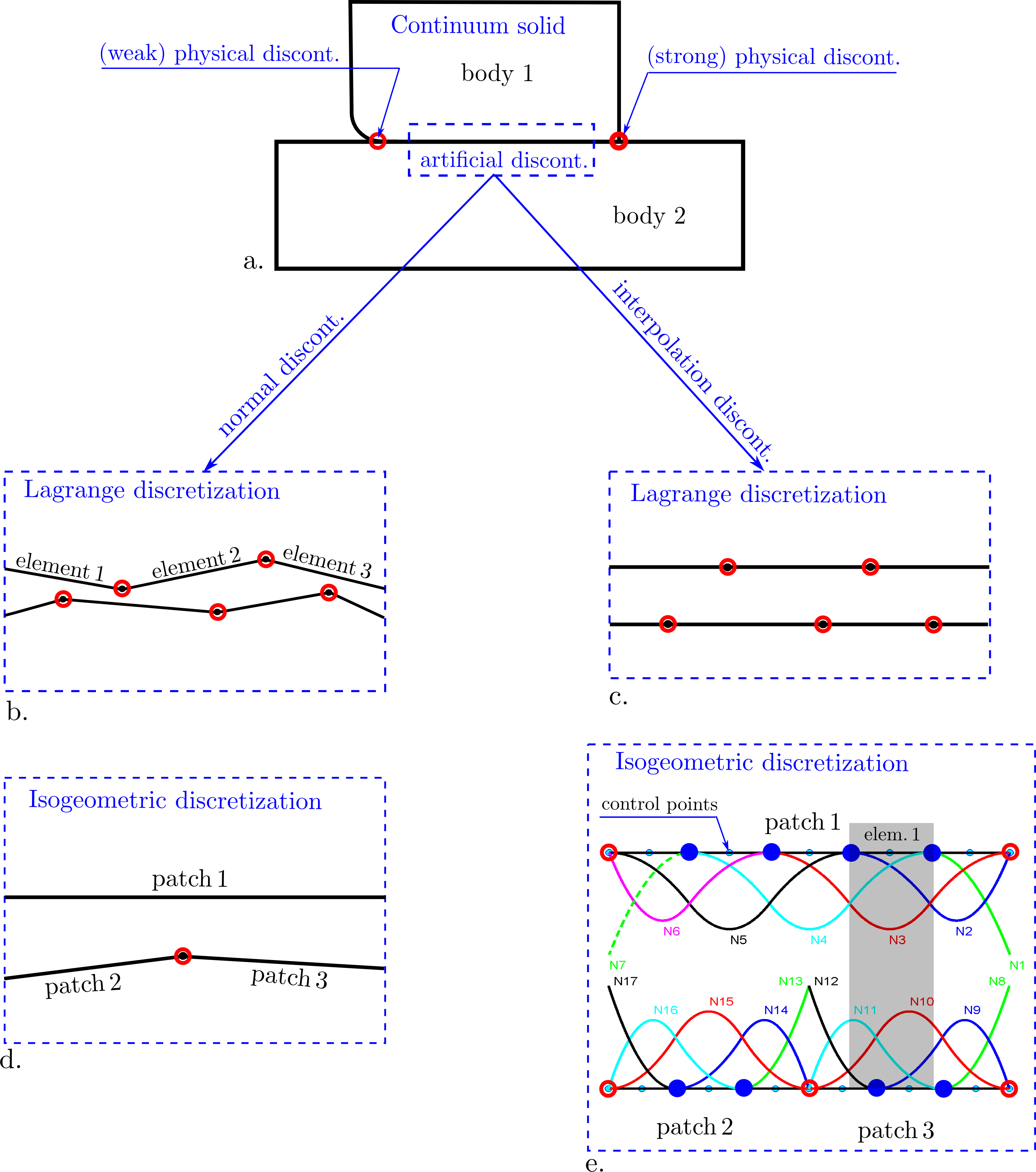}}
\end{picture}
\caption{\textcolor{black}{A classification of discontinuities of the contact pressure: The red circles indicate various locations where discontinuities may appear. The discontinuities include \textit{physical} discontinuities (a), and \textit{artificial} discontinuities that are further classified into (artificial) \textit{normal} discontinuities (b \& d) caused by mismatches in the normal vector field, and  \textit{interpolation} discontinuities (c \& e) caused by the FE shape functions or their products. In (e)~quadratic NURBS shape functions are plotted. Although the normal vector field is smooth everywhere, shape functions and their products are still discontinuous at the patch interfaces.  Besides, the blue bullets indicate the locations (knots) where piecewise shape functions and their products are joined. Since these functions are generally non-monotonic, in some cases, standard numerical quadrature of either shape functions (e.g.~$N_5$) over patches (e.g.~patch 1), or products of shape functions (e.g.~$N_3\!\cdot\!N_9$) over elements (e.g.~element 1), may not be able to reproduce exact integration with a finite number of quadrature points. Therefore, although quadratic NURBS shape functions are smooth at the knots, the inaccuracy of standard numerical quadrature behaves like a weak discontinuity there. } } 
\label{f:discont}
\end{center}
\end{figure}
%
\\[1.5mm]
On one hand,} artificial discontinuities usually lead to discretization \textcolor{black}{and/or} quadrature errors on discretized contact surfaces. In classical mortar methods, the segmentation approach can treat {artificial} discontinuities at the overlapping boundaries of slave and master elements, see e.g.~\citet{puso04a,hesch2011,Apop2013,Bischoff17}. However, segmentation  is usually expensive in terms of computation time, while it is still needed even for IGA discretizations \citep{Buffa15,Apop2016}.  \\[1,5mm]
On the other hand, {physical} discontinuities  may lead to interpolation and/or quadrature errors within element (i.e.~local) domains. In the classical Gauss-point-to-segment  (GPTS) contact formulation with penalty contact,  {physical} discontinuities are captured by constitutive laws and thus no interpolation error appears. However, the accuracy  may still reduce drastically at the contact boundaries, since {physical} discontinuities may not be resolved properly by standard quadrature. In order to tackle this problem, \citet{rbq} have proposed a so-called {\em refined boundary quadrature} (RBQ) technique, which is based on adaptive partitioning of the surface finite elements that contain the contact boundary.\\[1.5mm] 
For mortar methods, {physical} discontinuities can be captured by so-called consistent dual shape function \citep{Cichosz2011,Apop2012,Apop2016}. However, since the consistency treatment is originally motivated by ensuring the biorthogonality condition for the dual shape function on elements that are partially in contact, to the best of our knowledge, only \textit{strong} physical discontinuities are considered by the approach.
%

In our mind, the need for consistency not only concerns the loss of biorthogonality that follows from a particular choice of the mortar shape function. More importantly, consistency also concerns the recovery of the physical discontinuities of the contact pressure at the contact boundary.  In fact, in order to pass the patch test, recovering  the strong physical discontinuity is indispensable. For this reason, it should be noted that increasing the number of quadrature points does not solve the problem. 
Therefore, the consistency for \textit{weak} physical discontinuities may be required as well. Examination of examples~\ref{sec:cylexample} and \ref{sec:iron2D} in this paper shows that the error in the contact forces is quite significant for coarse meshes. \\[1.5mm]
%
In principle, the consistency of dual shape functions for weak physical discontinuities can also be obtained by the same approach as for strong physical discontinuities. 
Note, however, that a major deficiency of consistent dual shape functions is the ill-conditioning of the stiffness matrix when the contact boundary approaches any element boundary \citep{Cichosz2011,Apop2013}. This is due to the fact that dual shape functions are constructed from inverting mass matrices that are approaching  zero \citep{duong17a}. 

Further, it should be noted that both discretization  and interpolation errors due to the above-mentioned discontinuities can be reduced by mesh refinement. This approach, however, diminishes an important advantage of NURBS discretizations, i.e.~that the geometry can be represented accurately with coarse meshes. 

In this contribution, we thus propose an extended finite element (XFEM) interpolation to recover the physical discontinuities of the contact pressure. The new formulation is accordingly denoted the \textit{extended mortar method}.  The central issue of this approach is the explicit detection of the contact boundaries. Here, we will use the level-set method proposed by \citet{rbq}. \textcolor{black}{A similar level-set approach can also  be found in \cite{Graveleau2015}.} The level-set function is then used as the enrichment function to recover the physical discontinuities. Therefore, both weak and strong physical discontinuities are accounted for in the contact formulation. In order to reduce the quadrature error associated with recovering physical discontinuities, we employ the RBQ technique as done for GPTS in \citet{rbq}.   \\[1.5mm]
Besides, the artificial discontinuities are also treated. \textcolor{black}{Seamless NURBS discretizations are employed here to avoid normal discontinuities.  In order to further} eliminate the mortar coupling term that causes the interpolation discontinuities, we apply the so-called \textit{two-half-pass} approach \citep{spbc,spbf}.  In this approach, the two bodies in contact are treated equally, i.e.~there is no distinction between slave and master surfaces. This is done by considering two passes with alternating slave and master roles, but only accounting for the forces acting on the slave surface. 
This leads to an unbiased contact formulation 
that does not contain the mortar coupling term - responsible for the expensive segmentation -
in the contact force vectors. Segmentation can thus be avoided without any loss of accuracy.\\[1.5mm]
%
%
%
%
%
%
The remaining part of this paper is organized as follows: Sec.~\ref{s:gaps} presents the contact kinematics.  Sec.~\ref{s:presure} defines various contact pressures and discusses work-conjugated contact variables.
 In Sec.~\ref{s:stdmortar} The existing mortar formulations are summarized.  For comparison, we also present its two-half-pass version. Discontinuities and quadrature schemes are also discussed. Sec.~\ref{s:XFEM} presents the extended mortar contact in full-pass and two-half-pass version. Sec.~\ref{s:comibineMT} unifies contact formulations for GPTS, existing mortar methods, and the proposed extended mortar method.  Sec.~\ref{s:patchtesta} examines the patch test for various contact formulations. Several numerical examples are presented in Sec.~\ref{s:numex}. Finally, Sec.~\ref{s:conclude} concludes the paper.

\section{Contact kinematics}\label{s:gaps}
This section presents different measures for the normal contact gap and its variation. In order to account for non-smoothness due to discretization, kinematical variables are defined directly for the discretized system.

%
Consider two discretized surfaces in the current configuration that are coming into contact: slave surface $\Gamma^\mrs$ and  master surface $\Gamma^\mrm$. The two surfaces consist in total of  $n_{\mathrm{el}}$ elements,  numbered $e = 1,\,...,\,n_{\mathrm{el}}$. Each element $e$ contains $n_e$ nodes (or control points), and occupies the domain $\Gamma^e$ in the current configuration.  Any point $\bx\in \Gamma^e$  is associated with a coordinate $\bxi$ in the parameter domain $\sP$, and  can be interpolated from the nodal positions, $\bx_A$, by the standard shape functions $N_A$  as
\eqb{l}
\bx(\bxi) = \ds\sum_{A=1}^{n_e} N_A(\bxi)\, \bx_A  = \mN_e\, \mx_e~, \quad \bxi\in\sP~,

\label{e:interpxs}
\eqe
where $\mN_e:= [N_1\,\bI,~ N_2\,\bI,~..., N_{n_e}\,\bI]$ and $\mx_e := [\bx_1,~ \bx_2,~..., \bx_{n_e}]^\mrT$ denote the element shape function and elemental node arrays, respectively.

On the contact surfaces, a set of tangent vectors can be defined by
\eqb{l} 
 \ba_\alpha:=\ds\pa{\bx}{\xi^\alpha} = \mN_{e,\alpha}\,\mx_e~,\quad (\alpha=1,2)~,
 \label{e:tangentsl}
\eqe
where $\mN_{e,\alpha}:=\partial{\mN_e}/\partial{\xi^{\alpha}}$. The normal vector at $\bxi$ is given by
\eqb{l}
\bn := \ds\frac{\ba_1\times\ba_2}{\norm{\ba_1\times\ba_2}}~.
\label{e:normalmsl}
\eqe
With Eqs.~\eqref{e:tangentsl} and \eqref{e:normalmsl}, we can define the components of the metric tensor, 
\eqb{l} 
a_{\alpha\beta}=\ba_\alpha\cdot\ba_\beta~,
\label{e:a_ab}
\eqe
and those of the curvature tensor,
\eqb{l} 
b_{\alpha\beta}=\bn\cdot\ba_{\alpha,\beta} = \bn \cdot\mN_{e,\alpha\beta}\,\mx_e~.
\eqe
Similarly, in the reference configuration, tangent vectors $\bA_\alpha$ and metric tensor components $A_{\alpha\beta}$ can be defined. From the metric tensors, the surface stretch of the contact surfaces is given by
\eqb{l} 
 J:=\ds\frac{\sqrt{\det[a_{\alpha\beta}]}}{ \sqrt{\det[A_{\alpha\beta}]}} = \ds\frac{\norm{\ba_1\times\ba_2}}{\norm{\bA_1\times\bA_2}}~.
 \label{e:Jss}
\eqe

\subsection{Raw gap}
Let $\sE_\mrs$ and $\sE_\mrm$ be the  sets of the element numbers in the slave and master surfaces, respectively.  The pointwise normal {\textit{raw gap}} $g_\mrn$ at position $\bx\in\Gamma^e$, with $e\in\sE_\mrs$,  is defined as
\eqb{l}
g_\mrn (\bxi) := \bn_\mrp\cdot\big(\bx (\bxi) - \bx_\mrp\big)~, \quad \bxi\in\sP~,
\label{e:rawgn}
\eqe
where $\bx_\mrp:=\bx(\bxi_\mrp) \in \Gamma^{\hat{e}}$, with $\hat{e}\in\sE_\mrm$, denotes the closest projection point of $\bx$ onto the master surface with $\bxi_\mrp$ being the corresponding coordinate in $\sP$; $\bn_\mrp$  denotes the normal vector defined by $\bn_\mrp:=\bn(\bxi_\mrp)$. 
The variation of $g_\mrn$ follows from Eq.~\eqref{e:rawgn} and \eqref{e:interpxs} as \citep{wriggers-contact}
\eqb{l}
\delta g_\mrn = \bn_\mrp\cdot\big( \mN_e\, \delta\mx_e - \mN_\mrp\, \delta\mx_{\hat{e}} \big)~,
\label{e:vargn}
\eqe
where $\mN_\mrp:=\mN_{\hat{e}}(\bxi_\mrp)$.

\subsection{Weighted nodal gap}
The \textit{weighted nodal gap} is defined by
\eqb{l}
\tilde{g}_{\mrn A} := \ds \int_{\Gamma^A_0} M_A\,g_\mrn ~\dif A_\mrs~, 
\label{e:smop0}
\eqe
where $\Gamma^A_0\subseteq\Gamma^\mrs_0$ is the support area of node $A$ in the reference configuration, and $M_A$ is called {\em mortar shape function}. Eq.~\eqref{e:smop0} realizes the (mortar) projection of the raw gap \eqref{e:rawgn} onto the nodal degrees of freedom. Several choices of  $M_A$  are listed in Tab.~\ref{tab:MTshape} based on the least squares approach by \citet{duong17a}. 

\begin{table}[!htp]
\begin{small}
\begin{center}
\def\arraystretch{2.7}\tabcolsep=8pt
\begin{tabular}{|l|l|l|}
\hline 
 Mortar types & Mortar shape functions & Mass matrices\\  
\hline   
\parbox{3.9cm}{M-GLS\\[1mm]\footnotesize{(Global least-squares)} }  &   $M_A:=  \ds\sum_B\, N_B\,  [{L}_{BA}]^{-1}$& \multirow{3}{*}{ \parbox[t]{3.83cm} {${L}_{AB}:=\ds\int_{\Gamma^\mrs_0} N_A\,N_B~\dif A_\mrs$}} \\ 
\cline{1-2} 
\parbox{3.9cm}{M-GLS*\\[1mm]\footnotesize{(Global least-squares, PU)} }  & $ M_A:=  \ds\sum_{B,C}\,N_B\, [{L}_{BC}]^{-1} \,{W}_{CA}$& \multirow{3}{*}{ \parbox[t]{3.83cm} { ${W}_{CA}  : =  \delta_{CA} \ds \int_{\Gamma^\mrs_0}\,N_A ~\dif A_\mrs$}}\\ 
\cline{1-2} 
\parbox{4.0cm}{M-LmLS\\[0.8mm]\footnotesize{(Lumped least-squares;\\or weighted std. shape fnc.)} }   & $ M_A:=  \ds\sum_B\,N_B\, {W}_{BA}^{-1}$ &\\ 
\cline{1-2}  
\parbox{4.0cm}{M-LmLS*\\[0.8mm]\footnotesize{(Lumped least-squares, PU;\\or standard shape function)} } & $ M_A:= N_A$ & \\ 
\hline
\parbox{4.0cm}{M-LcLS\\[1mm]\footnotesize{(Local least-squares)} } & $M_A : =  \ds\sum_{B,C,D}\, N_B\,[L^e_{BC}]^{-1} \, W^e_{CD} \,{W}^{-1}_{DA}$ & $L^e_{AB}: =\ds\int_{\Gamma^e_0\subseteq\Gamma^\mrs_0} N_A\,N_B~\dif A_\mrs$\\ 
\cline{1-2} 
\parbox{4.0cm}{M-LcLS*\\[0.8mm]\footnotesize{(Local least-squares, PU;\\or dual shape functions)} }& $M_A : =\ds\sum_{B,C}\, N_B\,[L^e_{BC}]^{-1} W^e_{CA}$ & $W^e_{CA}  : = \ds \delta_{CA} \ds \int_{\Gamma^e_0\subseteq\Gamma^\mrs_0}\,N_A ~\dif A_\mrs$\\ 
\hline
\end{tabular} 
\end{center}
 \label{tab:MTshape}   
 \vspace{-3mm}
\caption{Summary of mortar shape functions $M_A$ based on various least-squares approaches \citep{duong17a}: (*) indicates satisfaction of the partition of unity (PU). Note that M-LmLS,  M-LmLS*, and M-LcLS* are equivalent to the shape functions used in existing formulations, i.e.~of \textcolor{black}{\citet{yang05} and \citet{Lorenzis2012}},  \cite{puso04a}, and  \cite{Apop2012}, respectively.}
 \end{small}
 \end{table}

Note that the integration domain here is the fixed reference configuration. This choice will significantly simplify the variation and linearization. Nevertheless, it would also be possible to define operator \eqref{e:smop0} in the current configuration, \textcolor{black}{or on an intermediate surface.} \\
Considering that $M_A$ is independent from the deformation, the variation of $\tilde{g}_{\mrn A}$  according to Eq.~\eqref{e:smop0} reads
\eqb{l}
\delta\tilde{g}_{\mrn A} = \ds \int_{\Gamma^A_0} M_A\,\delta g_\mrn ~\dif A_\mrs~.
\label{e:varsmop0b}
\eqe

\subsection{Smoothed versus mortar gap} \label{s:dualgap}
The so-called (pointwise) {\em{smoothed gap}}, $\hat{g}_\mrn$, is defined by
\eqb{l}
\hat{g}_\mrn(\bxi) := \ds\sum_{A=1}^{n_e} N_A(\bxi)\,\tilde{g}_{\mrn A} = \bar{\mN}_e\,\tilde\mg_\mrn^e~,
\label{e:smgap}
\eqe
where $\bar{\mN}_e := [N_1,~ N_2,~..., N_{n_e}]$ and $\tilde{\mg}_{\mrn}^e :=  [\tilde{g}_{\mrn 1},~\tilde{g}_{\mrn 2},~ ...,~\tilde{g}_{\mrn {n_e}}]^\mrT$ denote the elemental shape function array and the elemental weighted nodal gap array, respectively.

Similarly, the  so-called (pointwise)  {\em{mortar gap}}, $g^*_\mrn$, is defined by
\eqb{l}
g^*_\mrn(\bxi) := \ds\sum_{A=1}^{n_e} M_A(\bxi)\,\tilde{g}_{\mrn A} = {\mM_e}\,\tilde\mg_\mrn^e~,
\label{e:dualgap}
\eqe
where $\mM_e := [M_1,~ M_2,~..., M_{n_e}]$, and where we have assumed that $M_A$ has the same support area as $N_A$.\footnote{For the general case, $n_e$ would simply be extended to contain all nodes whose support areas cover $\bx(\bxi)$}  Thus, the corresponding variations of $\hat{g}_\mrn$ and   $g^*_\mrn$ read
\eqb{l}
\delta\hat{g}_\mrn = \ds\sum_{A=1}^{n_e} N_A\,\delta\tilde{g}_{\mrn A} = \bar{\mN}_e\,\delta\tilde\mg_\mrn^e~,
\label{e:varsmgap}
\eqe
\eqb{l}
\delta g^*_\mrn = \ds\sum_{A=1}^{n_e} M_A\,\delta\tilde{g}_{\mrn A} = {\mM_e}\,\delta\tilde\mg_\mrn^e~.
\label{e:vardualgap}
\eqe
\section{Contact pressures and work conjugate contact pairs} \label{s:presure}
In Sec.~\ref{s:gaps}, we have presented various contact gap measures. In this section we define the corresponding contact pressures and discuss work conjugate contact pairs in the context of mortar methods.
\subsection{Raw pressure}
For the penalty method, the pointwise \textit{raw contact pressure} at $\bxi\in\sP$ is defined by  
\eqb{l}
 p(\bxi) :=\epsilon_\mrn\, g_\mrn~,
 \label{e:rawpess}
\eqe
where the constant $\epsilon_\mrn$ is a deformation-independent penalty parameter. $p$ is introduced here as the nominal pressure, representing the contact force per reference area. It is related to the true contact pressure  \citep{puso04a, Apop2012,Apop2016} by
\eqb{l}
 p^\mathrm{true} = J^{-1}\, p = J^{-1}\,\epsilon_\mrn\, g_\mrn=:  \bar{\epsilon}_\mrn\, g_\mrn~.
  \label{e:rawpesstrue}
\eqe
Here, $J$ is the surface stretch defined by Eq.~\eqref{e:Jss}, and $\bar{\epsilon}_\mrn$ is now a deformation-dependent penalty parameter.
\subsection{Weighted nodal pressure}
For the penalty method, the  weighted nodal pressure, denoted $\tilde{p}_A$, can be defined by  (see e.g. \citet{temizer2011})
\eqb{l}
 \tilde{p}_A:=\epsilon_\mrn\, \chi_A\, \tilde{g}_{\mrn A} =  \chi_A\, \ds \int_{\Gamma^A_0} M_A\,\epsilon_\mrn\, g_\mrn\,~\dif A_\mrs~, 
  \label{e:smpressure}
\eqe
where  $\chi_A:= \frac{1}{2}\big[1 - \sign(\tilde{g}_{\mrn A} )\big]$ is an active variable at node $A$ that ensures that $\tilde{p}_A = 0$ for  $\tilde{g}_{\mrn A} > 0$.

%
\subsection{Smoothed and mortar contact pressure}
Analogous to Eqs.~\eqref{e:smgap} and \eqref{e:dualgap}, the so-called {\em smoothed contact pressure} $\hat{p}$, and the {\em mortar contact pressure} $p^*$ can be defined by 
\eqb{l}
 \hat{p}(\bxi) := \ds\sum_{A=1}^{n_e} N_A(\bxi)\,\tilde{p}_A= \bar{\mN}_e\,\tilde\mpp^e~,
 \label{e:pstd}
\eqe
and
\eqb{l}
 {p^*}(\bxi) := \ds\sum_{A=1}^{n_e} M_A(\bxi)\,\tilde{p}_A= {\mM_e}\,\tilde\mpp^e~,
 \label{e:pdual}
\eqe
respectively, where $\tilde{\mpp}^e := [\tilde{p}_{1},~\tilde{p}_{ 2},~ ...,~\tilde{p}_{ n_e}]^\mrT$ denotes the elemental weighted nodal pressure array.
\subsection{Contact work conjungation}\label{s:Vir}
Each pair of gap and pressure measures results in a different contact formulation. Tab.~\ref{t:contwork}  lists the work-conjugate pairs that are used in classical Node-to-segment (NTS), GPTS, and mortar contact formulations.

\begin{table}[!htp]
\begin{center}
\def\arraystretch{1.5}\tabcolsep=6pt
\begin{tabular}{|c|l|l|c|}
\hline 
contact formulation & work-conjugate pairs & \textcolor{black}{KKT} cond.  &contact pressure  \\ 
\hline 
 GPTS &   $(p,g_\mrn)$ & $p\,g_\mrn = 0$ & $ p = p(g_\mrn)$  \\ 
 \hline 
   NTS & $(p_A:=p(\bxi_A),g_{\mrn A}:=g_\mrn(\bxi_A))$ &  $p_A\,g_{\mrn A} = 0$ &  $p_A=p_A(g_{\mrn A})$    \\ 
   \hline 
 Generalized mortar: & $(\hat{p} , \hat{g}_\mrn)$ & $\hat{p}\,\hat{g}_{\mrn} =0$  & $\hat{p} = \ds\sum_{A=1}^{n_e}\,N_A\,\tilde{p}_A$   \\ 
    (a) penalty method~~~~~ &  &     &   (a)  $\tilde{p}_A=\tilde{p}_A(\tilde{g}_{\mrn A})$   \\ 
    (b) Lagrange multiplier  &  &     &   (b)  $\tilde{p}_A=$\,unknown   \\
 \hline 
   Standard mortar: &  equivalent pairs:  &  $\tilde{p}_A\,\tilde{g}_{\mrn A} = 0 $; & $p^* = \ds\sum_{A=1}^{n_e}\,M_A\,\tilde{p}_A$  \\
(a) penalty method~~~~~ & $(\tilde{p}_A , \tilde{g}_{\mrn A}) $ or $(p^*, g_\mrn)$ or $(p, g^*_\mrn) $&  or $p^*\,g_\mrn = 0$;  &  (a) $\tilde{p}_A=\tilde{p}_A(\tilde{g}_{\mrn A})$    \\ 
 (b)  Lagrange multiplier&    & or  $p\,{g}_{\mrn}^* = 0$.  & (b)  $\tilde{p}_A=$\,unknown   \\    
\hline 
\end{tabular} 
\label{t:contwork}
 \vspace{-3mm}
\end{center}
\caption{Contact work-conjugate pairs and corresponding contact formulations: The \textcolor{black}{Karush-Kuhn-Tucker (KKT)} conditions for the node-to-segment (NTS) can be obtained by requiring the KKT conditions from the GPTS  formulation to be satisfied discretely at nodes/knots. Likewise, the standard mortar formulation is a special case of the generalized mortar formulation.}
\end{table}
In particular, for the standard mortar formulation (e.g.~by \citet{puso04a,Lorenzis2012}), the third  \textcolor{black}{Karush-Kuhn-Tucker} condition is enforced in a weak sense by the discrete equation $\tilde{p}_A\,\tilde{g}_{\mrn A} = 0\,,$ for each $A=1,~...,~n_\mrs$, where $n_\mrs$ denotes the number of slave nodes. These equations can also be obtained by \textcolor{black}{requiring}  the KKT condition from the so-called generalized mortar formulation to be satisfied only at nodes or knots (see Tab.~\ref{t:contwork}).  
Besides, by means of Eqs.~\eqref{e:smop0} and \eqref{e:smpressure}, their contact potential can be shown to satisfy
\eqb{l}
\Pi_\mrc =  \ds\sum_{A=1}^{n_\mrs} \tilde{p}_A\,\tilde{g}_{\mrn A} = \ds\int_{\Gamma^\mrs_0} p^*\,g_\mrn\,\dif A_\mrs = \ds\int_{\Gamma^\mrs_0} p\,g^*_\mrn\,\dif A_\mrs~.
\label{e:contpoten}
\eqe 
Thus,  $(p^*,g_\mrn)$ and $(p,g^*_\mrn)$ are equivalent work-conjugate pairs to $(\tilde{p}_A,\tilde{g}_{\mrn A})$. 
%
%
In view of \eqref{e:contpoten}, the mortar pressure $p^*$ is always the (approximated) nominal pressure  for any mortar shape function $M_A$. 
Furthermore, in view of Eqs.~\eqref{e:smop0} and \eqref{e:smpressure}, the mortar method can be interpreted as a smoothing of the raw contact gap $g_n$. In principle, the concept of the smoothing can be extended further by considering $\hat{g}_\mrn$ as the raw contact gap and performing the smoothing process once again. Similarly, new mortar formulations can be derived by combining different work-conjugate pairs. 
\section{Standard mortar contact} \label{s:stdmortar}

In this section, we first summary the finite element contact forces for the standard mortar formulation of Eq.~\eqref{e:contpoten}.  Since this formulation employs a full-pass algorithm, we denote it \textit{standard mortar full-pass (SMFP)} formulation. Next, since this formulation is biased w.r.t.~the choice of slave and master surfaces, we will modify it by the so-called two-half-pass approach \citep{spbc,spbf}. The modified formulation is thus called \textit{standard mortar two-half-pass  (SM2HP)} formulation. Further, discontinuities in the considered contact formulations are also discussed. It will also be shown that the SM2HP formulation can avoid the segmentation as is required in SMFP.


\subsection{Standard mortar full-pass contact (SMFP)} \label{s:fpmt}

\subsubsection{Finite element contact forces}
The contact virtual work for SMFP can be obtained by taking the variation of  contact potential \eqref{e:contpoten} w.r.t. the displacement. That is,
\eqb{l}
\delta\Pi_\mrc = \ds\sum_{e\in\sE_\mrs}\, \int_{\Gamma^e_0}  p^*\, \delta{g}_{\mrn} \,\dif A_e~.
\label{e:MTvirworkstd}
\eqe 
By taking Eq.~\eqref{e:vargn} into account, we obtain 
\eqb{l}
\delta\Pi_\mrc =   \ds\sum_{e\in\sE_\mrs}\,( \delta \mx_e\cdot\mf_e +  \delta \mx_{\hat{e}}\cdot\mf_{\hat{e}} )~,
\label{e:disvirc}
\eqe
where $\hat{e}\in \sE_\mrm$ denotes the master element(s) that contain closest projection points emanating from the slave element $e$, and where
\eqb{lll}
\mf_e := \ds\int_{\Gamma^e_0} \mN_e^\mrT\, \bT^*_\mrc\,\dif A_e~ \quad $and$\quad \mf_{\hat{e}} := - \ds\int_{\Gamma^e_0} \mN_\mrp^\mrT\, \bT^*_\mrc\,\dif A_e~
\label{e:fcsm} 
\eqe 
are the finite element contact forces with the nominal contact traction defined by
\eqb{l}
\quad \bT^*_\mrc:= p^*\,\bn_\mrp~.
\eqe 

%
\subsubsection{Discontinuities}\label{s:discont}
Global contact equilibrium requires $\delta\Pi_\mrc=0$ and follows from Eq.~\eqref{e:disvirc} as
\eqb{lll}
 \ds\sum^{n_\mrs}_{A=1}\int_{\Gamma^\mrs_0} N_{A} \,\, p^*\,\bn_\mrp \,\dif A_\mrs~  =   \ds\sum^{n_\mrm}_{B=1} \ds\int_{\Gamma^\mrs_0} N_{B}(\bxi_\mrp)\,\, p^*\,\bn_\mrp \, \dif A_\mrs~, 
\label{e:fcsmb}
\eqe 
where $n_\mrm$ denotes the number of master nodes. Eq.~\eqref{e:fcsmb} is satisfied when the integration on both sides is exact. However, 
there may exist discontinuities in the integrands: physical discontinuities of  $p^*$ and/or $\bn_\mrp$ at the contact boundaries, \textcolor{black}{and artificial discontinuities  of  $\bn_\mrp$  and  $N_B(\bxi_\mrp)\cdot M_{A}(\bxi)$ within the contact surface}. Therefore, two aspects must be considered here: (1) accurately capturing the physical discontinuities and (2) accurate quadrature. In classical approaches, the so-called consistent mortar shape function can be used to capture the  (strong) physical discontinuity \citep{Apop2012,Apop2016}, while segmentation is used  for quadrature (see e.g.~\citet{puso04a}). Quadrature is discussed in the following.

\subsubsection{Numerical quadrature }\label{s:numintFP}
Numerical quadrature is used for the evaluation of the contact forces \eqref{e:fcsm}. Since discontinuities may appear in integrands, a suitable quadrature scheme should be chosen for accurate integration. The simplest method is using a large number of quadrature points (see \cite{Lorenzis2012}). Such a scheme, however, is not robust and accurate since strong discontinuities cannot be captured well by this method. \\[1.5mm]
Alternatively, a more robust and accurate approach is segmentation, which requires  to split (or segment)  the integration domain at interfaces where the integrands are discontinuous e.g.~at  the overlapping boundaries of master-slave elements  \citep{puso04a}. 
Note that at  the overlapping boundaries for Lagrange elements, both \textcolor{black}{normal and interpolation} discontinuities coincide. \\[1.5mm]
Moreover, for NURBS, since the \textcolor{black}{artificial} discontinuities are present at overlapping boundaries of master-slave elements, in principle, the segmentation is required there as well. The requirement is especially important at the patch boundaries if multiple patches are used, since additional \textcolor{black}{normal} discontinuities appear in this case. \\[1.5mm] 
Segmentation,  however,  reduces the efficiency of the contact formulation significantly. In this paper, we thus also aim at developing a contact formulation that avoids costly segmentation. With this in mind, we consider the so-called {\em two-half-pass} approach, which is extended to mortar contact in the following section.

\subsection{Standard mortar  two-half-pass contact (SM2HP)}\label{s:2HP}
%
\subsubsection{Finite element contact forces}
In this section, an unbiased mortar formulation is obtained by applying the so-called \textit{two-half-pass} algorithm introduced by \cite{spbc}. 
Accordingly, the two contact surfaces are now treated equally and thus denoted by numbering ($1$ and $2$) instead of calling them master and slave surfaces. On each contact surface,  the contact pressures and  the contact gaps are defined. We redefine the contact potential \eqref{e:contpoten} for each contact surface separately, and consider the other surface as \textcolor{black}{variationally (var.)} fixed.\footnote{\textcolor{black}{I.e.~the configuration is fixed when taking variation, but it is not necessarily fixed when linearizing.}} That is,
\eqb{l}
\Pi_\mrc  := \ds \left. \ds\int_{\Gamma^1_0} p^*_1\,g_\mrn^1\,\dif A_1 \right|_{\Gamma^2 \mathrm{~var.\,fixed}} + \left. \ds\int_{\Gamma^2_0} p^*_2\,g_\mrn^2\,\dif A_2 \right|_{\Gamma^1 \mathrm{~var.\,fixed}}~,
\eqe 
where $\Gamma^k$, $p^*_k$, and $g_\mrn^k$ denote the domain, the mortar pressure \eqref{e:pdual}, and the raw contact gap \eqref{e:rawgn} of surface $k=1,2$, respectively. Taking the variation of contact potential above gives
\eqb{l}
\delta\Pi_\mrc = \ds\sum_{e\in\sE} \delta\mx_e \cdot \mf_e~,
\label{e:Pi2HP}
\eqe
where $\sE:=\sE_\mrs\cup\sE_\mrm$, and 
\eqb{lll}
\mf_e = \ds\int_{\Gamma^e_0} \mN_e^\mrT\, p^*\,\bn_\mrp \,\dif A_e~.
\label{e:fcs2HP}
\eqe 

\subsubsection{Discontinuities}\label{s:discont2HP}
From Eq.~\eqref{e:Pi2HP} follows the global contact equilibrium for SM2HP
\eqb{lll}
 \ds\sum^{n_1}_{A=1} \ds\int_{\Gamma^1_0} N_{A}^1\,\,  p^*_1\,\bn_\mrp^1\,\dif A_1 = \ds\sum^{n_2}_{B=1}  \ds\int_{\Gamma^2_0} N_B^2\,\,  p^*_2\,\bn_\mrp^2\,\dif A_2~,
\label{e:momentum2HP}
\eqe 
where $n_k$, $\bn^k_\mrp$ and $N^k_A$  denote the number of nodes, the normal vector, and the shape functions on contact surface $k$, respectively.
As seen, since there is no coupling term as for SMFP, the \textcolor{black}{interpolation} discontinuities are avoided in SM2HP.
\textcolor{black}{In Eqs.~\eqref{e:fcs2HP} and \eqref{e:momentum2HP}, the weighted nodal pressure $\tilde{\mpp}^e$ usually causes no discontinuity of the contact pressure $p_k^*$, since $p_k^*$ is interpolated from $\tilde{\mpp}^e$ (see Eq.~\eqref{e:pdual}).}\\[1.5mm]
%
Note that the coupling term still appears in the tangent matrices (see Appendix~\ref{sap:contact}), but not in the finite element forces \eqref{e:fcs2HP}. The quadrature of the tangent matrices does not affect accuracy. Instead, it only affects the rate of convergence of the Newton-Raphson iteration.\\[1.5mm]
\textcolor{black}{However, like for SMFP, we still must account for any physical discontinuities of $p^*_k$ and/or  $\bn_\mrp^k$ at contact boundaries. }
Therefore, segmentation is only required at the (physical) contact boundary for an accurate  quadrature in SM2HP.  We thus refer  to this as {\em contact-boundary partitioning} in order to distinguish it from segmentation at overlapping boundaries of contact elements. The level-set approach presented in \cite{rbq} can  be used for this purpose.\\[1.5mm]
In terms of accuracy,  it turns out that SM2HP is quite inaccurate. The main reason is that linear momentum balance holds only in a weak (global) sense (see Eq.~\eqref{e:momentum2HP}), while  $p^*_1=p^*_2$ is not satisfied in a pointwise sense.  The condition $p^*_1=p^*_2$ strongly depends on the relative descretization error between the two contact surfaces.  In fact, this issue leads to the failure of the contact patch test for dissimilar meshes as will be shown later. This is in contrast to the GPTS two-half-pass approach (see \cite{spbc,spbf}) where the contact constraints are enforced pointwise. \\[1.5mm]
Therefore, a further advancement of the mortar formulation is required, so that the pointwise condition  $p^*_1=p^*_2$  is somehow bounded for the two-half-pass approach.  At the same time, both weak and strong {physical discontinuities} at the contact boundaries still need to be  treated. These issues motivate the development of the so-called {\em extended mortar method}, which will be presented next.

\section{Extended mortar contact}\label{s:XFEM}
This section proposes a new isogeometric mortar contact formulation based on an XFEM interpolation in order to recover the physical discontinuities of the contact pressure at contact boundaries. 
Therefore, the proposed contact formulation is denoted {\em extended mortar method}. Both its full-pass and two-half-pass version are presented. 


In the following we first recapitulate the basic extended isogeometric approximation. Then this approximation is applied to the contact pressure.
\subsection{Basic extended isogeometric approximation}
 The basic idea of extended isogeometric analysis is to enrich the original (e.g.~NURBS) approximation space such that certain features like discontinuities (e.g.~crack or interfaces) can be reproduced without modifying the mesh \citep{Moes99,Luycker11}. This can be done by the so-called partition of unity concept. That is, 
\eqb{l}
\ds\sum_{A=1}^{n_s} N_A(\bxi) \psi(\bxi) = \psi(\bxi)~, 
\label{e:enrterm}
\eqe
where $N_A$ are the NURBS shape functions, and $\psi$ is an enrichment function characterizing the discontinuities of the approximated field.\\[1.5mm]
We apply this approach to capture discontinuities of the mortar contact pressure \eqref{e:pdual} at contact boundaries, i.e.
\eqb{l}
 p^* =  \ds\sum_{A=1}^{n_\mathrm{s}} M_A \,\tilde{p}_A^\mathrm{std} +  \ds\sum_{A=1}^{n_\mathrm{s}} M_A\, \psi(\phi)\,\tilde{p}_A^\mathrm{enr} ~.
 \label{e:XFEMp}
\eqe
Here,   $\tilde{p}_A^\mathrm{std}$ and $\tilde{p}_A^\mathrm{enr}$ are nodal coefficients to be determined later, and $\psi$ is defined by
\eqb{l}
 \psi(\phi) := \sign(\phi) = \left\{\begin{array}{rl} 1 & $ if $ \phi > 0 ~,\\
                    0 & $ if $ \phi = 0~,\\
                    -1 & $ if $ \phi < 0~,\\
                    \end{array} \right.
                 \label{e:pjump}  
\eqe
where $\phi(\bx)$ is a level-set function defined by \citep{rbq}
     \eqb{l}
        \phi(\bx) \left\{\begin{array}{rl} <0 & $ in $ \Gamma^e_\mrin~,\\
                    =0 & $ on $ \mcalC^e~,\\
                    >0 & $ in $ \Gamma^e_\mro~.\\
                    \end{array} \right.
        \label{e:defgN}
    \eqe 

\textcolor{black}{Here, $\Gamma^e_\mrin$ and $\Gamma^e_\mro$ denote in-contact and out-of-contact domains that are separated by the contact boundary $\mcalC^e$ on the contact element $\Gamma^e$}. Note that the first term in Eq.~\eqref{e:XFEMp} is the standard interpolation \eqref{e:pdual}, while the second term is an enrichment term according to \eqref{e:enrterm} that is included in order to reproduce  a discontinuity.  \\[1.5mm]
Further, in order to compute the nodal coefficients $\tilde{p}_A^\mathrm{std}$ and $\tilde{p}_A^\mathrm{enr}$, one distinguishes between extrinsic and intrinsic enrichments (see e.g.~\cite{Fries2006}). With  extrinsic enrichment, the unknown  coefficients $\tilde{p}_A^\mathrm{enr}$ associated with the enriched solution add extra degrees of freedom to the global system, while with  intrinsic enrichment, $\tilde{p}_A^\mathrm{std}$ and $\tilde{p}_A^\mathrm{enr}$ are determined from a certain reproducing condition. Here, we will follow the latter approach, presented in detail in the following.
\subsection{A least-squares approach}
In order to determine the extra unknowns  $\tilde{p}_A^\mathrm{enr}$  in Eq.~\eqref{e:XFEMp}, we consider the reproducing condition
\eqb{l}
g_r: = p^* - \bar{p} \approx 0,
\label{e:reproducing}
\eqe
where $\bar{p}$ is the nominal, pointwise contact pressure given by
\eqb{l}
\bar{p} = \epsilon_\mrn\,H(\phi)\,g_\mrn  =: J\,\bar{p}^{\mathrm{true}}~.
\label{e:pressH}
\eqe
Here, $\bar{p}^{\mathrm{true}} = \bar{\epsilon}_\mrn\,H(\phi)\,g_\mrn$, and $H$ is the Heaviside function,
\eqb{l}
 H(\phi) = \left\{\begin{array}{rl} 0 & $ if $ \phi\geq 0 ~,\\
                    1 & $ if $ \phi < 0~.\\
                    \end{array} \right.
                    \label{e:hevisde}
\eqe
To enforce condition~\eqref{e:reproducing}, a simple method is to minimize the least-squares potential,
\eqb{l}
\Pi_{LS} := \ds\int_{\Gamma^\mrs_0} \frac{1}{2} ( p^* - \bar{p} )^2\,\dif A_\mrs~.
\label{e:PiLS}
\eqe
By  inserting Eq.~\eqref{e:XFEMp} into \eqref{e:PiLS} and setting $\delta\Pi_{LS} = 0$, we obtain the following system of equations
\eqb{l}
\left\{\begin{array}{l} \mn^{\mrs\mrs}\,\tilde{\mpp}^\mathrm{std}  +  \mn^{\mrs\mrx}\,\tilde{\mpp}^\mathrm{enr} = \boldsymbol{\lambda}~,\\
                         \mn^{\mrs\mrx}\,\tilde{\mpp}^\mathrm{std} +  \mn^{\mrs\mrs}\,\tilde{\mpp}^\mathrm{enr} = \boldsymbol{\lambda}^\mrx ~,\\
                    \end{array} \right.
  \label{e:sysLS}                  
 \eqe                   
 where $\tilde{\mpp}^\mathrm{std} := [\tilde{p}_A^\mathrm{std}]$, $\tilde{\mpp}^\mathrm{enr} := [\tilde{p}_A^\mathrm{enr}]$, 
 
 \eqb{l}
 \mn^{\mrs\mrs}:= \ds\int_{\Gamma^\mrs_0}  \mM_e^\mrT\,\mM_e \,\dif A_\mrs~,\quad  \mn^{\mrs\mrx}:= \ds\int_{\Gamma^\mrs_0} \psi \,\mM_e^\mrT\,\mM_e \,\dif A_\mrs~,
\eqe
and
 \eqb{l}
 \boldsymbol{\lambda}:= \ds\int_{\Gamma^\mrs_0}  \mM_e^\mrT\, \bar{p} \,\dif A_\mrs~,\quad   \boldsymbol{\lambda}^\mrx:= \ds\int_{\Gamma^\mrs_0} \psi \,\mM_e^\mrT\, \bar{p} \,\dif A_\mrs~.
 \label{e:lambdax}
\eqe
By inserting Eq.~\eqref{e:pressH} into  Eqs.~\eqref{e:lambdax}, we find $\boldsymbol{\lambda} =  \boldsymbol{\lambda}^\mrx$. Thus, it follows from Eq.~\eqref{e:sysLS}  that  $2\tilde{\mpp}^\mathrm{std} = 2\tilde{\mpp}^\mathrm{enr} =:\tilde{\mpp}$, and
\eqb{l}
\tilde{\mpp}= \mW_\mrx^{-1}  \boldsymbol{\lambda} ~,
\label{e:pxa}
\eqe
with 
\eqb{l}
\mW_\mrx :=  \mn^{\mrs\mrs}+ \mn^{\mrs\mrx} = \ds\int_{\Gamma^\mrs_0}  H(\phi)\,\mM_e^\mrT\,\mM_e \,\dif A_\mrs~.
\label{e:Wxfem}
\eqe
As seen, $\mW_\mrx$ is symmetric and invertible  within the space spanned by the nodal active set ($\chi_A>0$) defined by
\eqb{l}
[\chi_A]: = H([h_A]);\quad [h_A]: =-\ds\int_{\Gamma^\mrs_0} M_A \, H(\phi) \,\dif A_\mrs~.
\label{e:actsetN}
\eqe
For inactive nodes, i.e.~$\chi_A=0$, we have $\tilde{p}_A^\mathrm{std} = \tilde{p}_A^\mathrm{enr} = 0$.

\subsection{Extended mortar contact forces}\label{s:MTxig2HPfp}
With the ingredients above, we are now able to derive an extended mortar contact formulation. For concise presentation, we assume that $M_A$ has the same support area as $N_A$. Expression \eqref{e:XFEMp} then can be rewritten as
\eqb{l}
 p^* = \ds\sum_{A=1}^{\,n_e} M_A^\mrx \,\tilde{p}_A = \mM_\mrx \,\tilde{\mpp}_e~,
 \label{e:XFEMpTen}
\eqe
since $\psi(\phi) = 2H(\phi) - 1$.  Here, $\mM_\mrx:=H(\phi)\mM_e $ denotes the extended shape function array, and $\tilde{\mpp}_e$ is the nodal pressure vector determined from Eq.~\eqref{e:pxa}. With expression \eqref{e:XFEMpTen}, the variation of the  potential~\eqref{e:contpoten} at the equilibrium state \textcolor{black}{(i.e.~$g_\mrn\approx 0$)} reads 
\eqb{l}
\delta\Pi_\mrc = \ds \int_{\Gamma^\mrs_0}  p^*\, \delta{g}_{\mrn} \,\dif A_\mrs ~,
\label{e:virWb2}
\eqe
which has the same form as the standard case \eqref{e:MTvirworkstd}. Therefore, expressions (\ref{e:disvirc}--\ref{e:fcsm}) are still valid if $\tilde{p}_A$ is computed by Eq.~\eqref{e:pxa}, and $\mM_e$  is  replaced by $\mM_\mrx$. We thus obtain the so-called \textit{extended mortar full-pass} (XMFP) as 
\eqb{l}
\delta\Pi_\mrc =  \ds\sum_{e\in\sE_\mrs}\,( \delta \mx_e \cdot\mf_e +  \delta \mx_{\hat{e}}\cdot\mf_{\hat{e}})~,
\label{e:disvircx}
\eqe
where
\eqb{lll}
\mf_e \dis \ds\int_{\Gamma^e_0} \mN_e^\mrT\, p^*\,\bn_\mrp\,\dif A_e = \ds\int_{\Gamma^e_0} H(\phi)\, \mN_e^\mrT\, \bn_\mrp\, \mM_e\,\dif A_e~\tilde{\mpp}^e~ , \\[2mm]
\mf_{\hat{e}} \dis - \ds\int_{\Gamma^e_0} \mN_\mrp^\mrT\, p^*\,\bn_\mrp\,\dif A_e =  - \ds\int_{\Gamma^e_0}H(\phi) \,\mN_\mrp^\mrT\, \bn_\mrp \,\mM_e\,\dif A_e~\tilde{\mpp}^e ~.
\label{e:fcsmx}
\eqe 
Similar to Sec.~\ref{s:2HP}, the \textit{extended mortar two-half-pass} (XM2HP) reads
\eqb{l}
\delta\Pi_\mrc = \ds\sum_{e\in\sE} \delta\mx_e \cdot \mf_e~,
\label{e:Pi2HPXIG}
\eqe
where $\sE := \sE_\mrs\cup\sE_\mrm$ and $\mf_e$ are computed by Eq.~\eqref{e:fcsmx}a.

%
\textbf{Remark: } Since the matrix $\mW_\mrx$ is a global quantity, no ill-conditioning problem  is to be expected when the contact boundaries approach the element boundaries. This is in contrast to standard mortar methods based on consistent dual shape functions \citep{Apop2012}. 

\subsection{Numerical quadrature}
Since the proposed extended mortar formulation only modifies the interpolation of the contact pressure to account for physical discontinuities,  the discussion of the discontinuities in Sec.~\ref{s:discont} and  \ref{s:discont2HP} is still valid for XMFP and XM2HP, respectively. Thus, for XMFP, apart from the physical discontinuities appearing at the contact boundary, artificial (\textcolor{black}{interpolation}) discontinuities can appear at overlapping element boundaries. To treat those, either segmentation or many quadrature points is needed there. For XM2HP on the other hand, 
discontinuities only appear at the contact boundary. In order to reduce the quadrature error associated with those, we employ RBQ \citep{rbq}.  
Strictly, RBQ is only needed for strong discontinuities, e.g.~at sharp corners. For weak discontinuities many quadrature points can be used instead of RBQ. But this can decease efficiency and robustness.  
Since there are no discontinuities within the contact area\footnote{In case of multiple NURBS patches, there may still be \textcolor{black}{normal} discontinuities due to mismatch $\bn_\mrp^k$ at patch interfaces during deformation. This can be easily treated by applying a continuity constraint (see e.g.~\cite{solidshell}) instead of segmentation.},  a special quadrature treatment like segmentation is not needed for XM2HP.

\section{A unified contact formulation}\label{s:comibineMT}
The contact virtual work for GPTS, standard mortar~\eqref{e:MTvirworkstd}, and extended mortar~\eqref{e:virWb2} can be written in the combined expression
\eqb{l}
\delta\Pi_\mrc = \ds \int_{\Gamma^\mrs_0}  p^*\, \delta{g}_{\mrn} \,\dif A_\mrs ~,
\label{e:vircontactW}
\eqe
with a suitable definition of the mortar contact pressure $p^*$. Tab.~\ref{t:vircontactW} summarizes the definitions of $p^*$ for the considered contact formulations. 
\begin{table}[!htp]
\begin{center}
\def\arraystretch{2}\tabcolsep=4pt
\begin{tabular}{|l|l|l|c|}
\hline 
Contact formulation & Contact pressure & Auxiliary parameters & Finite element forces \\ 
\hline 
GPTS & $p^*:= \chi\,p$  & $p  = \epsilon_\mrn\, g_\mrn$ & \parbox[p]{3.7cm} {\small{see e.g. \cite{wriggers-contact}}}\\ 
\hline 
\parbox{3.4cm}{Standard mortar} &  \multirow{1}{*}{$p^* := \ds\sum_{A=1}^{n_e}\,M_A\,\tilde{p}_A$ ~\eqref{e:pdual}} & $\tilde{\mpp} := \mA\,\ds\int_{\Gamma^\mrs_0}  \mM_e^\mrT \,p\,\dif A_\mrs~$ & \multirow{1}{*}{see Eq.~\eqref{e:fcsm}} \\[2mm] 
\cline{1-1}\cline{3-3}
\hline 
Extended mortar & $p^* := \ds\sum_{A=1}^{n_e}\,M^\mrx_A\,\tilde{p}_A$ ~\eqref{e:XFEMpTen} &  $\tilde{\mpp}:= \mW_\mrx^{-1} \ds\int_{\Gamma^\mrs_0}  \mM_\mrx^{\mrT}\, p\,\dif A_\mrs$ & see Eq.~\eqref{e:fcsmx}\\ 
\hline 
\end{tabular} 
\label{t:vircontactW}
\caption{Contact pressure for various contact formulations. In the table $\mA:=\mathrm{diag}([\chi_A])$.} 
\vspace{-0.5cm}
\end{center}  
\end{table}
For  linearization, the interpolation for $p^*$ in the mortar formulations can be further written into the unified expression
\eqb{l}
p^* := \ds\sum_{A=1}^{n_e}\,\Phi_A\,\tilde{p}_A~,
\label{e:unifiedpdual}
\eqe
where $\Phi_A$ denote the unified shape functions. That is, $\Phi_A=M_A$ for standard mortar and $\Phi_A=M^\mrx_A$ for extended mortar.  The expression of the nodal parameter  $\tilde{\mpp}$ can also be unified, 
\eqb{l}
\tilde{\mpp}:= \mG\, \ds\int_{\Gamma^\mrs_0}  \mPhi^\mrT \,p\,\dif A_\mrs~.
\label{e:unifiedP}
\eqe
where $\mG=\mA$,
and $\mG=\mW_\mrx^{-1}$ for standard mortar 
and extended mortar, respectively. 
Details on the linearization of Eq.~\eqref{e:vircontactW} can be found in Appendix~\ref{sap:contact}.
\section{Contact patch test}\label{s:patchtesta}
\begin{figure}[!htp]
\begin{center} \unitlength1cm
\begin{picture}(0,6.3)
\put(-6.5,-.1){\includegraphics[height=70mm]{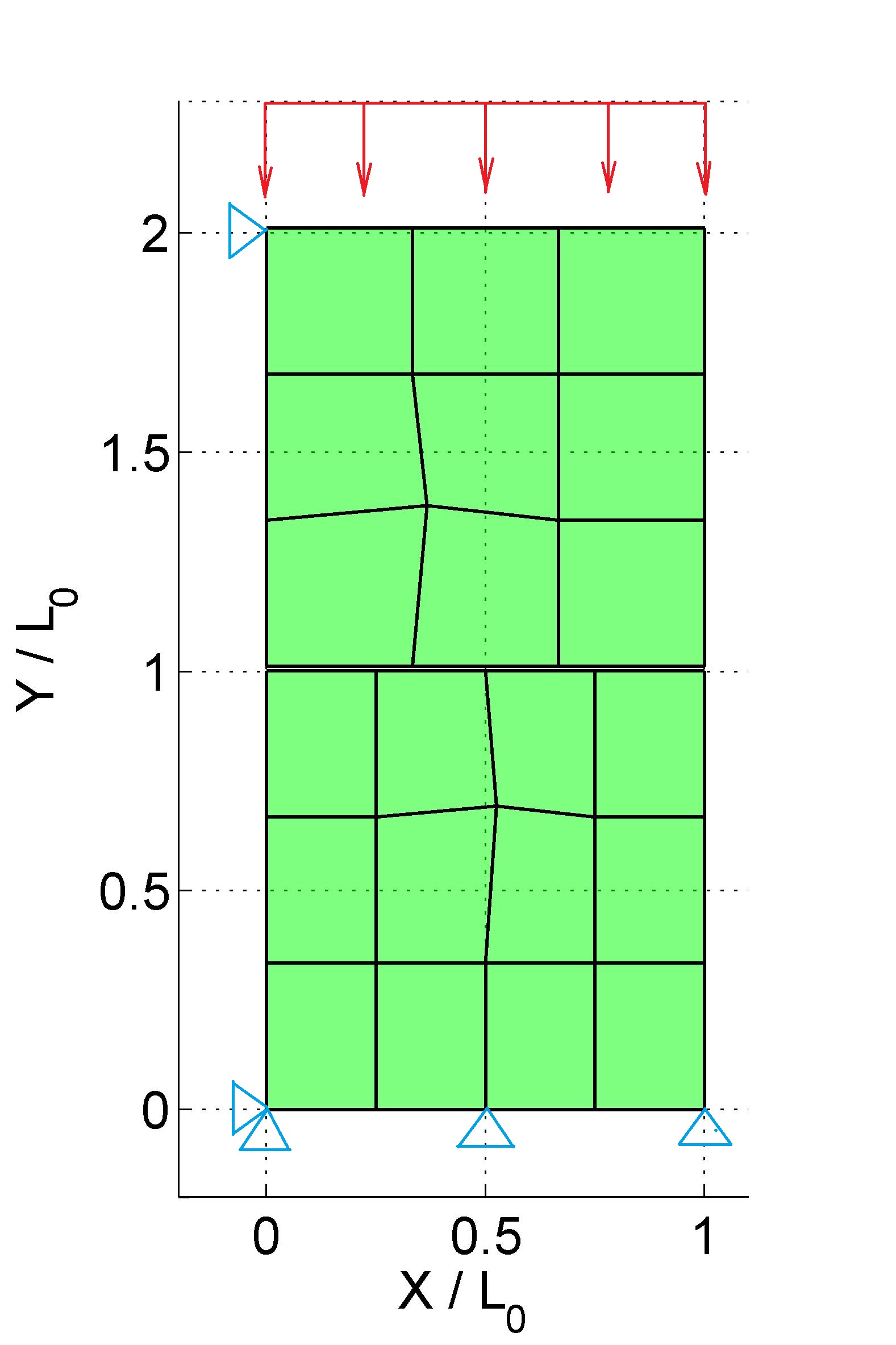}}
\put(0,-.1){\includegraphics[height=70mm]{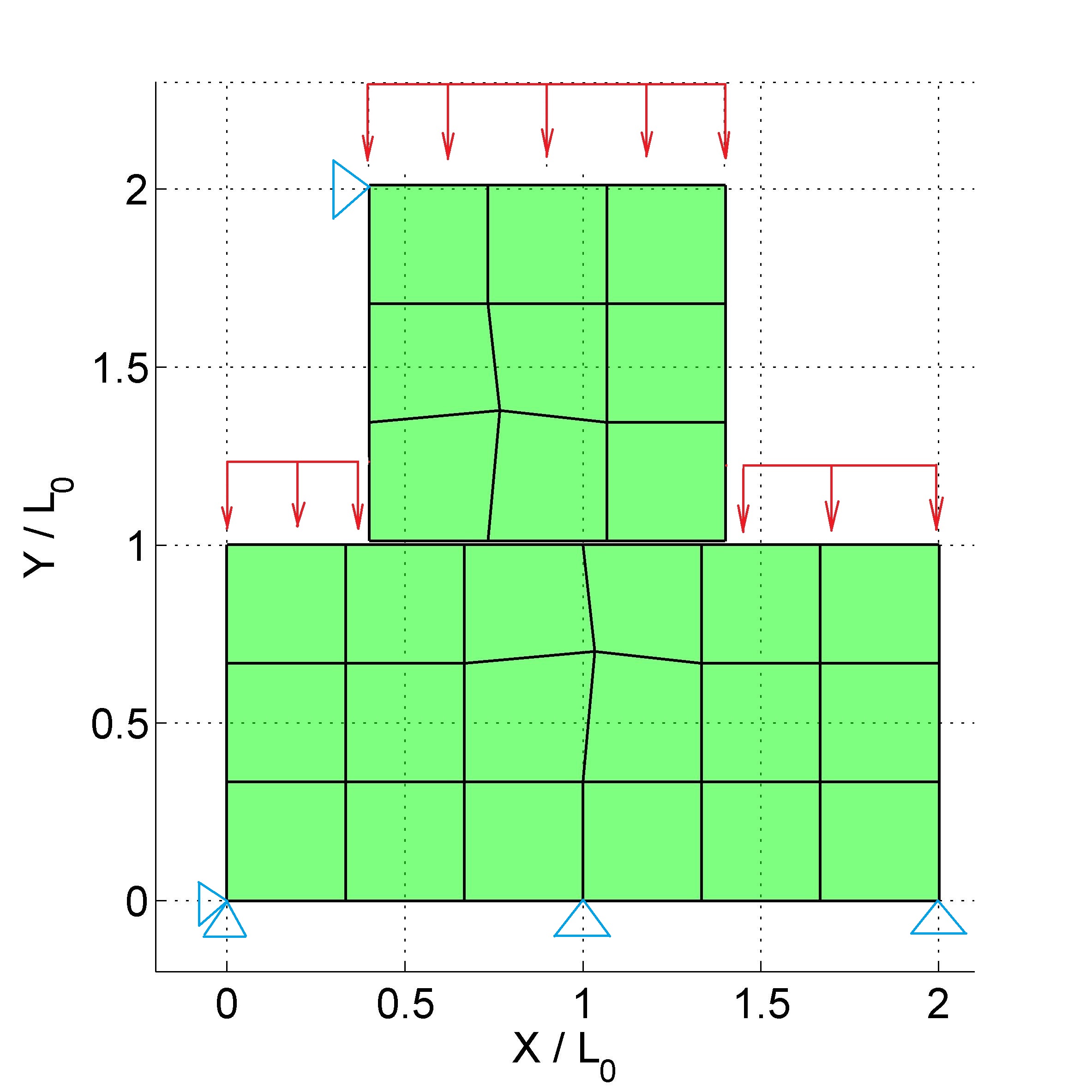}} 
\put(-6.0,0){a.}
\put(-2.5,6){$\bar{p}$}

\put(0.7,0){b.}

\put(2.0,6){$\bar{p}$}
\end{picture}
\caption{\textcolor{black}{Contact patch tests: 
a.~case~1: non-conforming mesh, b.~case~2: generalized patch test.
 $\bar{p}$ denotes an applied uniform pressure.} } 
\label{f:patchINI}
\end{center}
\end{figure}
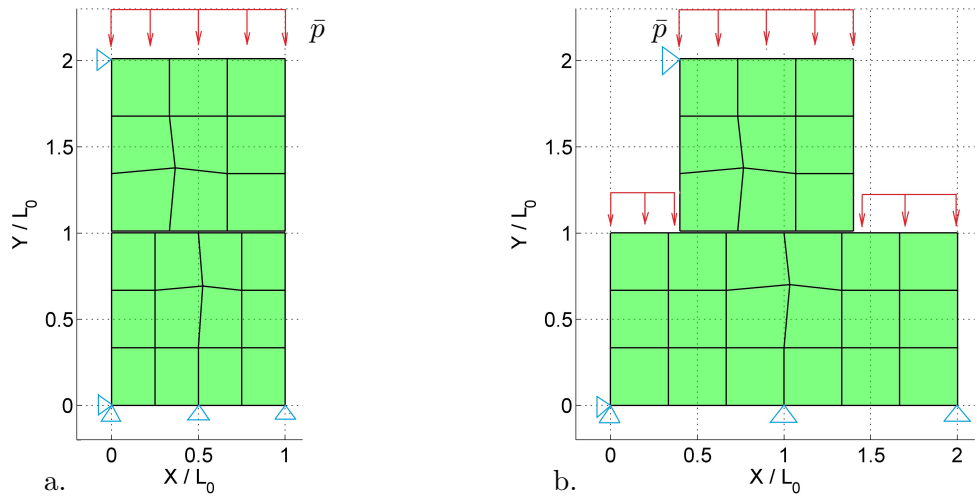
\vspace{-0.3cm}
This section presents two-dimensional patch tests 
for various contact formulations including GPTS  full-pass (GPFP), standard mortar two-half-pass (SM2HP),  extended mortar full-pass (XMFP) and extended mortar two-half-pass (XM2HP).
\textcolor{black}{The standard mortar full-pass formulation (SMFP) has already been demonstrated to pass the contact patch test  \citep{McDevitt2000,puso04a,yang05,Apop2012,Kim12,Dittmann14}. Also the GPTS two-half-pass formulation (GP2HP) has already been shown to pass the contact patch test \citep{spbc}. } 
 \\[1.5mm]
%
%
%
Two test cases are examined as is shown in Fig.~\ref{f:patchINI}. 
Test case~1 (Fig.~\ref{f:patchINI}a) is used to investigate  quadrature errors due to non-conforming meshes (see Sec.~\ref{s:case2}). 
Test case~2 (Fig.~\ref{f:patchINI}b) is used to assesses the effectiveness of RBQ in both the GPTS and the proposed mortar formulation (see Sec.~\ref{s:case3}). The RBQ technique is employed in test case 2 since there are contact elements that are partially in contact. \textcolor{black}{All simulations are based on the penalty method and frictionless contact.}

%
%

\subsection{Contact patch \textcolor{black}{test case 1}: non-conforming mesh}\label{s:case2}

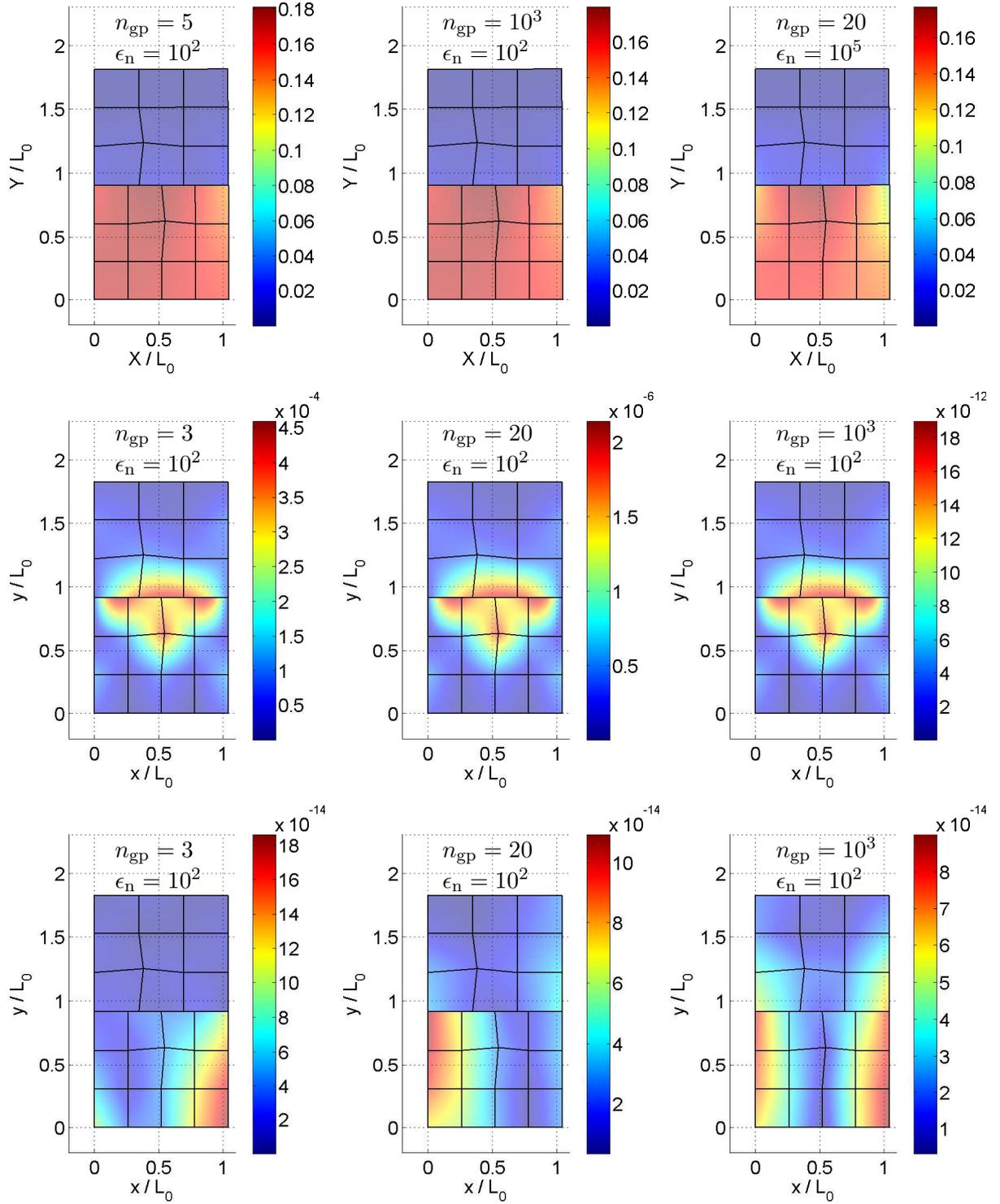
\begin{figure}[!htp]
\begin{center} \unitlength1cm
\begin{picture}(0,20.1)

\put(-7.9,13.5){\includegraphics[height=68mm]{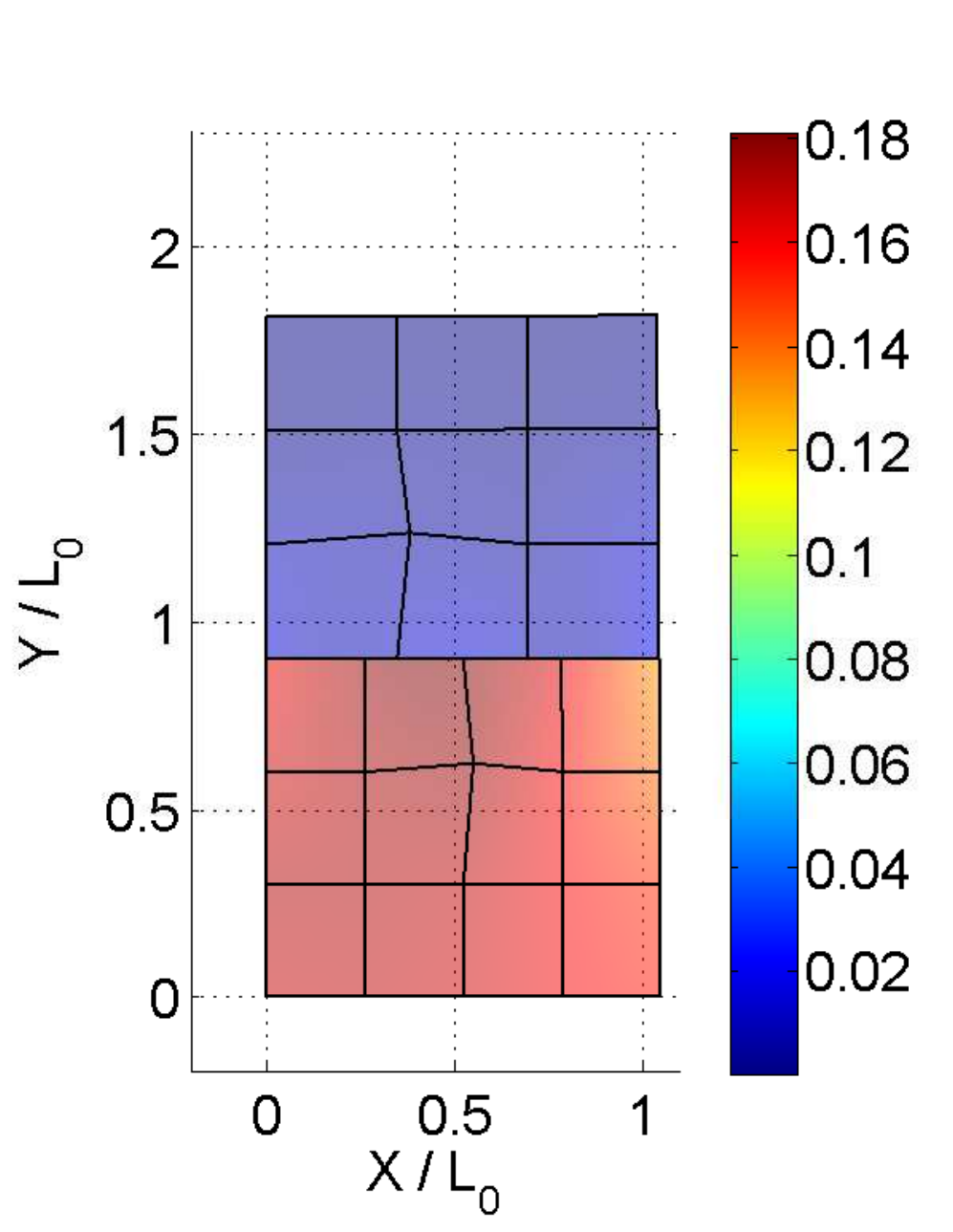}}
\put(-2.5,13.5){\includegraphics[height=68mm]{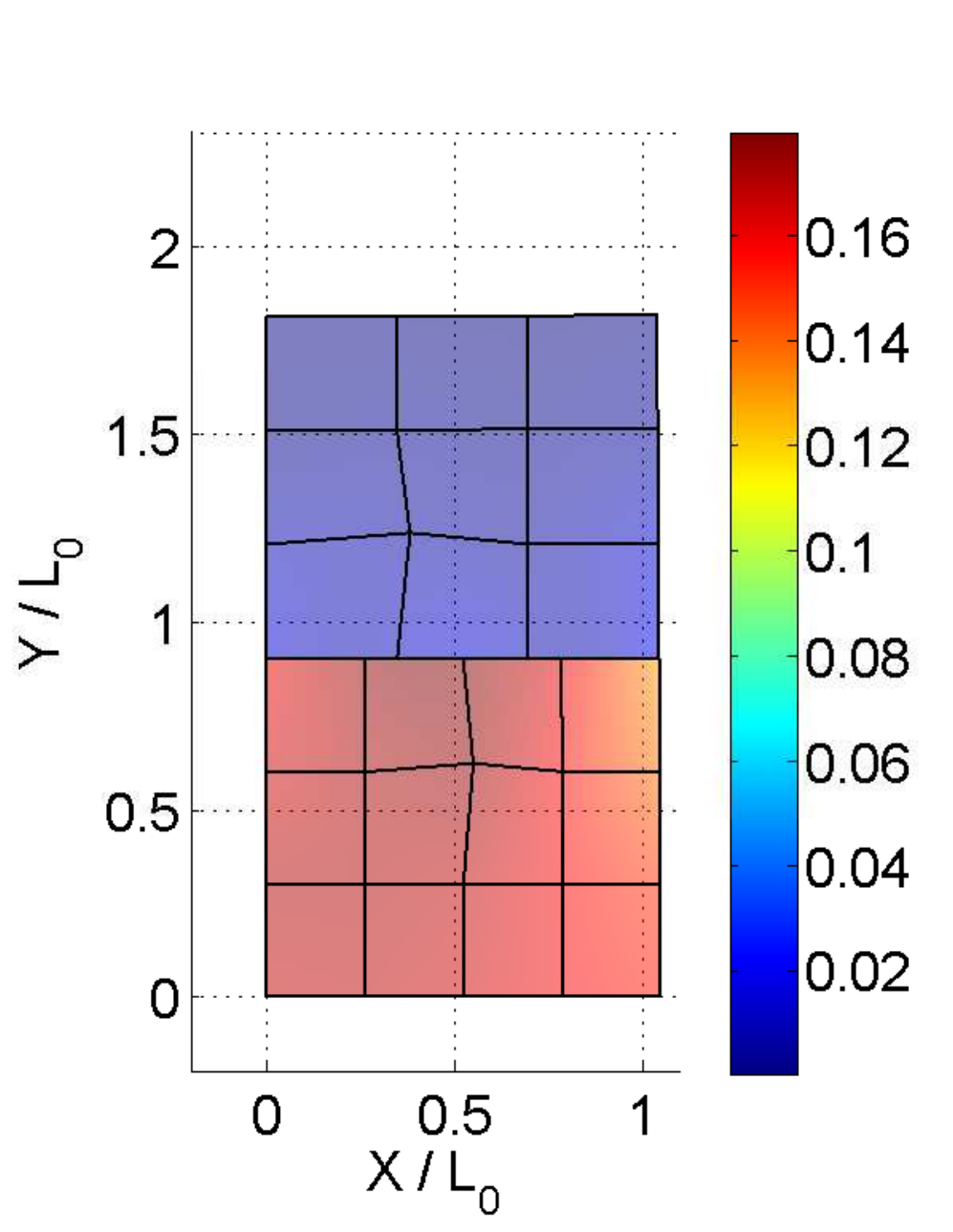}}
\put(2.8,13.5){\includegraphics[height=68mm]{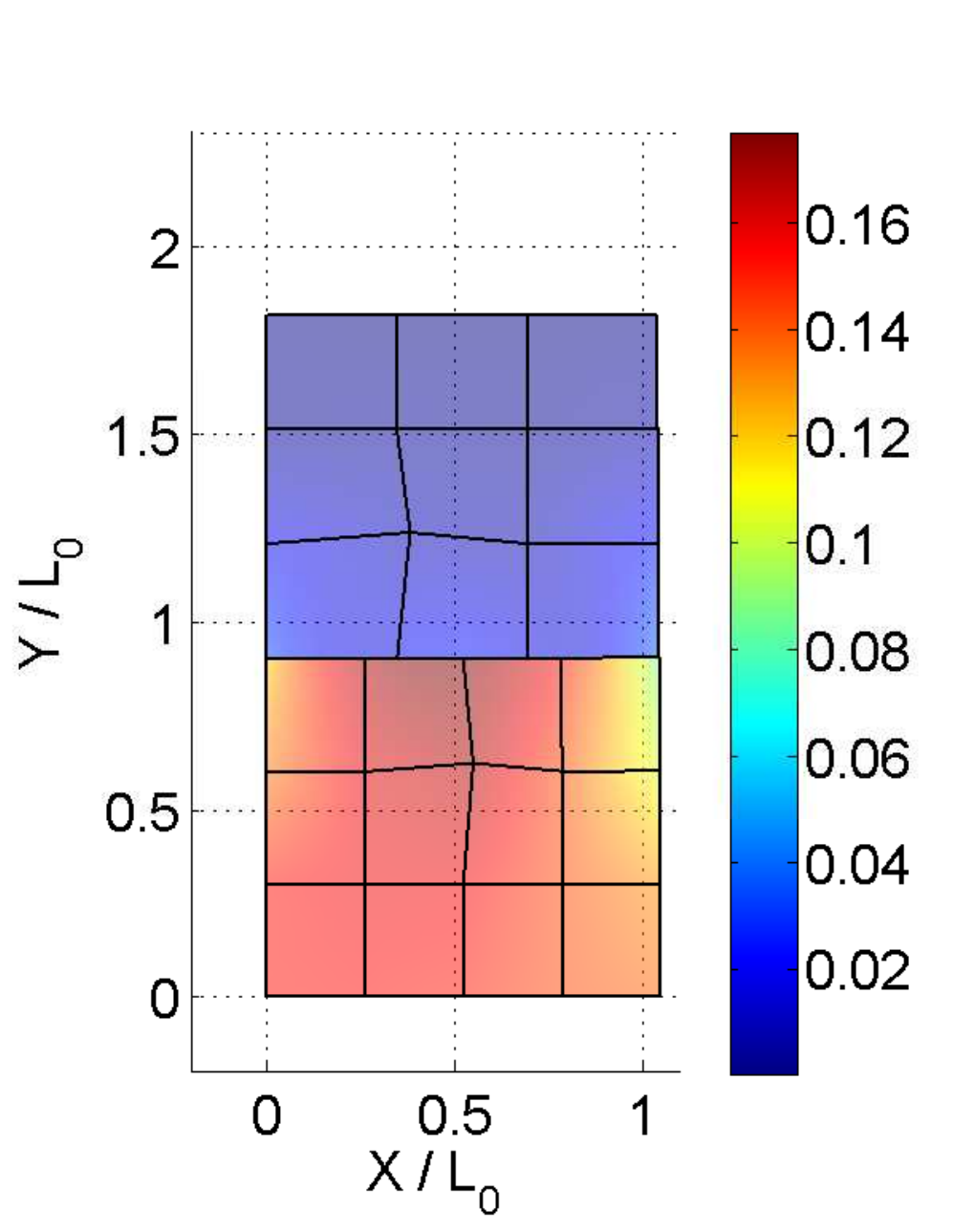}} 

\put(-7.9,6.75){\includegraphics[height=68mm]{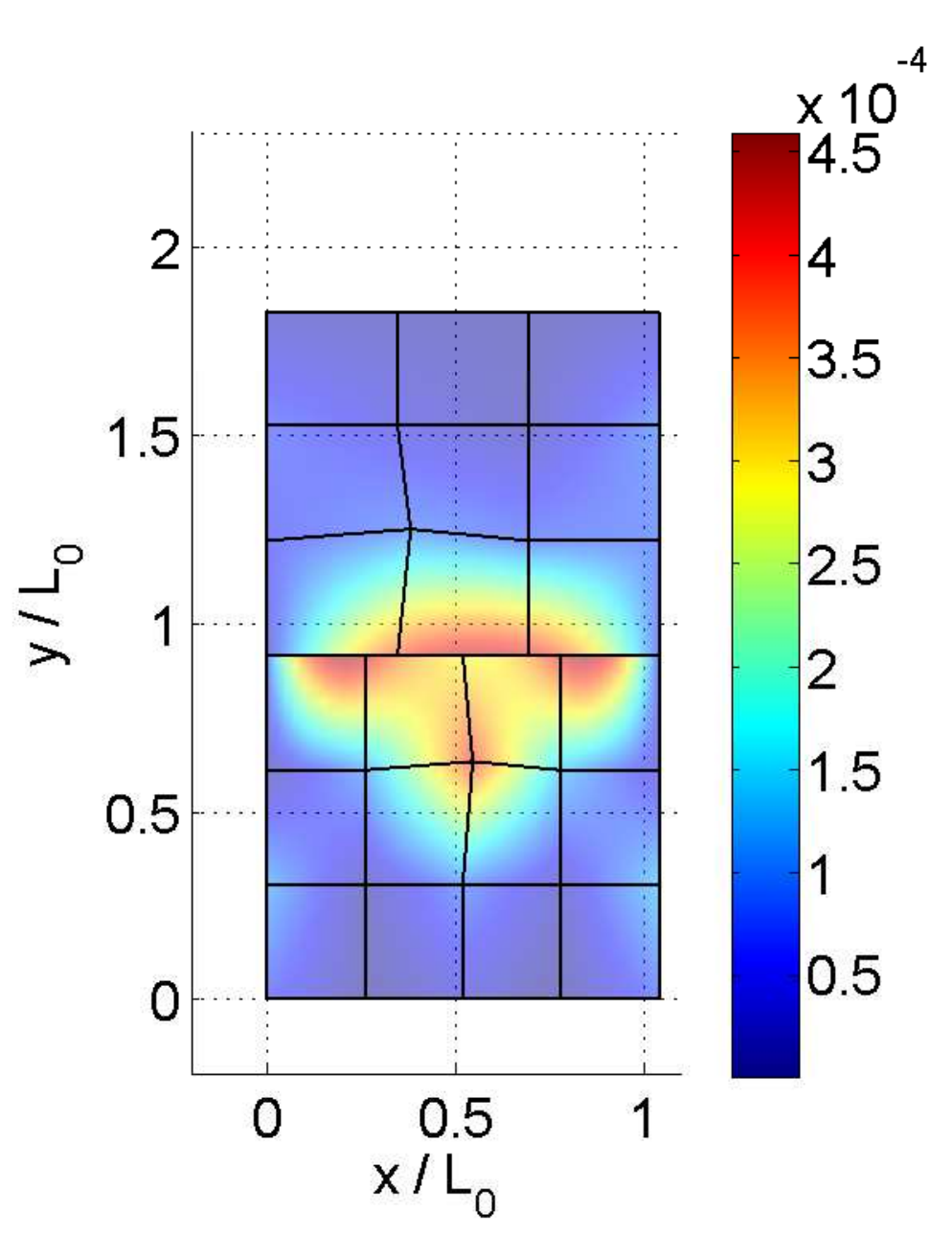}}
\put(-2.5,6.75){\includegraphics[height=68mm]{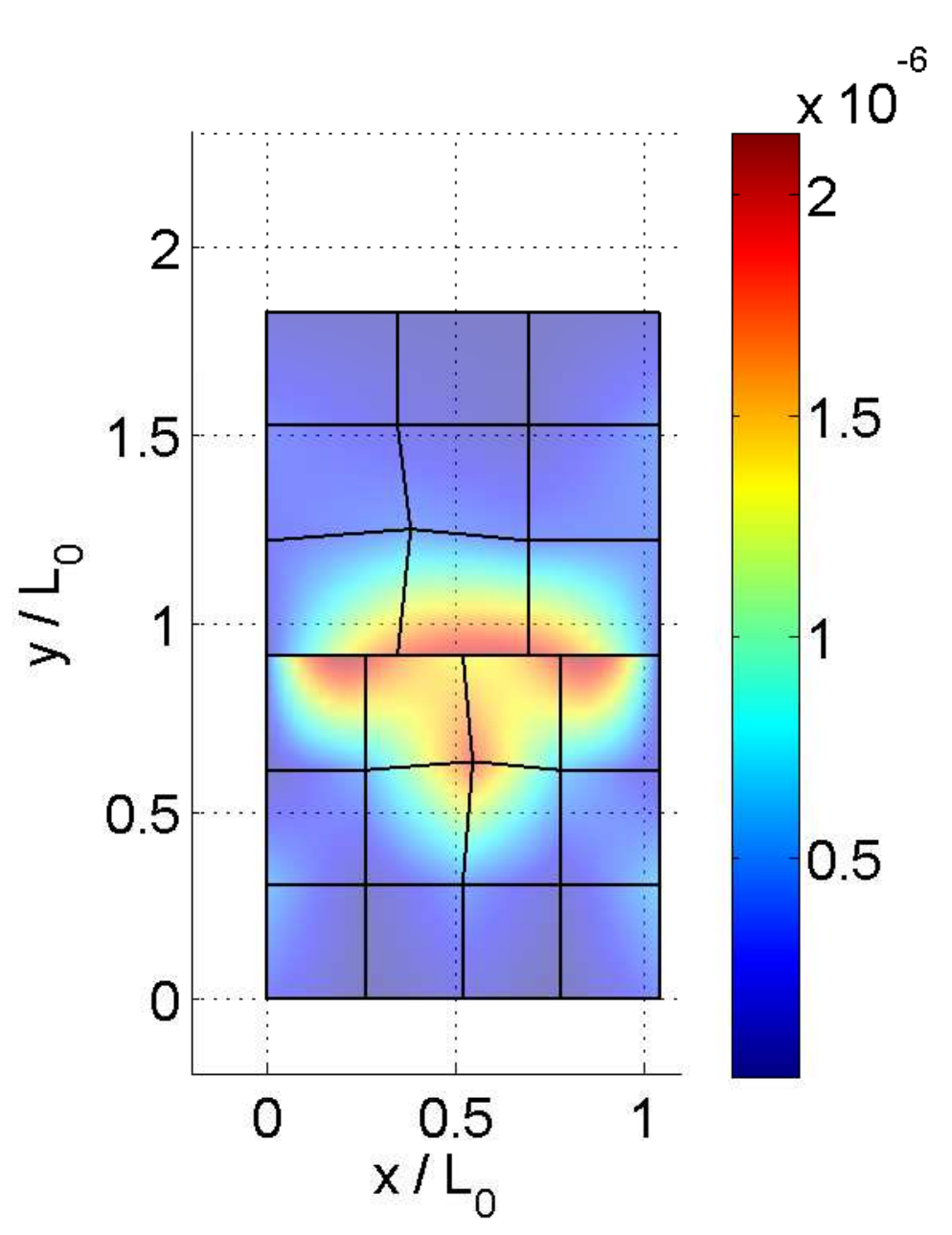}}
\put(2.8,6.75){\includegraphics[height=68mm]{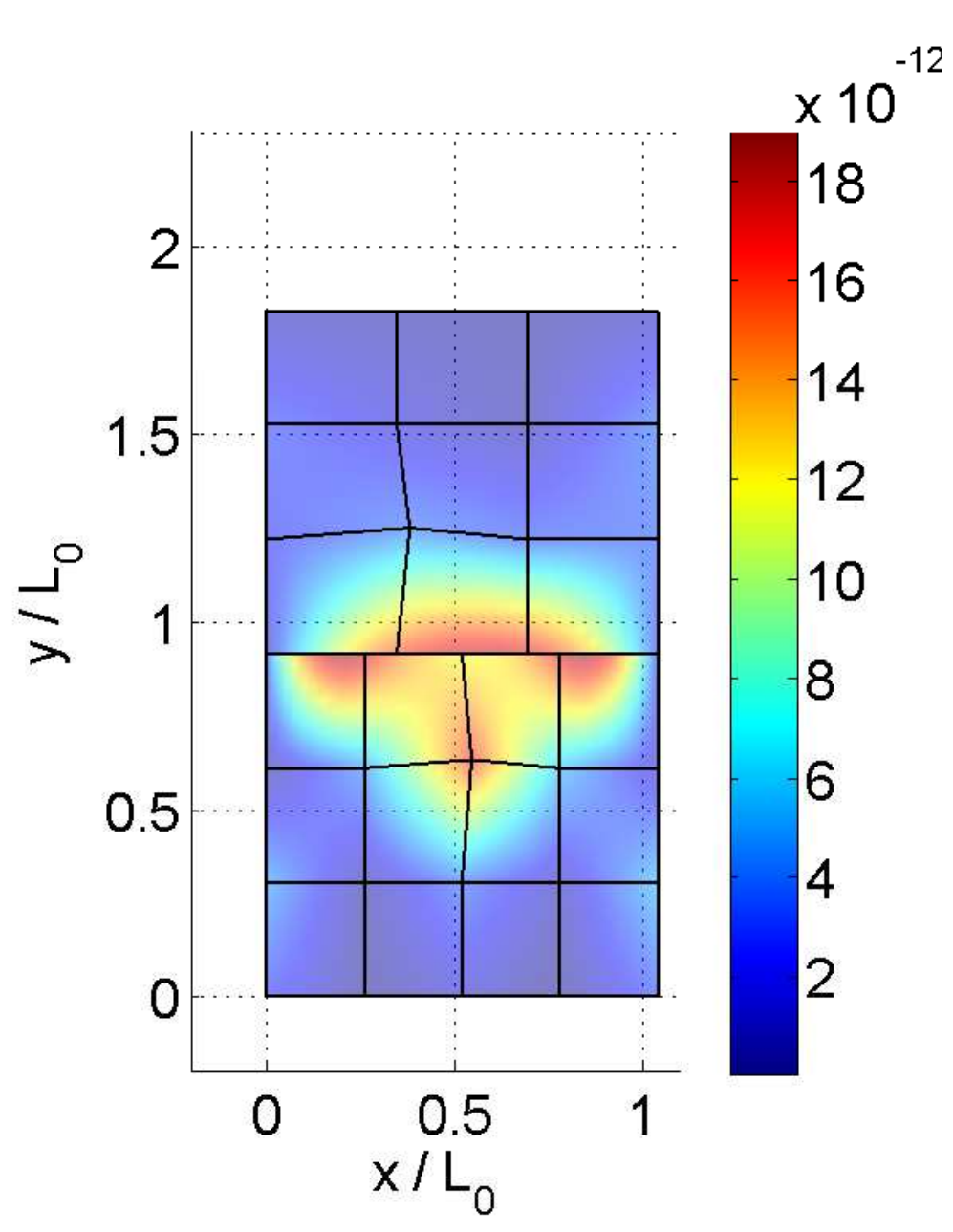}} 

\put(-7.9,0){\includegraphics[height=68mm]{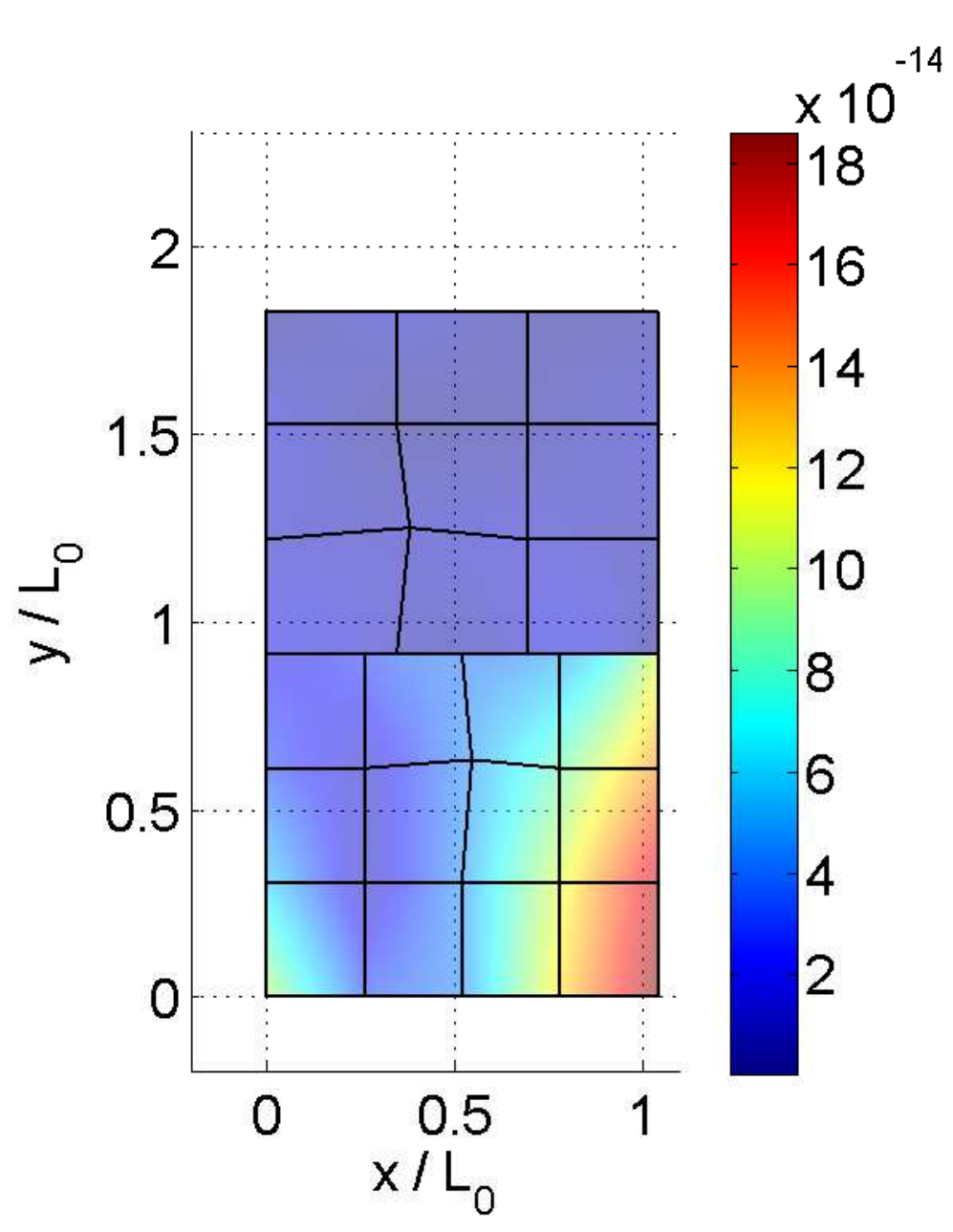}}
\put(-2.5,0){\includegraphics[height=68mm]{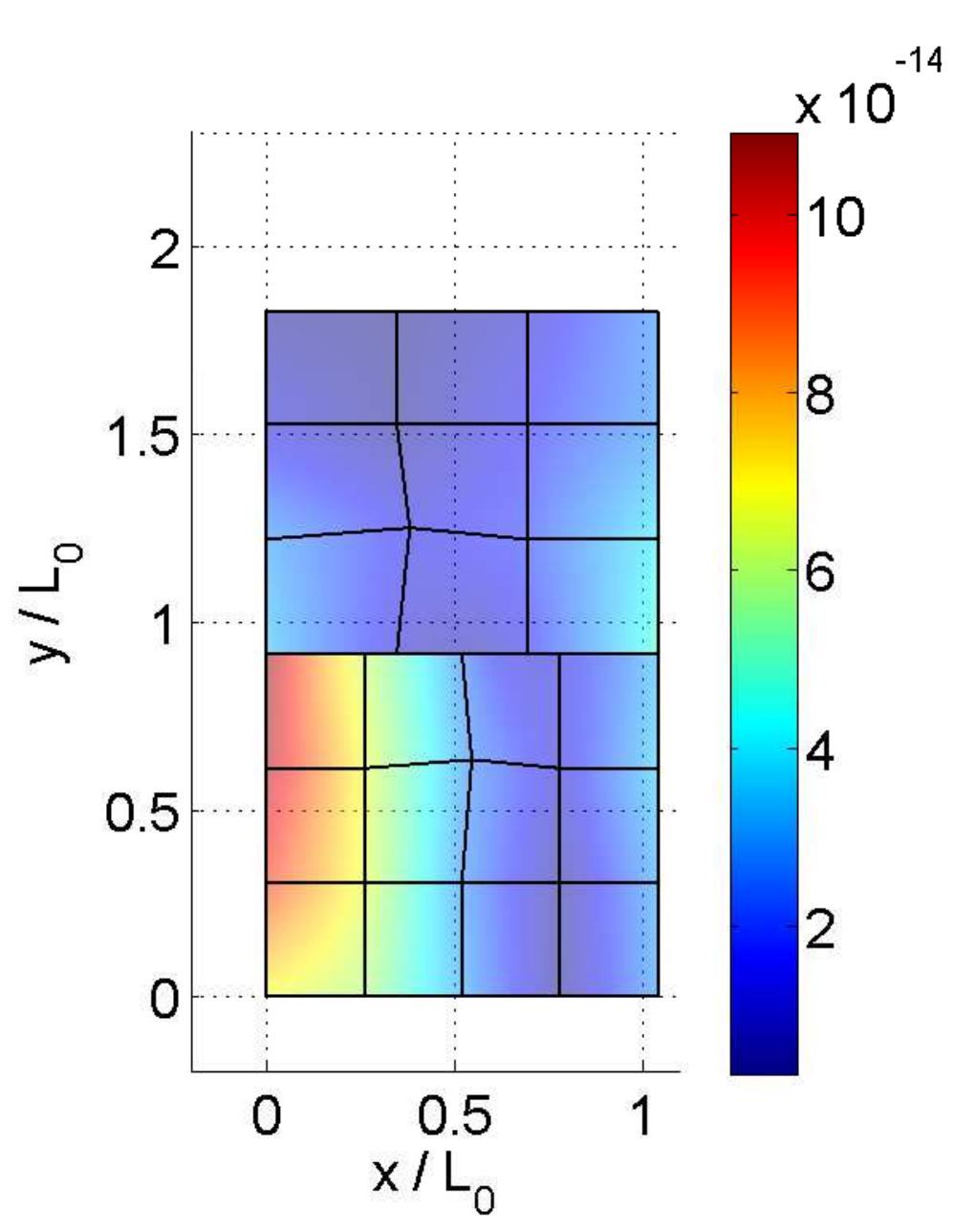}}
\put(2.8,0){\includegraphics[height=68mm]{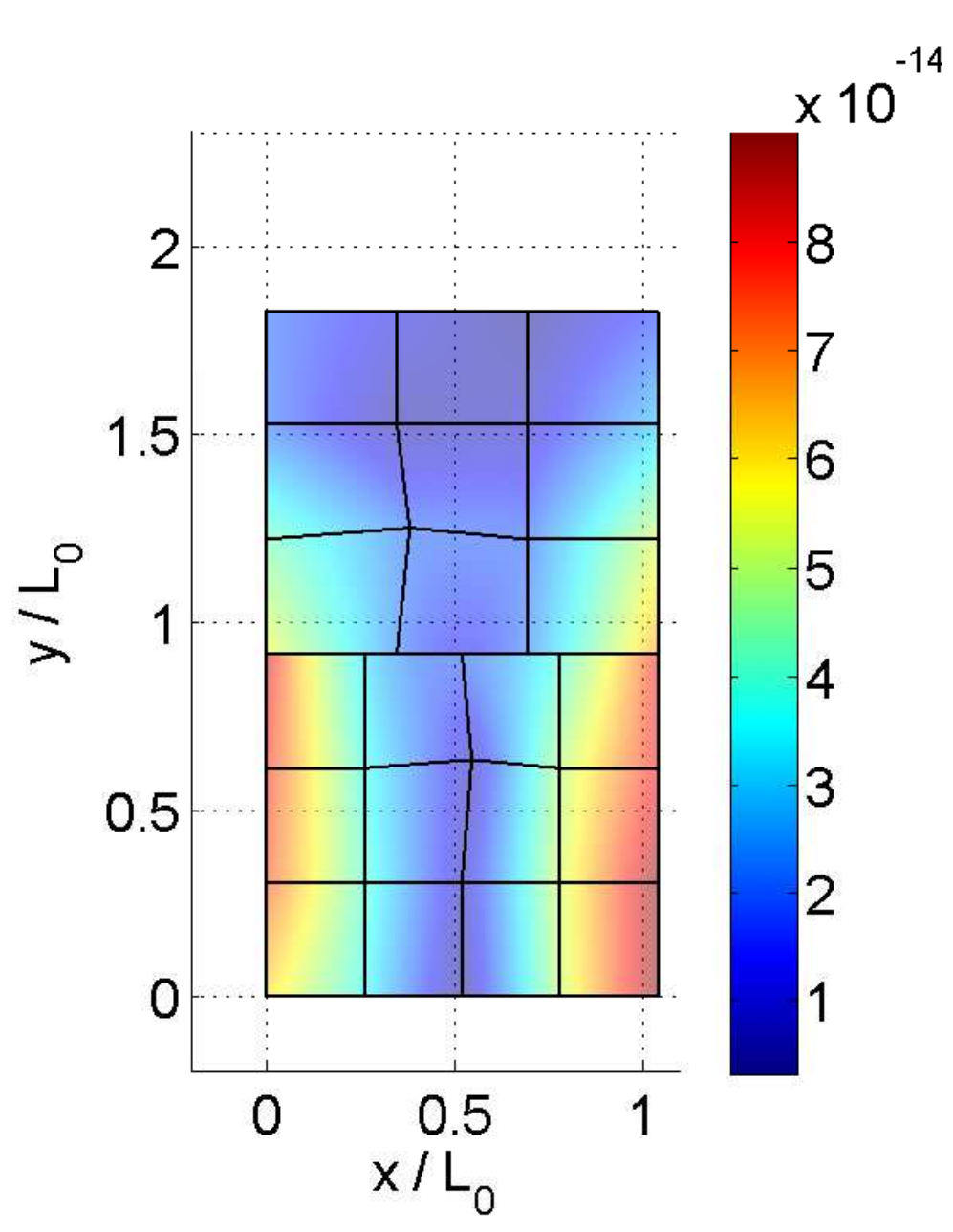}} 




\put(-6.1,19.2){$n_\mathrm{gp}=5$}
\put(-6.1,18.7){$\epsilon_\mrn = 10^2$}

\put(-0.8,19.2){$n_\mathrm{gp}=10^3$}
\put(-0.8,18.7){$\epsilon_\mrn = 10^2$}

\put(4.6,19.2){$n_\mathrm{gp}=20$}
\put(4.6,18.7){$\epsilon_\mrn = 10^5$}
\put(-6.1,12.5){$n_\mathrm{gp}=3$ }
\put(-6.1,12){$\epsilon_\mrn=10^2$ }

\put(-0.8,12.5){$n_\mathrm{gp}=20$ }
\put(-0.8,12){$\epsilon_\mrn=10^2$ }

\put(4.6,12.5){$n_\mathrm{gp}=10^3$ }
\put(4.6,12){$\epsilon_\mrn=10^2$ }
\put(-6.1,5.7){$n_\mathrm{gp}=3$}
\put(-6.1,5.2){$\epsilon_\mrn=10^2$ }
\put(-0.8,5.7){$n_\mathrm{gp}=20$ }
\put(-0.8,5.2){$\epsilon_\mrn=10^2$ }
\put(4.6,5.7){$n_\mathrm{gp}=10^3$ }
\put(4.6,5.2){$\epsilon_\mrn=10^2$ }

\end{picture}
\caption{Patch test \textcolor{black}{case~1} (non-conforming meshes): Standard mortar two-half-pass \textbf{(SM2HP)} (top row) versus  extended mortar contact formulations with full-pass \textbf{(XMFP)} (middle row) and two-half-pass \textbf{(XM2HP)} (bottom row) varying  $n_\mathrm{gp}$ and $\epsilon_\mrn\,[E/L]$. 
The color shows relative errors in the vertical stress. } 
\label{f:case2_XFEMmt}
\end{center}
\end{figure}

\begin{figure}[!htp]
\begin{center} \unitlength1cm
\unitlength1cm
\setlength{\belowcaptionskip}{-0.4cm}
\begin{picture}(0,5.1)

\put(-8.0,-0.1){\includegraphics[width=75mm]{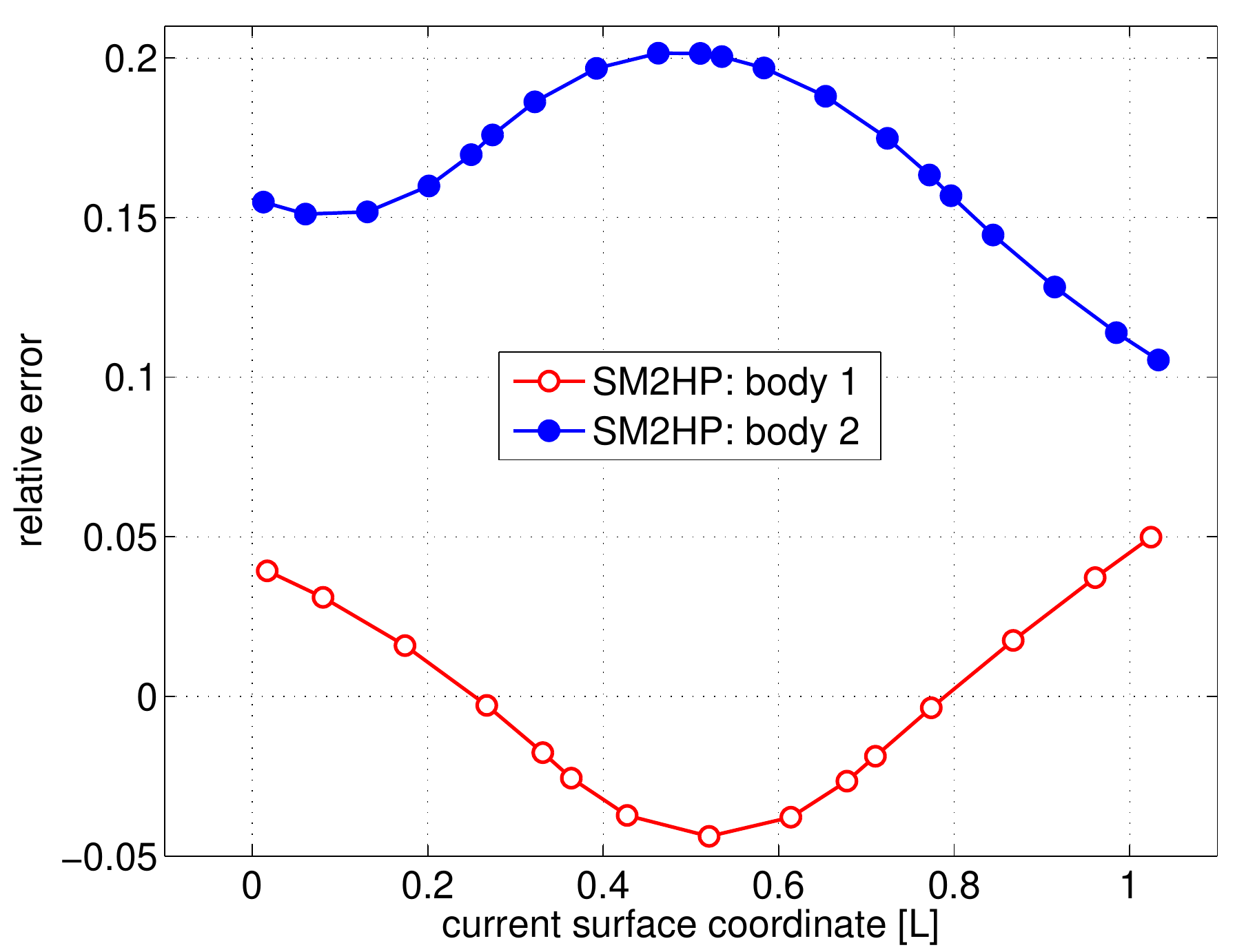}}
\put(0.5,-0.1){\includegraphics[width=70mm]{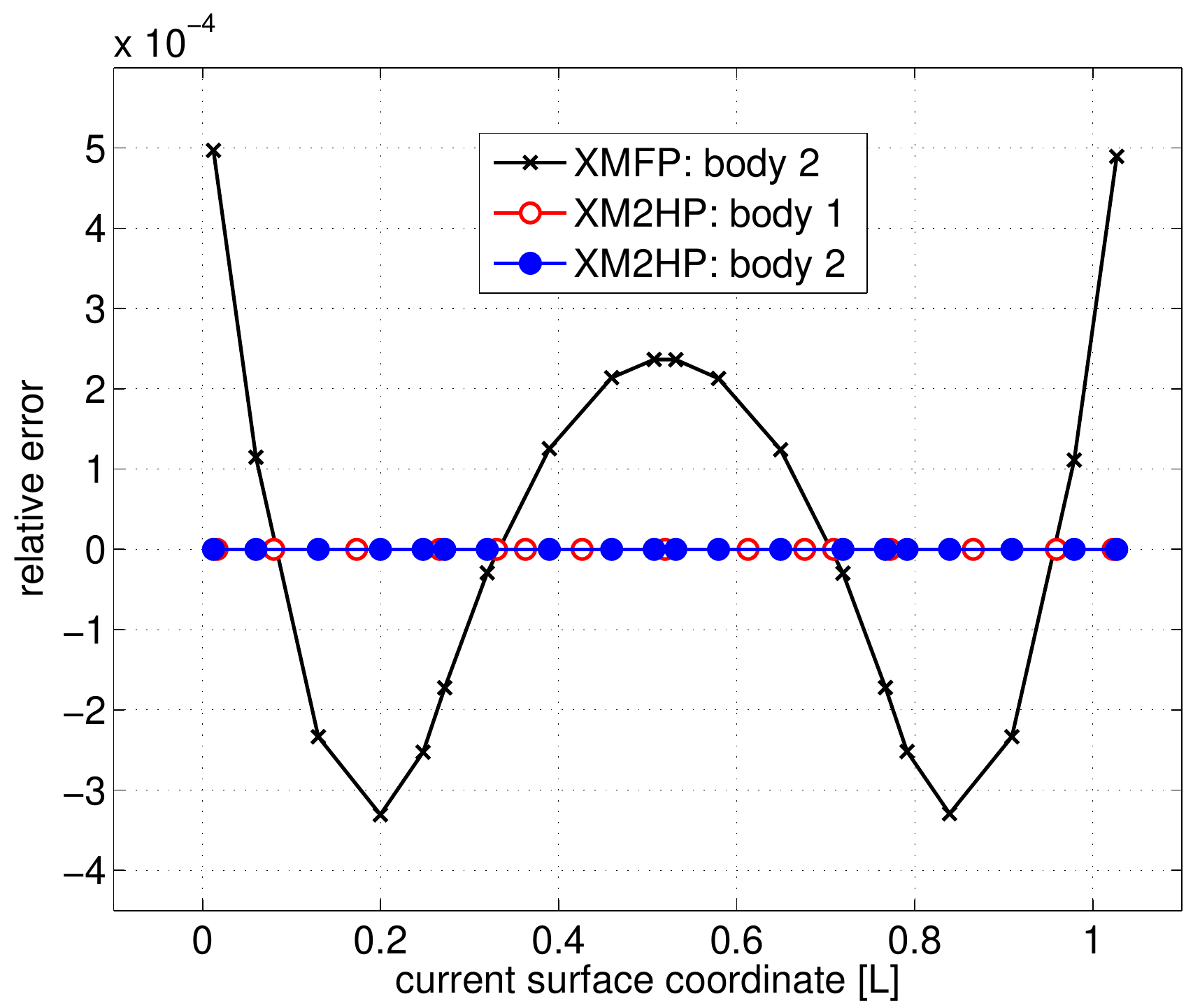}}

\put(-7.65,-0.1){a. }
\put(.8,-0.1){b.}

\end{picture}
\caption{Patch test \textcolor{black}{case~1} (non-conforming meshes): relative error of the   (true) mortar contact pressure \textcolor{black}{ (given in Eqs.~\eqref{e:pdual} and \eqref{e:rawpesstrue})} w.r.t.~the exact solution along the contact interface for a.~standard mortar two-half-pass (\textbf{SM2HP}) \textcolor{black}{versus b.~extended mortar full-pass (\textbf{XMFP}) and two-half-pass (\textbf{XM2HP})}.  Here, $n_\mathrm{gp}=5$,  $\epsilon_\mrn = 10^2~[E/L]$ is kept fixed \textcolor{black}{and no quadrature segmentation is used here}. The error of XM2HP is of the order of $10^{-12}$.} 
\label{f:case2_XFEMmt_press}
\end{center}
\end{figure}

For non-conforming meshes,  \textcolor{black}{errors} due to the mesh dissimilarity  appear and affect the contact formulations.  \textcolor{black}{Fig.~\ref{f:case2_XFEMmt} shows the relative error in  the vertical stress  while varying the number of quadrature points  $n_\mathrm{gp}$.}
%
\textcolor{black}{We observe that  SM2HP cannot pass this patch test (see the top row of Fig.~\ref{f:case2_XFEMmt}).} A detailed investigation of the contact interface, as shown in Fig.~\ref{f:case2_XFEMmt_press}a, reveals  for SM2HP a significant error in the local momentum balance due to the mesh dissimilarities. \\[1.5mm]
\textcolor{black}{On the other hand,} as the middle row of Fig.~\ref{f:case2_XFEMmt} shows, XMFP can pass the patch test  \textcolor{black}{only} within quadrature \textcolor{black}{accuracy. This implies}  that the quadrature segmentation in the contact domain plays a significant role in XMFP \textcolor{black}{(as is also seen in Fig.~\ref{f:case3_XFEMmt2})}.   Besides, the  two-half-pass version, XM2HP, can pass the test even for a small number of quadrature points (see the bottom row of Fig.~\ref{f:case2_XFEMmt} and Fig.~\ref{f:case2_XFEMmt_press}b). Therefore, this confirms that XM2HP indeed works as a segmentation-free formulation.


\subsection{Contact patch \textcolor{black}{test case 2}:  generalized case}\label{s:case3}
\begin{figure}[!hp]
\begin{center} \unitlength1cm
\unitlength1cm
\begin{picture}(0,21.1)

\put(-8.0,14.0){\includegraphics[width=75mm]{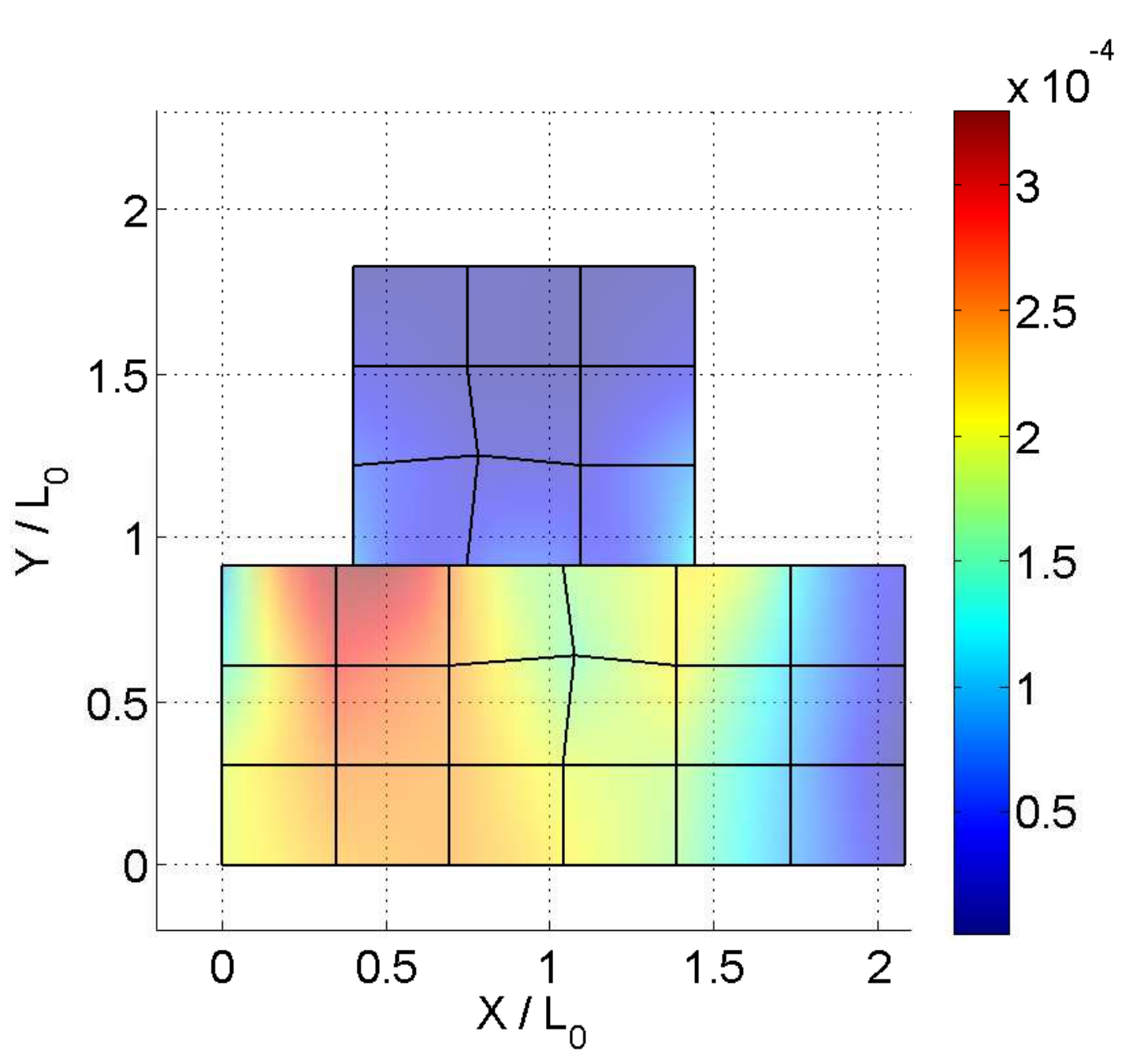}}
\put(0.0,14.0){\includegraphics[width=75mm]{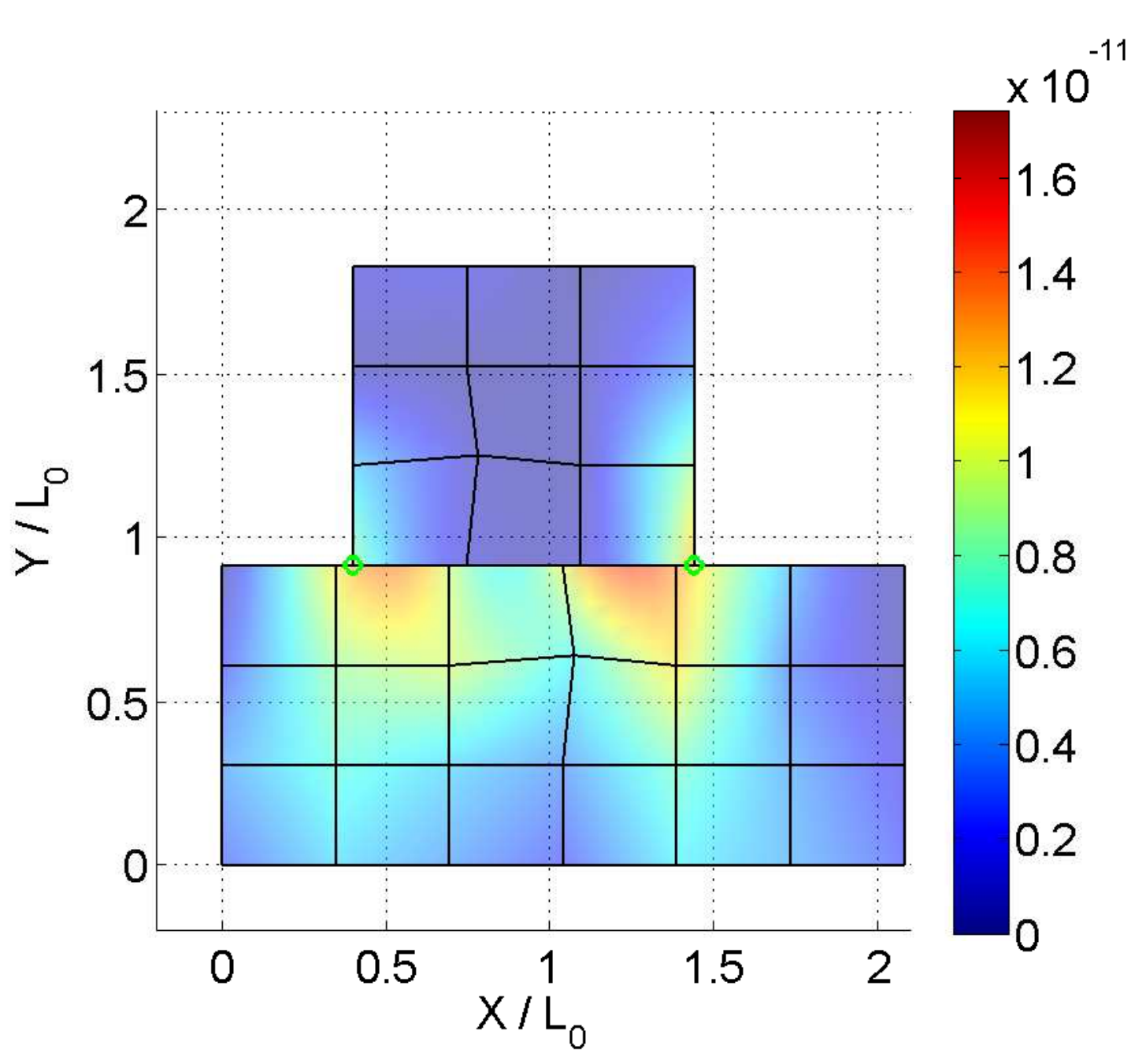}}

\put(-8.0,7.0){\includegraphics[width=75mm]{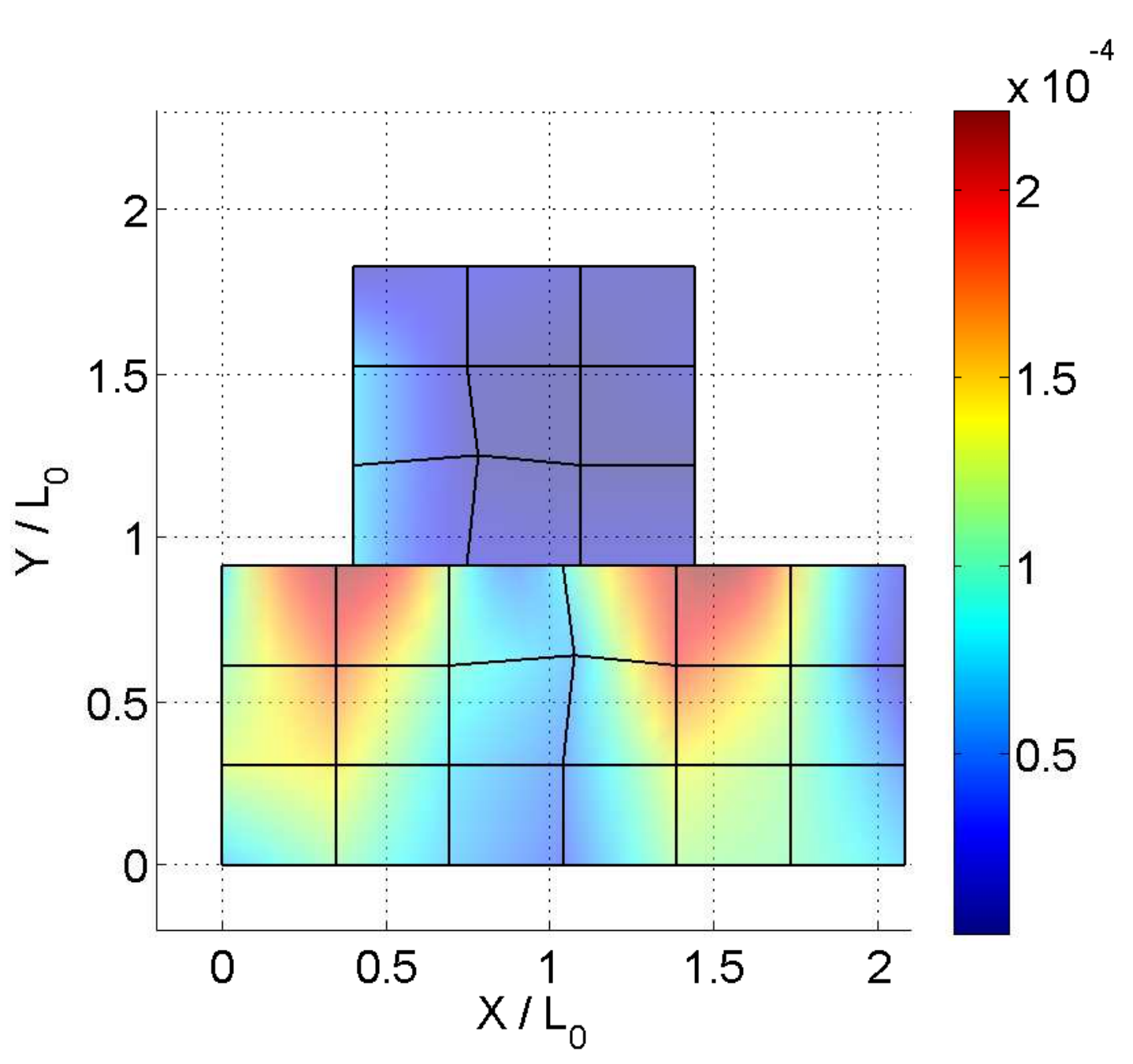}}
\put(0.0,7.0){\includegraphics[width=75mm]{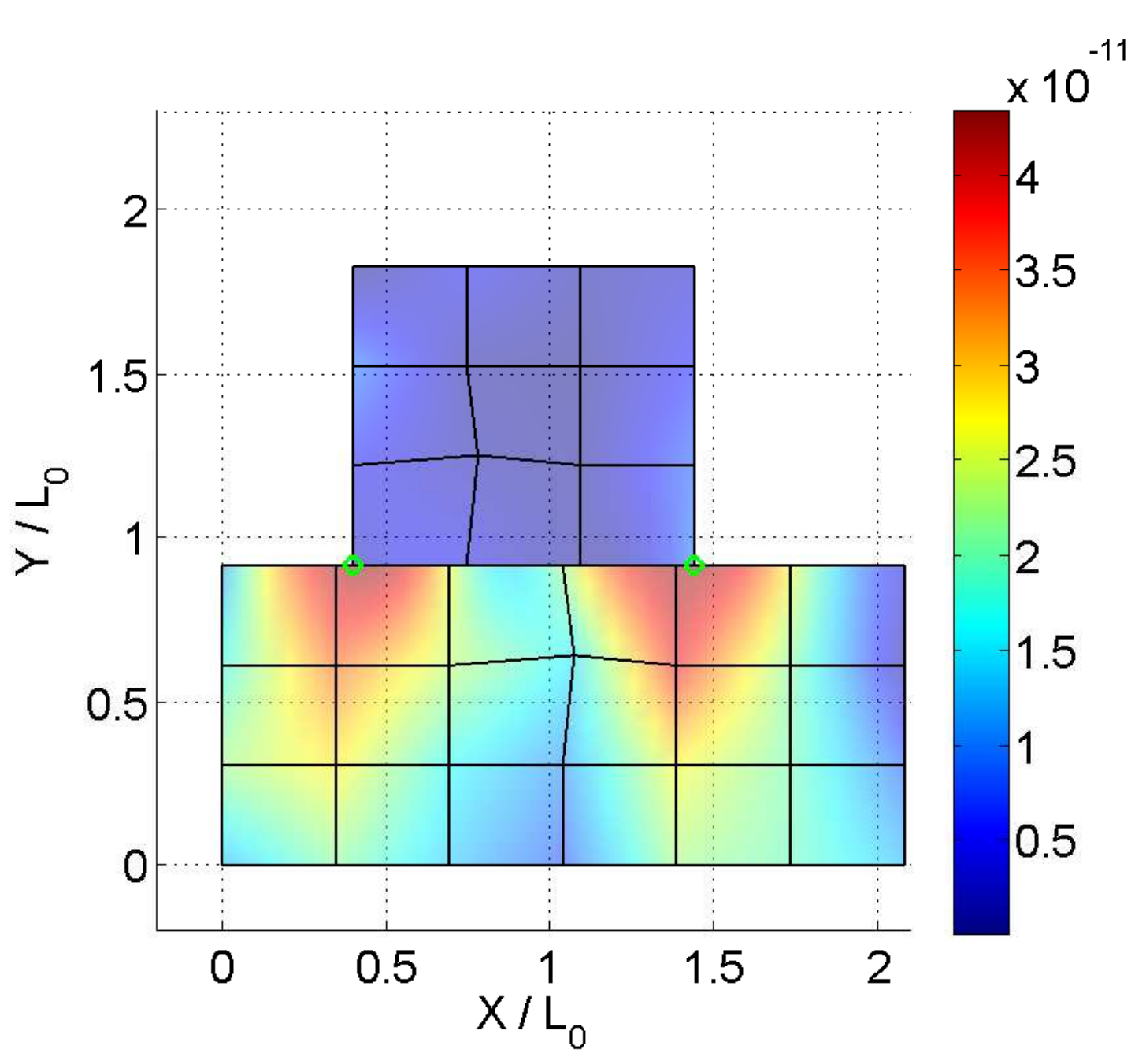}}

\put(-8.0,0){\includegraphics[width=75mm]{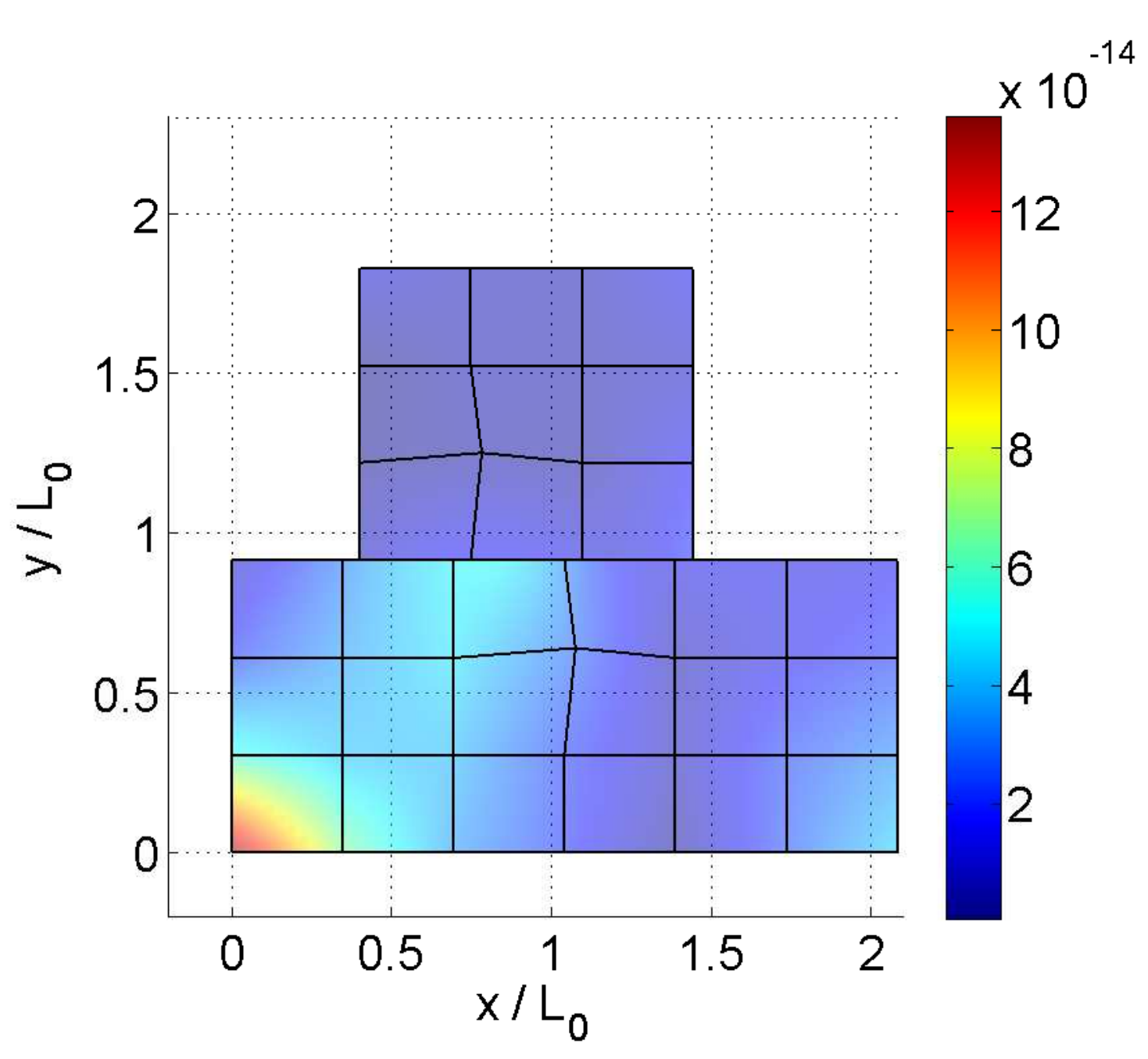}}
\put(0.0,0){\includegraphics[width=75mm]{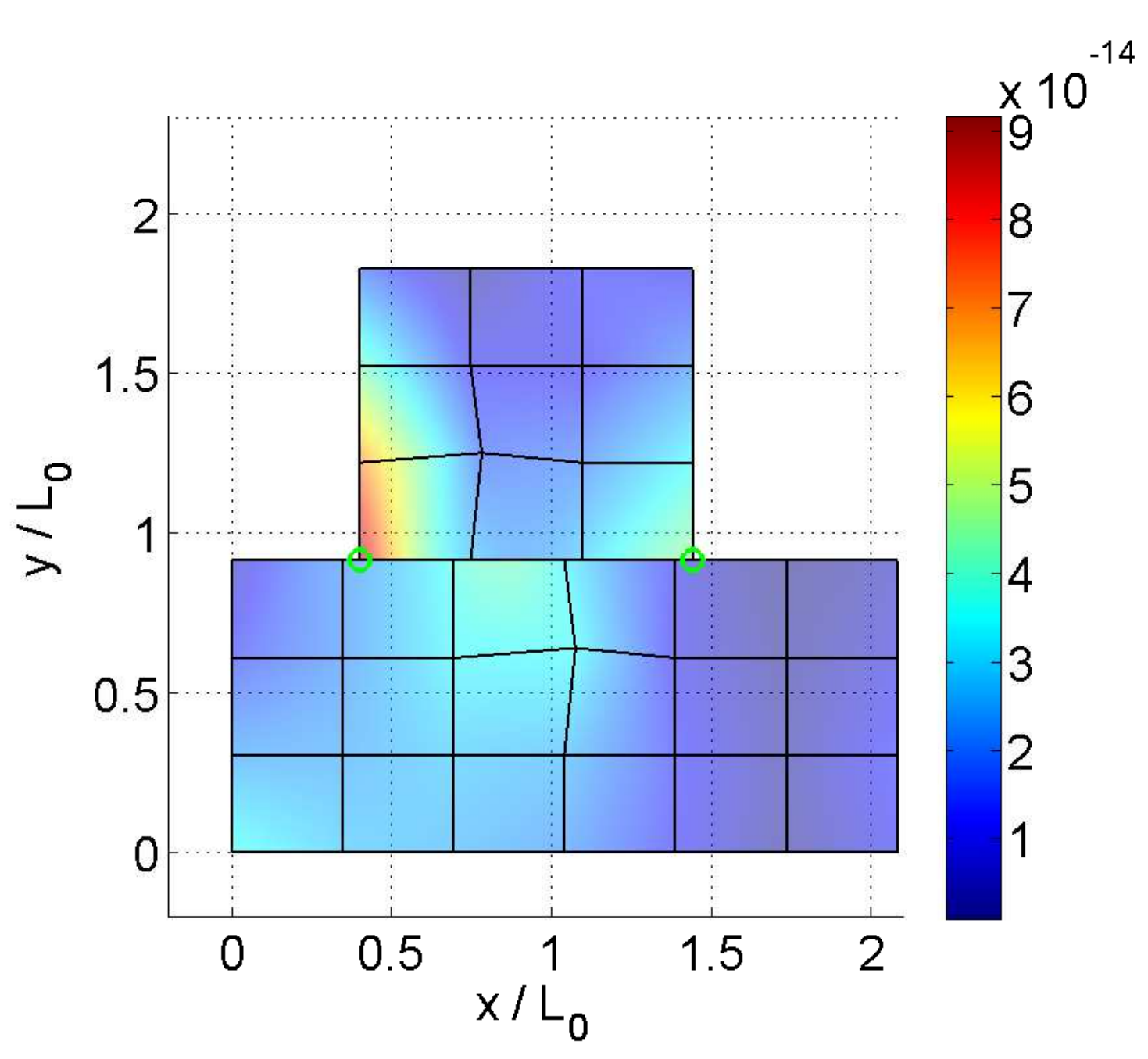}}

\put(-5.2,19.95){GPFP }
\put(-5.5,19.45){$n_\mathrm{gp}=10^3$ }

\put(2.3,19.9){GPFP--RBQ}
\put(2.5,19.4){$n_\mathrm{gp}=10^3$ }

\put(-5.2,13){XMFP }
\put(-5.5,12.5){$n_\mathrm{gp}=10^3$ }

\put(2.3,12.9){XMFP--RBQ}
\put(2.5,12.4){$n_\mathrm{gp}=10^3$ }

\put(-5.4,5.9){XM2HP}
\put(-5.4,5.4){$n_\mathrm{gp}=5$ }

\put(2.3,5.9){XM2HP--RBQ}
\put(2.5,5.4){$n_\mathrm{gp}=5$}

\end{picture}
\caption{Patch test \textcolor{black}{case~2} (generalized case): rows from top to bottom consider Gauss-point-to-segment full-pass (\textbf{GPFP}), and extended mortar with full-pass (\textbf{XMFP}) and two-half-pass (\textbf{XM2HP}).  Columns consider the quadrature without RBQ (left) and with RBQ (right). The green circles show the contact boundary of the lower block as it is obtained by the RBQ algorithm. $\epsilon_\mrn = \textcolor{black}{10^2}~[E/L]$. The color shows relative errors in the vertical stress. 
} 
\label{f:case3_XFEMmt}
\end{center}
\end{figure}

\begin{figure}[!hpb]
\begin{center} \unitlength1cm
\unitlength1cm
\begin{picture}(0,6.5)

\put(-8.0,0){\includegraphics[width=75mm]{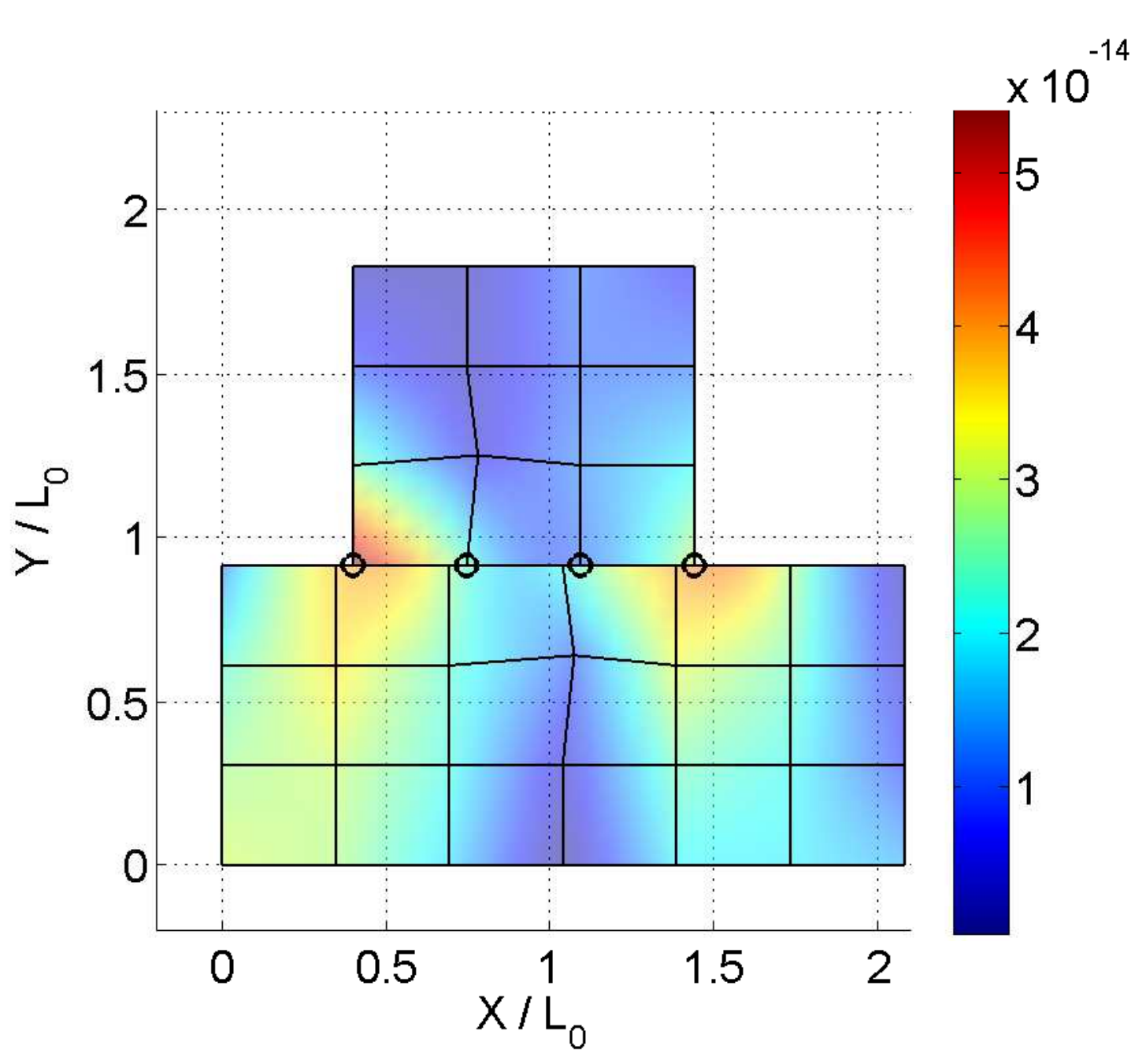}}
\put(0.0,0){\includegraphics[width=75mm]{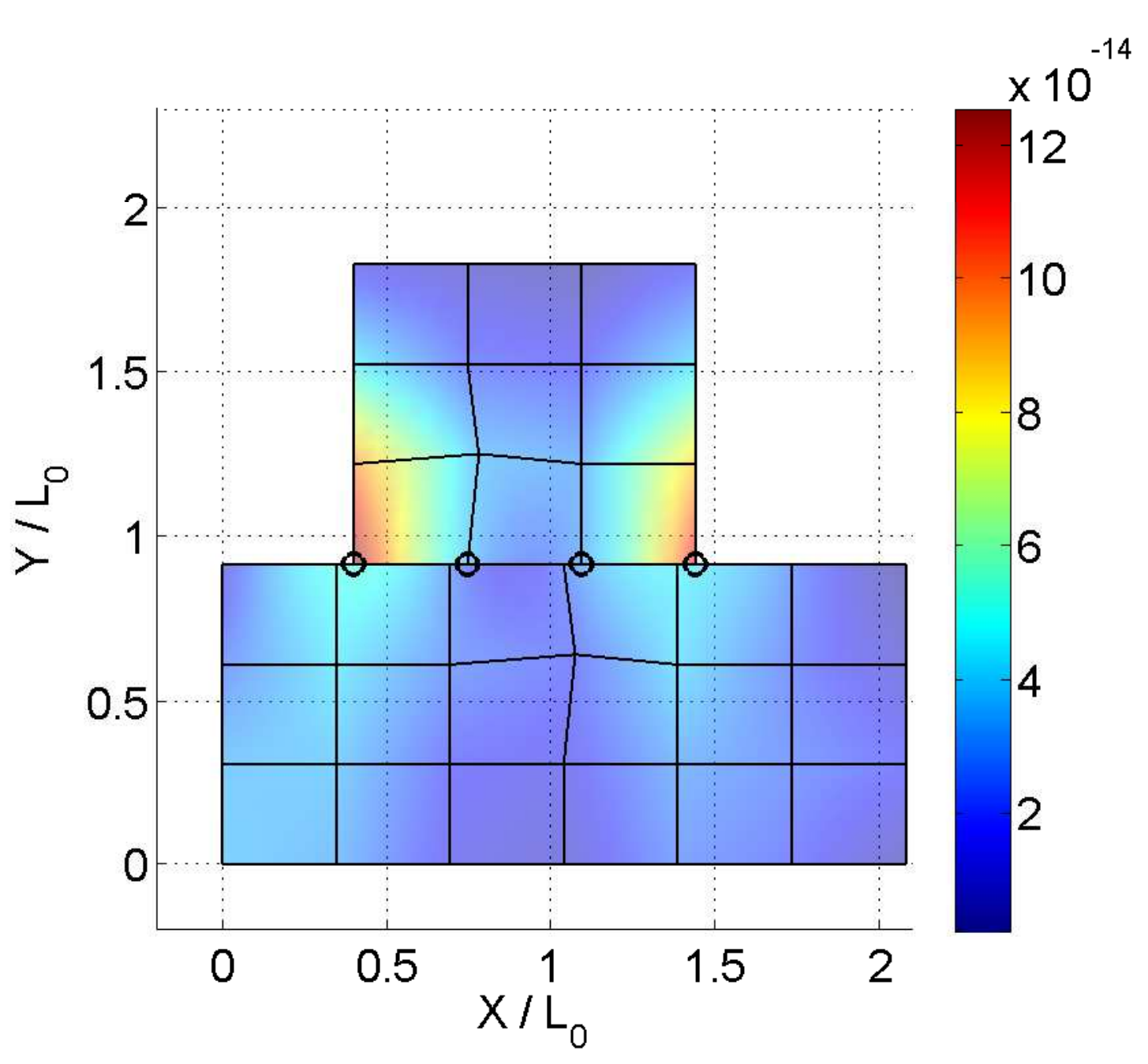}}

\put(-5.4,5.9){GPFP--segm.}
\put(-5.4,5.4){$n_\mathrm{gp}=5$ }

\put(2.3,5.9){XMFP--segm.}
\put(2.5,5.4){$n_\mathrm{gp}=5$}

\end{picture}
\caption{\textcolor{black}{Patch test case~2 (generalized case) with quadrature segmentation for Gauss-point-to-segment full-pass (\textbf{GPFP}) (left), and extended mortar full-pass (\textbf{XMFP}) (right).  The slave contact surface is from the lower body.  The black circles indicate segmented positions for numerical quadrature. $\epsilon_\mrn = 10^2~[E/L]$. The color shows relative errors in the vertical stress. }
} 
\label{f:case3_XFEMmt2}
\end{center}
\end{figure}

\textcolor{black}{This} patch test examines quadrature errors at elements that are partially in contact. We focus on the influence of RBQ \citep{rbq} on GPFP, XMFP, and XM2HP. \\[1.5mm]
RQB is distinct from \textcolor{black}{quadrature} segmentation (see e.g.~\citet{puso04a}), since RQB is based on an adaptive partitioning along the contact boundary on partial contact elements \citep{rbq}. The efficiency of RBQ in comparison with segmentation thus depends on the ratio of the number of partial contact elements to the total number of contact elements. \\[1.5mm]
As seen in Fig.~\ref{f:case3_XFEMmt}, without \textcolor{black}{special quadrature treatments,} both XMFP and GPFP cannot pass the generalized patch test even when $1000$ quadrature points per element are used.  \textcolor{black}{With RBQ,  XMFP and GPFP can reach high accuracy. In order to reach machine precision, quadrature segmentation of interior contact domain is additionally required as is shown in Fig.~\ref{f:case3_XFEMmt2}.}\\[1.5mm]
In contrast to this, XM2HP \textcolor{black}{can pass the patch test} with only $5$ Gauss points per element (both for the finite element force and tangent evaluation) without requiring any \textcolor{black}{special quadrature treatment (see the bottom row of Fig.~\ref{f:case3_XFEMmt}).} 
\vspace{-0.3cm}

\section{Numerical examples}\label{s:numex}

This section presents \textcolor{black}{four} examples to illustrate the robustness and accuracy of the extended mortar formulation developed in Sec.~\ref{s:XFEM}. We will compare extended mortar with standard mortar and classical GPTS methods. For all contact examples, quadratic NURBS will be used for the contact surfaces. \textcolor{black}{A} (plane strain) Neo-Hookean material model (see e.g.~\citet{ogden}) is used for (2D) 3D solids, respectively. For shells, we employ the Koiter model of \citet{solidshell}.

\subsection{2D indentation} \label{sec:cylexample}
\begin{figure}[!htp]
\begin{center} \unitlength1cm
\begin{picture}(0,10.2)
\put(-8.1,5.5){\includegraphics[height=52mm]{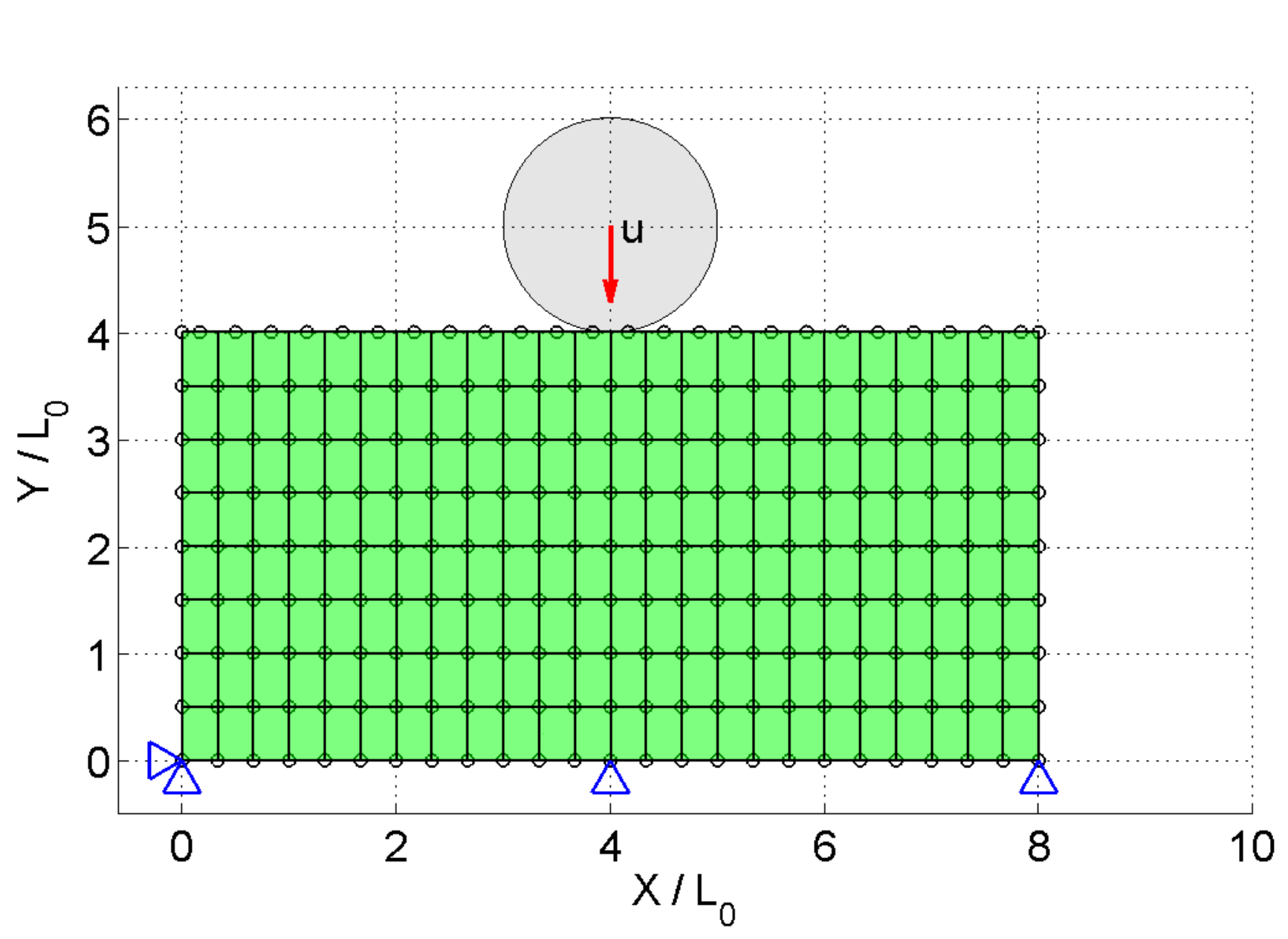}}
\put(0.1,5.5){\includegraphics[height=52mm]{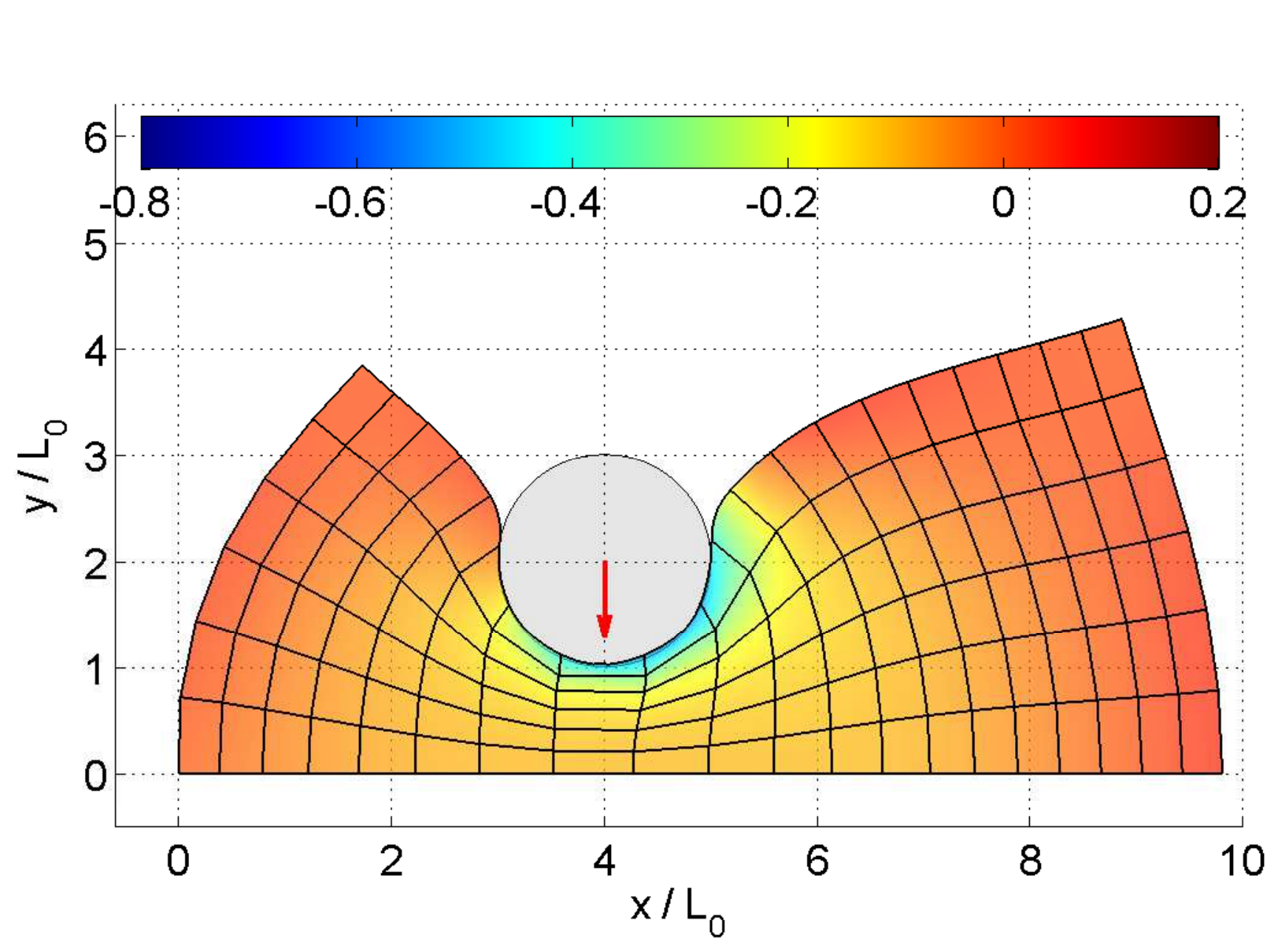}}
\put(-8.1,0){\includegraphics[width=70mm]{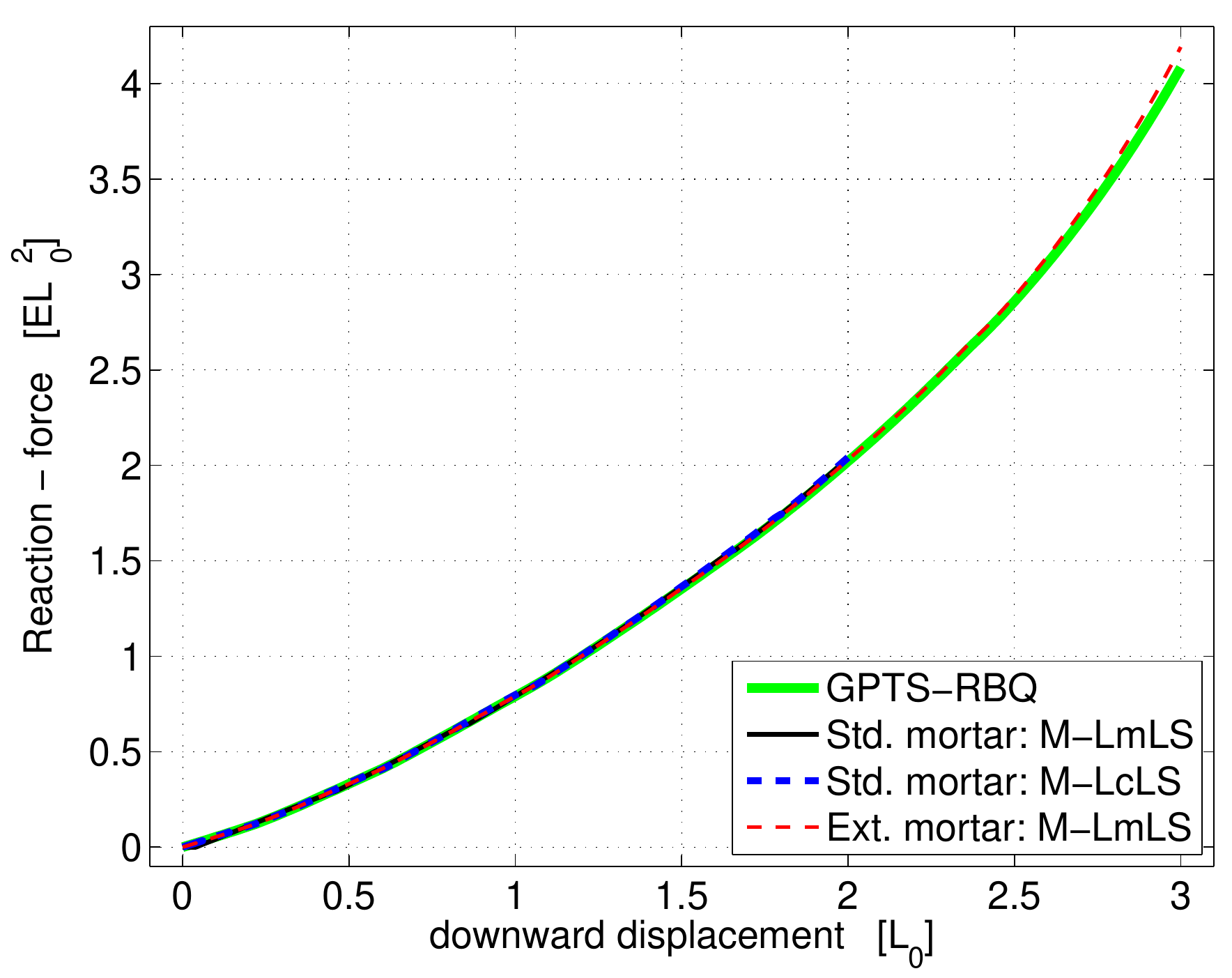}}
\put(-7,2.6){\includegraphics[height=25mm]{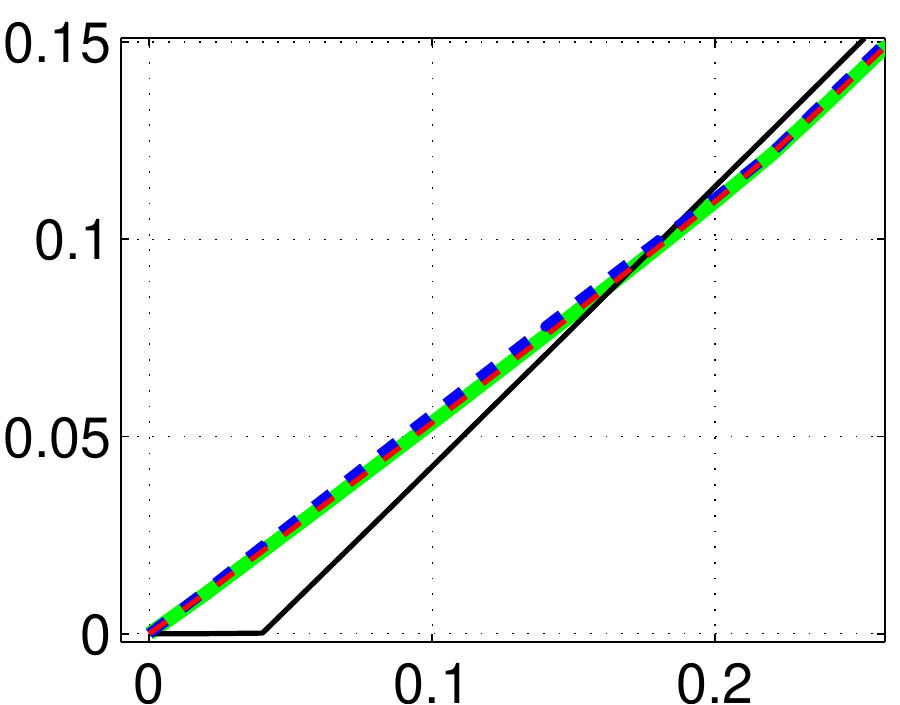}} 
\put(0.1,0){\includegraphics[width=70mm]{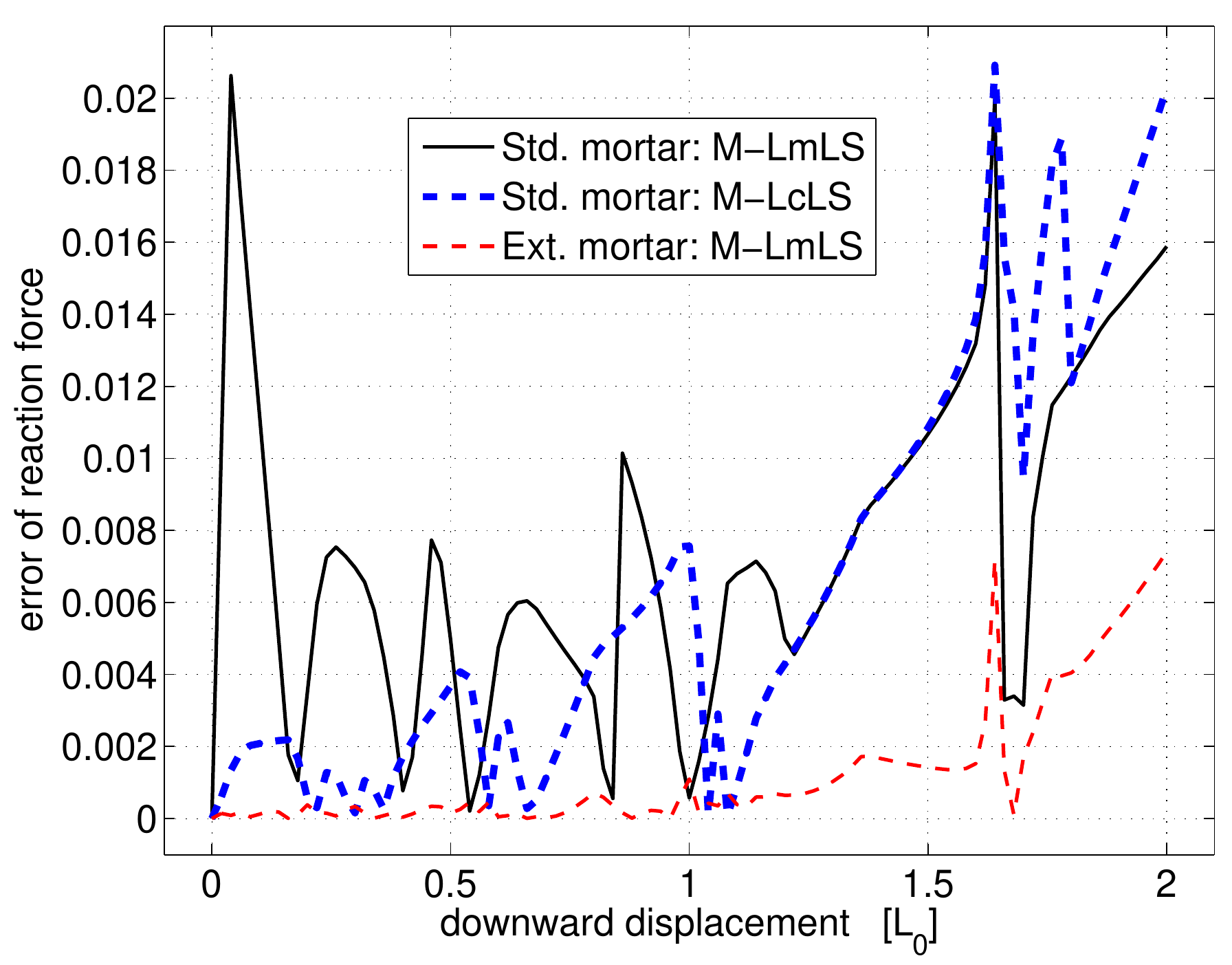}}
%
%
\put(-7.5,5.5){a. }
\put(0.8,5.5){b.}
\put(-7.5,0.0){c. }
\put(0.8,0.0){d.}
\end{picture}
\vspace{-2mm}
\caption{2D indentation: a.~Initial mesh and boundary conditions; b.~Deformation computed with extended mortar for $\bar{u} = -3\,L_0$. c.~Load--displacement curves for various contact formulations. d.~Absolute error in the reaction force w.r.t the GPTS--RBQ formulation (GPTS--RBQ is used here as a reference since it is shown to be more accurate than GPTS without RBQ \citep{rbq}). $\epsilon_\mrn = 100\,E_0/L_0$.} 
\label{f:indencyl}
\end{center}

\begin{center} \unitlength1cm
\begin{picture}(0,9.5)
\put(-7.5,4.7){\includegraphics[width=65mm]{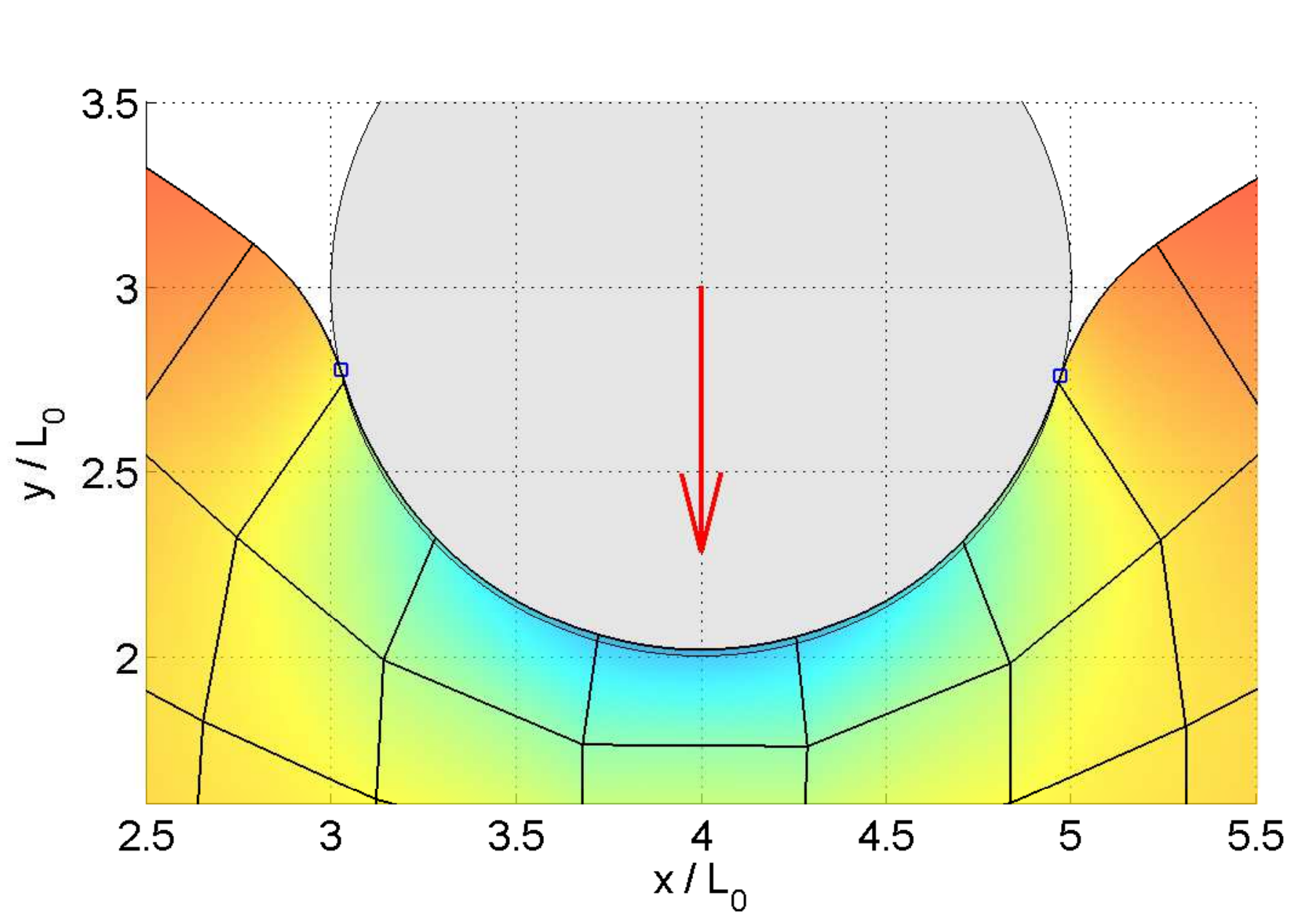}}  
\put(0.5,4.7){\includegraphics[width=65mm]{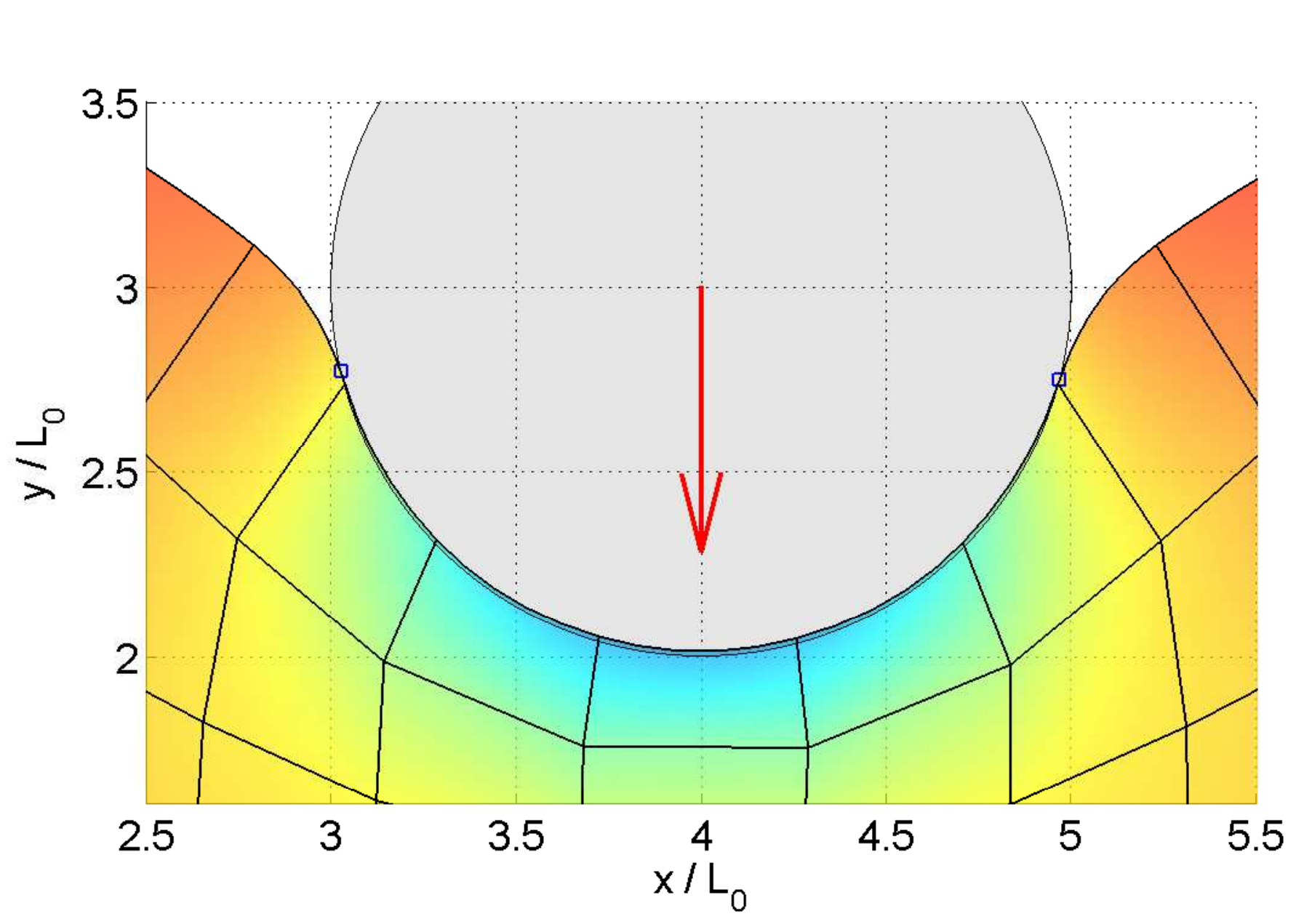}} 
%
\put(-7.5,0.1){\includegraphics[width=65mm]{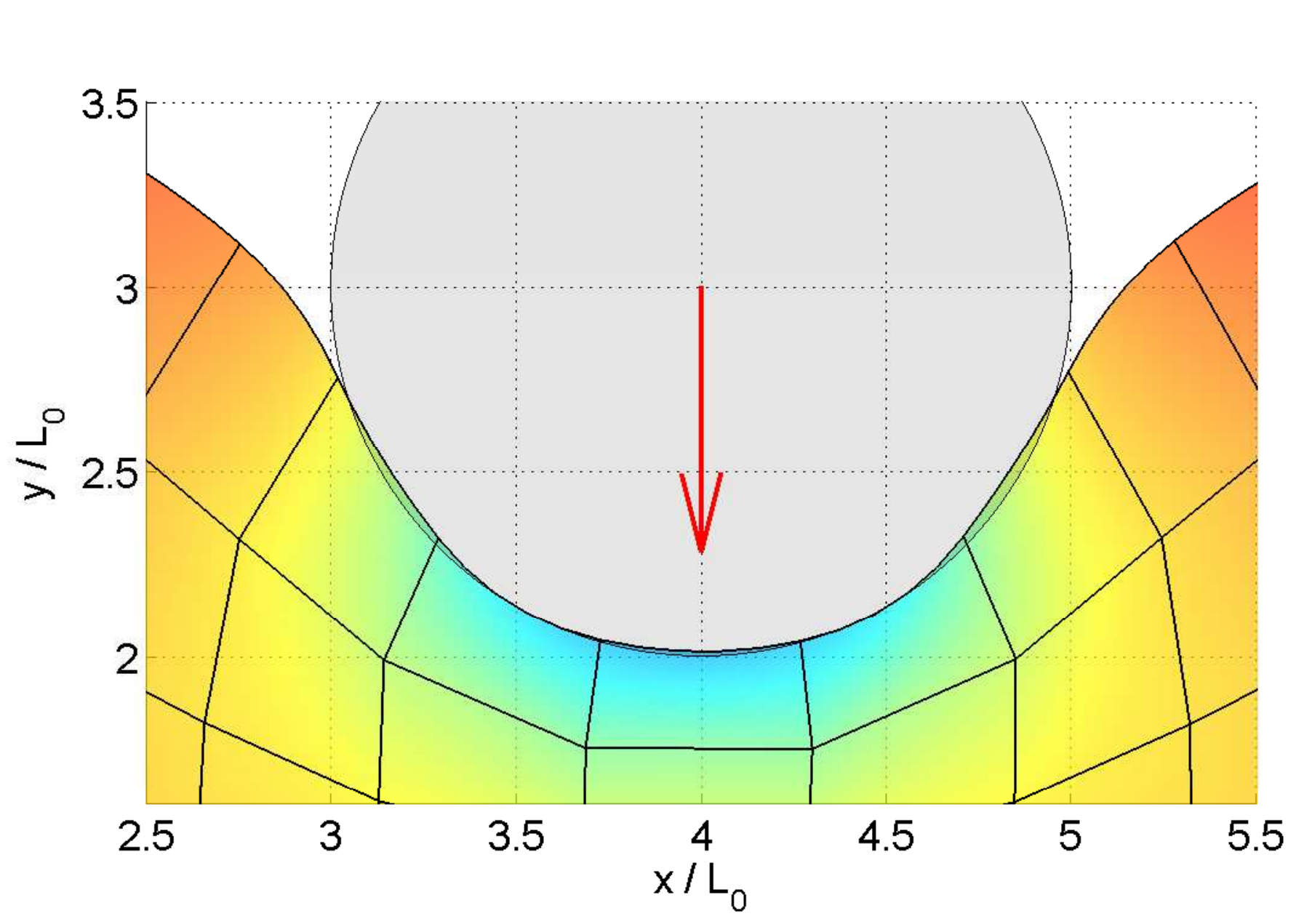}}
\put(0.5,0.1){\includegraphics[width=65mm]{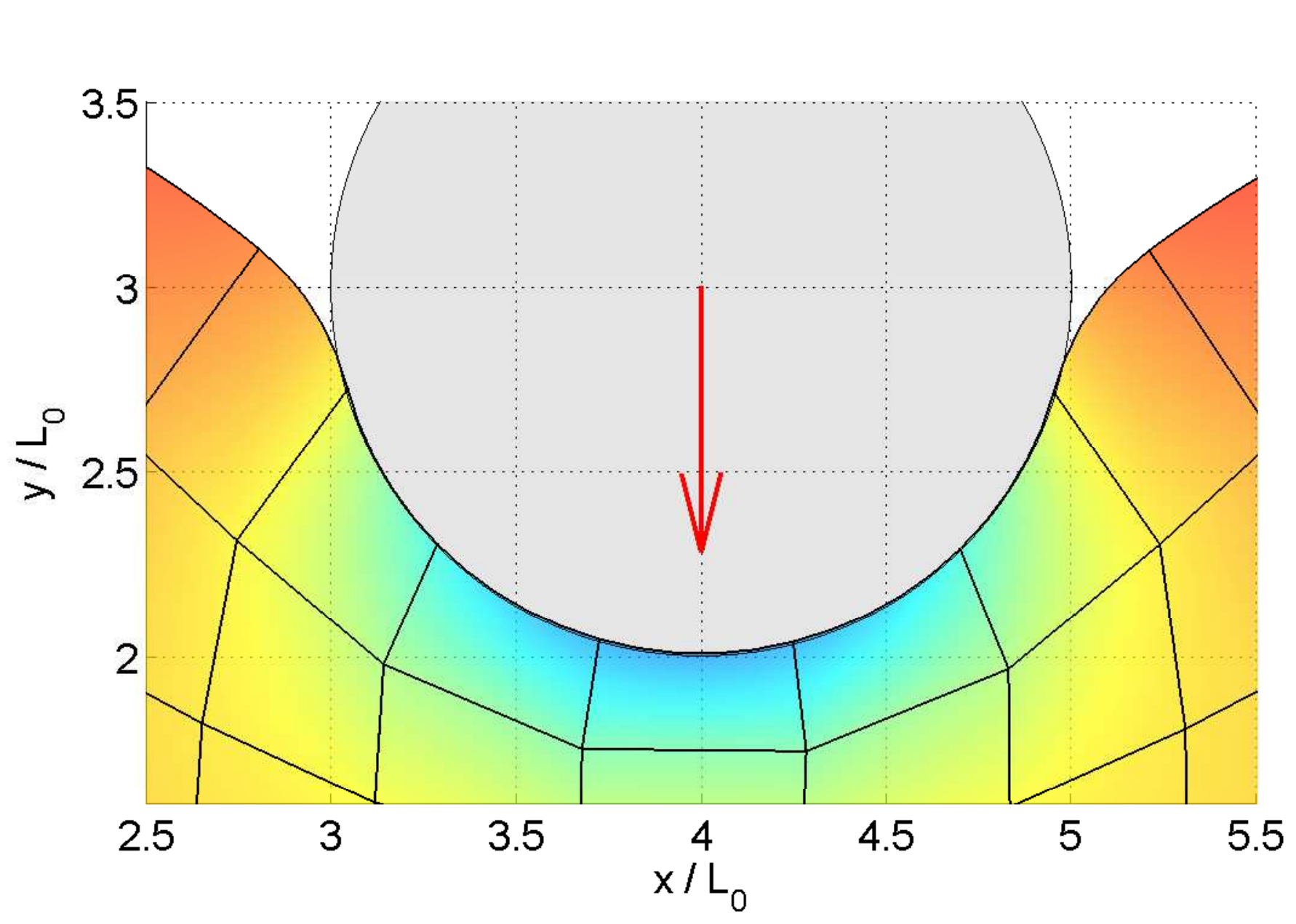}}
 %
%
%
\put(-6.9,4.7){a. }
%
\put(1,4.7){b.}
%
\put(-6.9,0.1){c. }
\put(1,0.1){d.}
\end{picture}
\vspace{-0.5cm}
\caption{ 2D indentation: enlargement of the contact area at $\bar{u} = -2\,L_0$ considering various contact formulations: a.~GPTS with RBQ, b.~Extended mortar with M-LmLS, c.~Standard mortar with M-LmLS, and d.~Standard mortar with M-LcLS. }
\label{f:cylpen}
\end{center}
\end{figure} 
\begin{figure}[!htp]
\begin{center} \unitlength1cm
\begin{picture}(0,20.5)
\put(-8.1,15){\includegraphics[width=52mm]{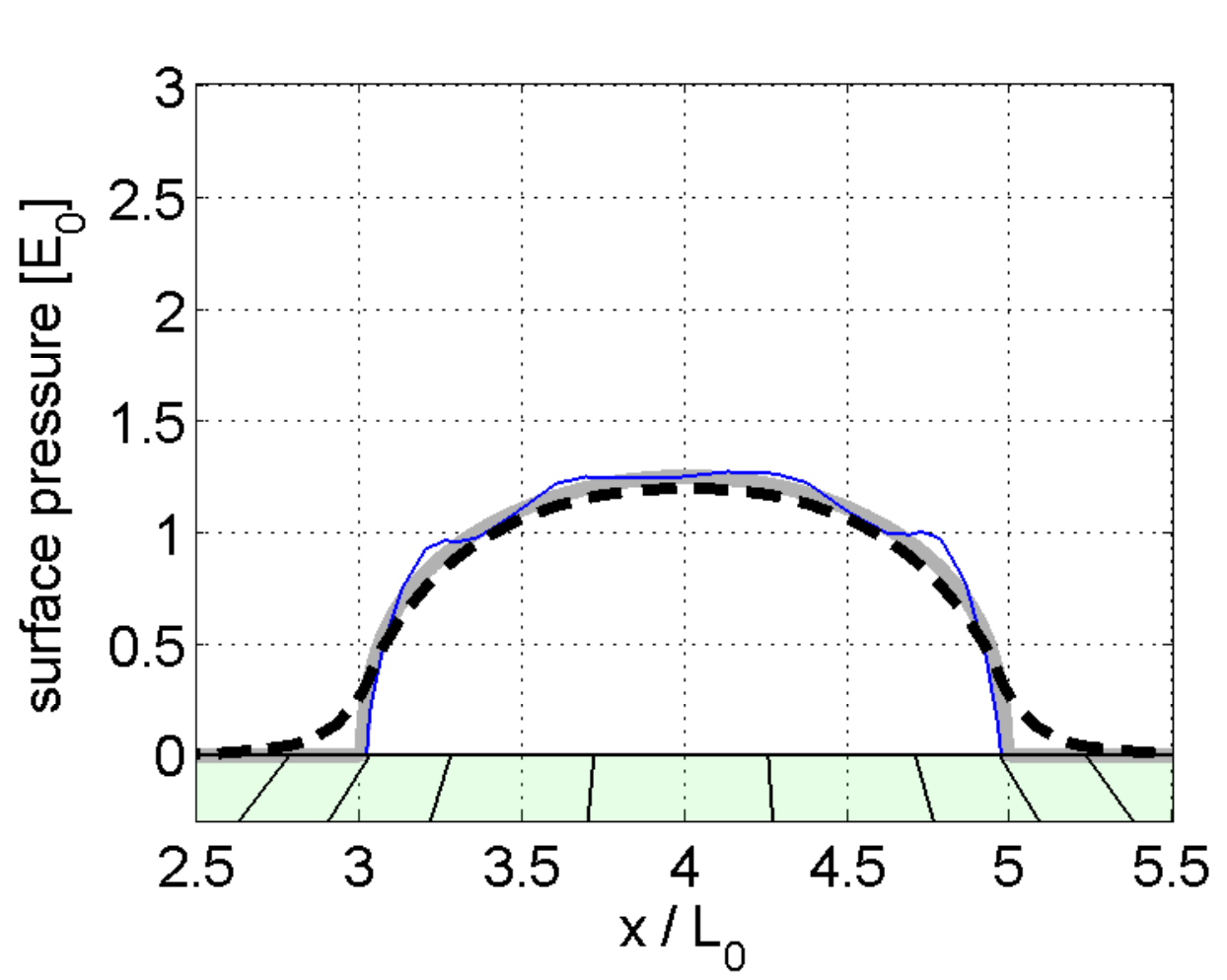}}  
\put(-2.6,15){\includegraphics[width=52mm]{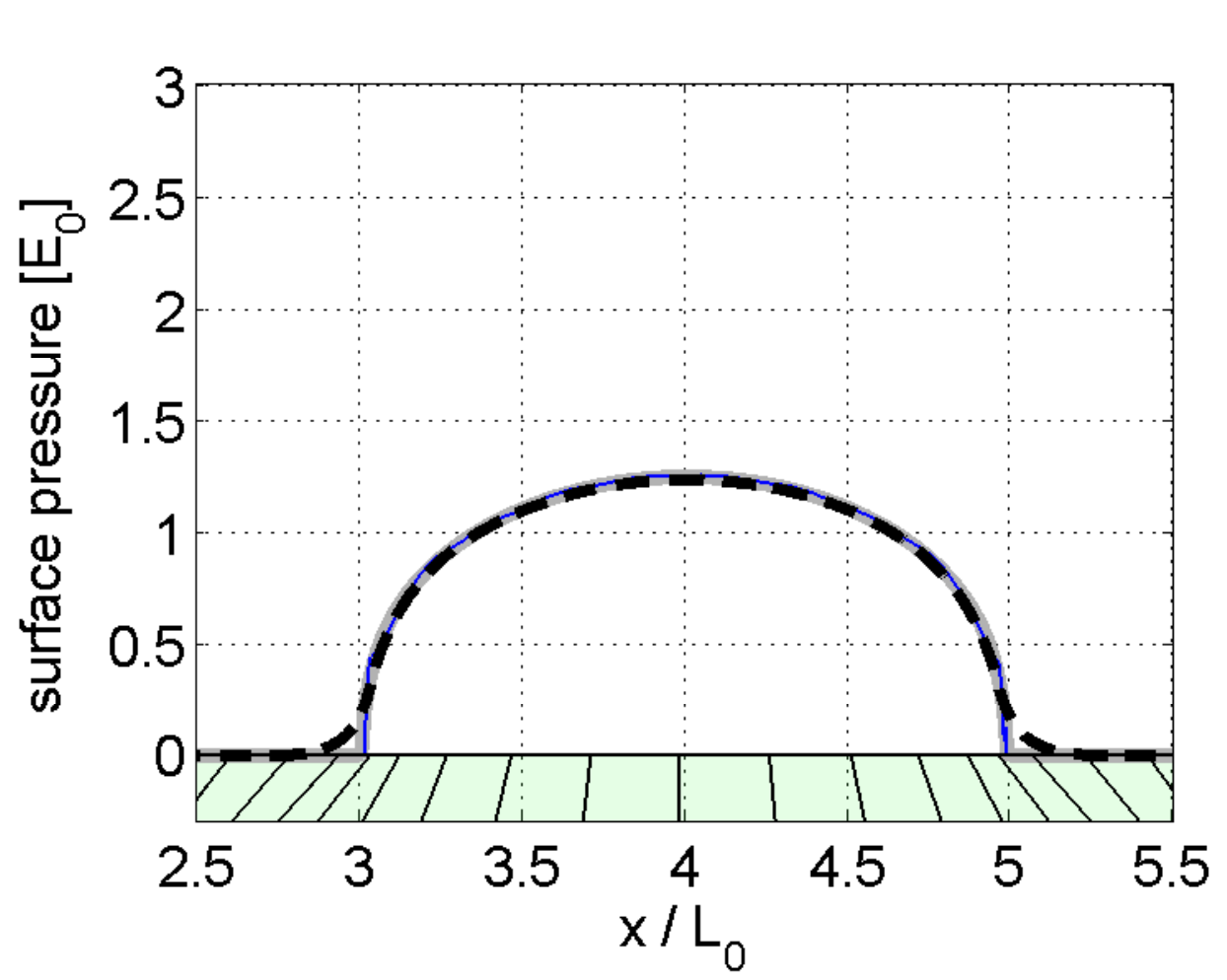}} 
\put(2.8,15){\includegraphics[width=52mm]{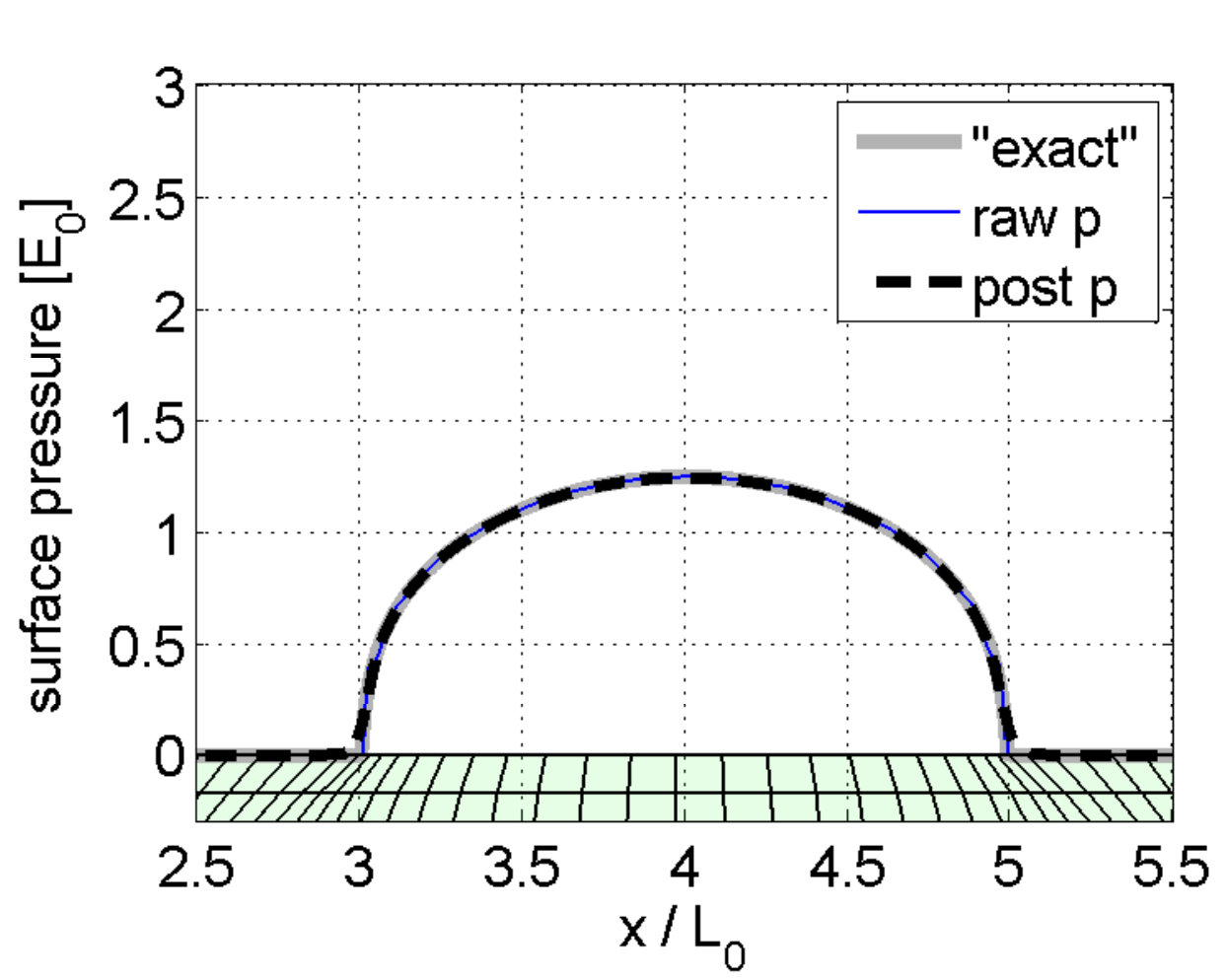}} 
\put(-8.1,10){\includegraphics[width=52mm]{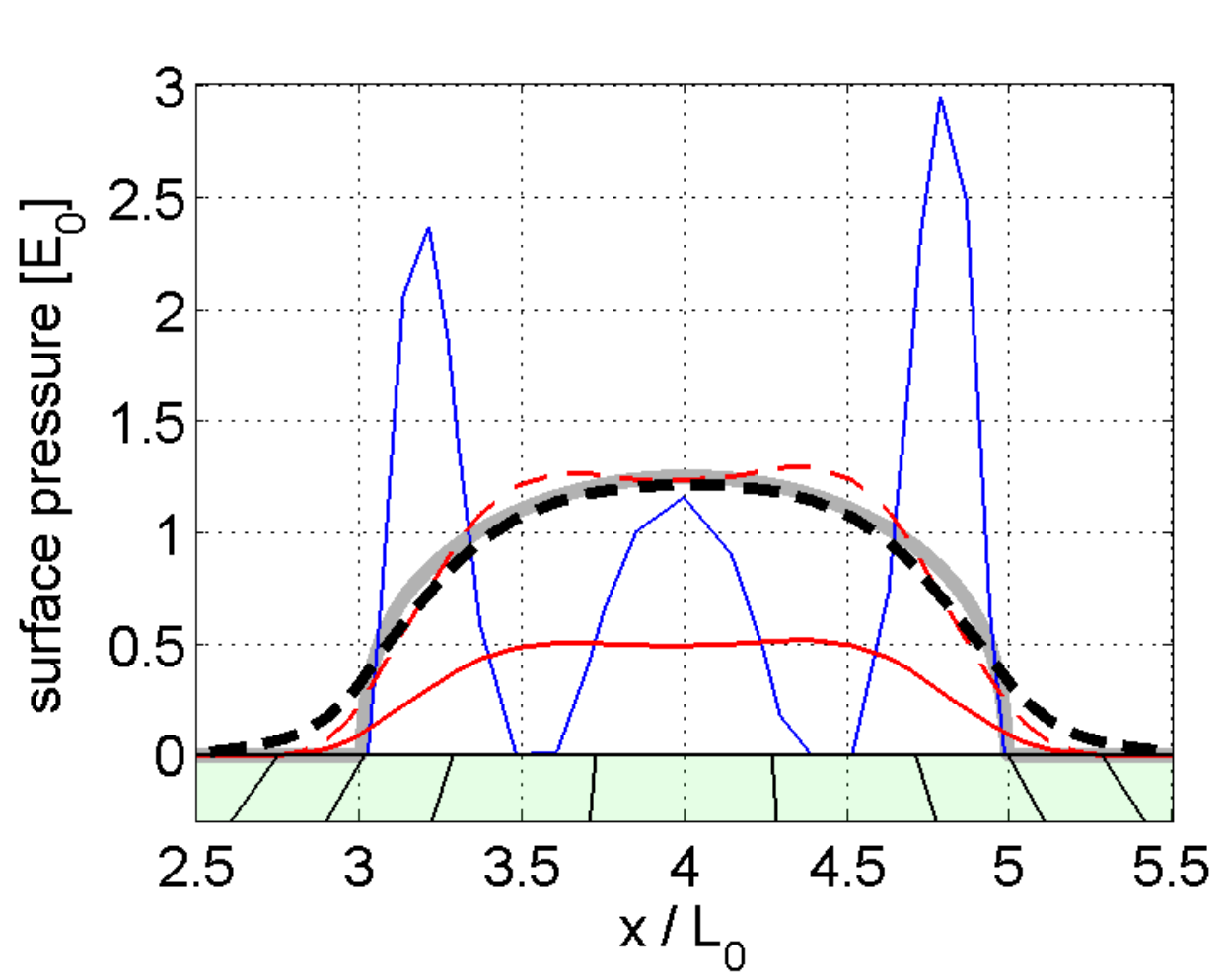}}  
\put(-2.6,10){\includegraphics[width=52mm]{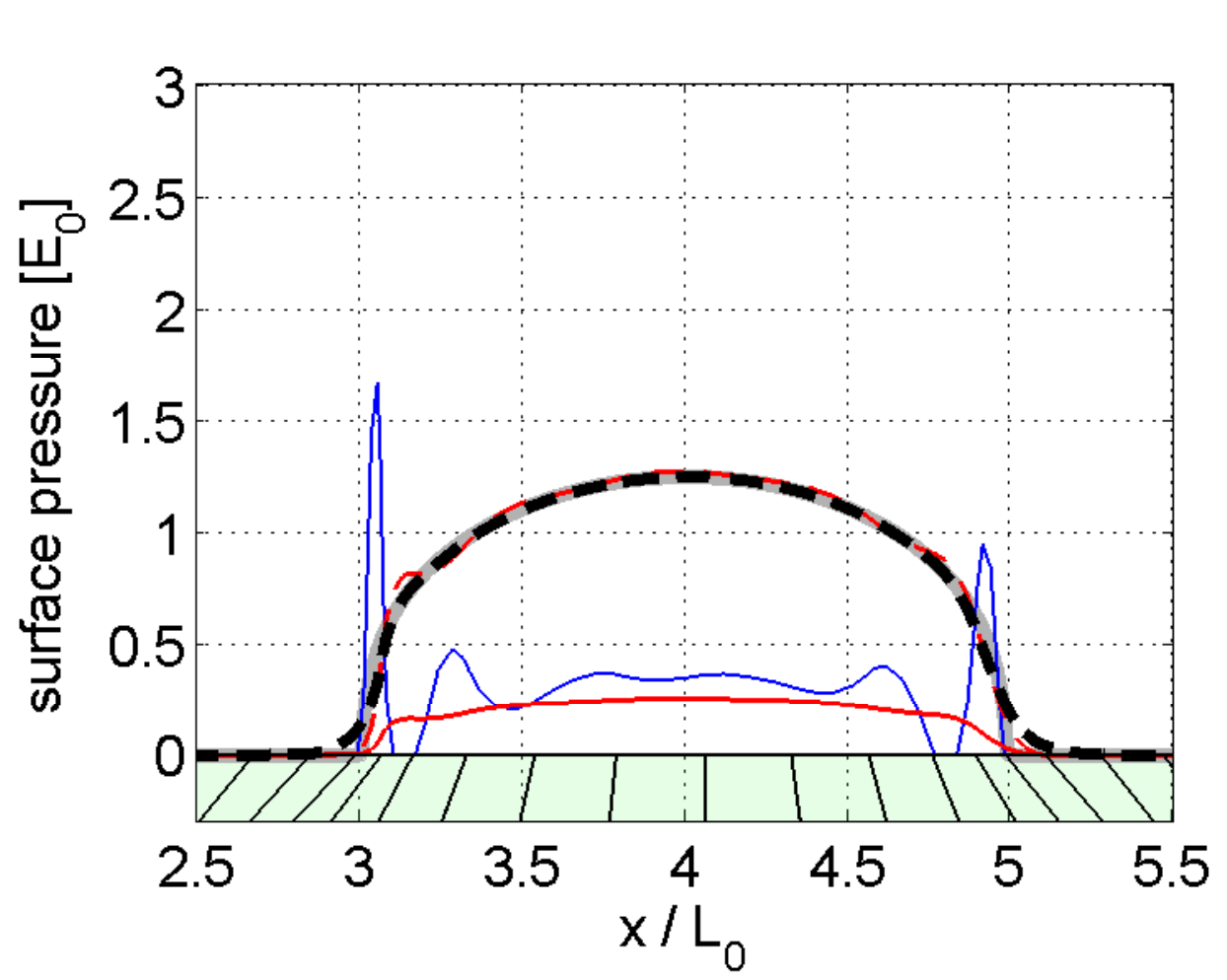}} 
\put(2.8,10){\includegraphics[width=52mm]{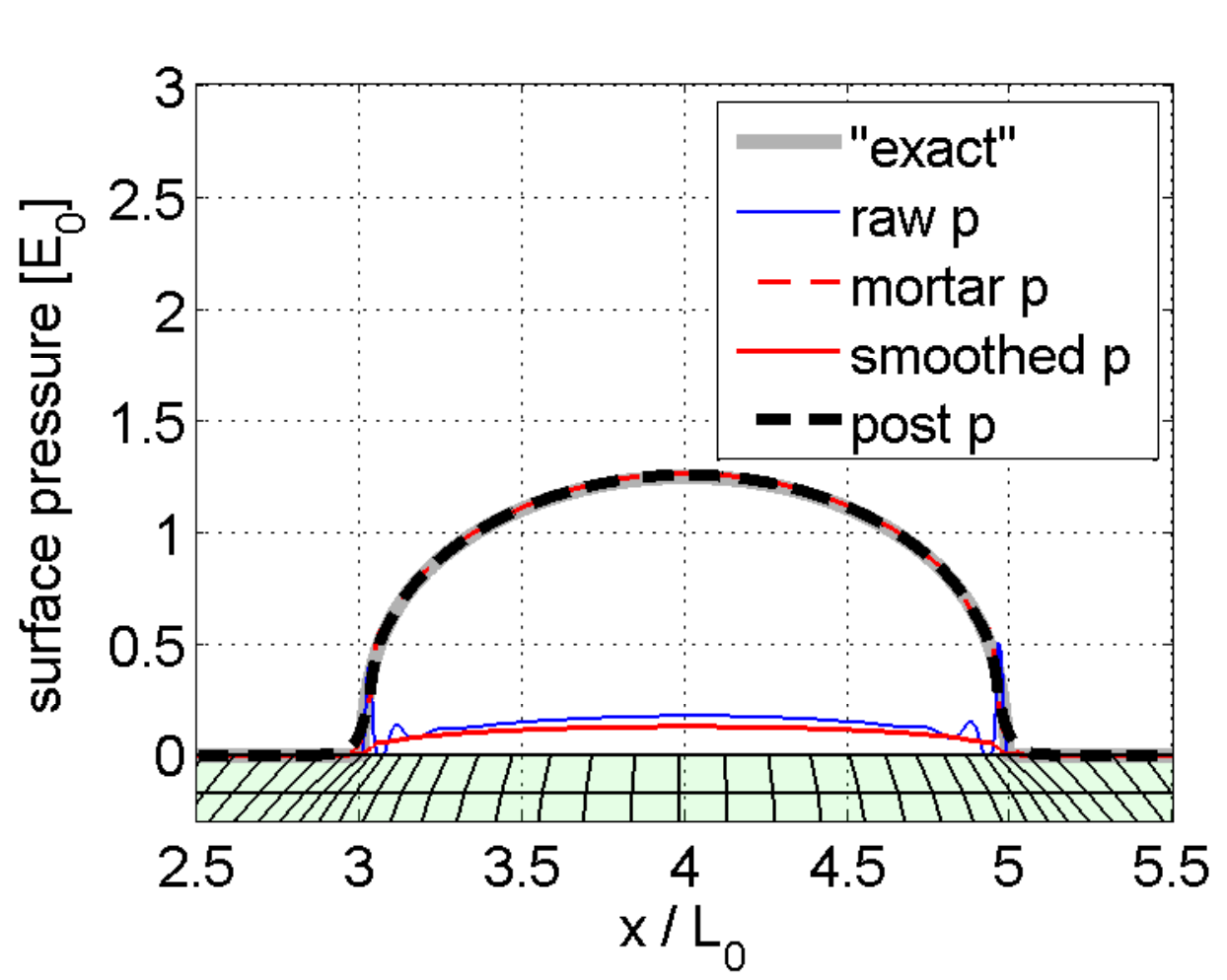}} 
\put(-8.1,5){\includegraphics[width=52mm]{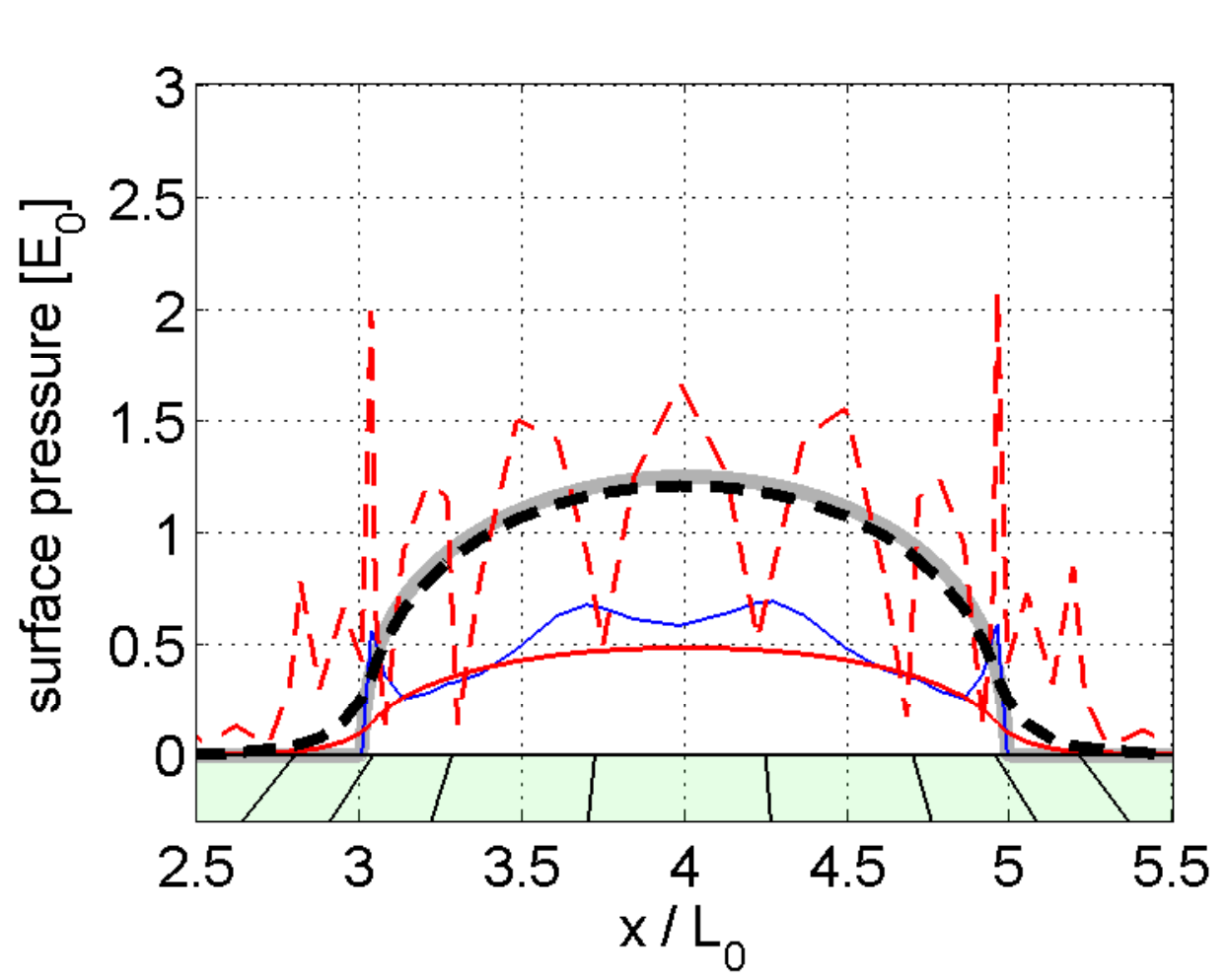}}  
\put(-2.6,5){\includegraphics[width=52mm]{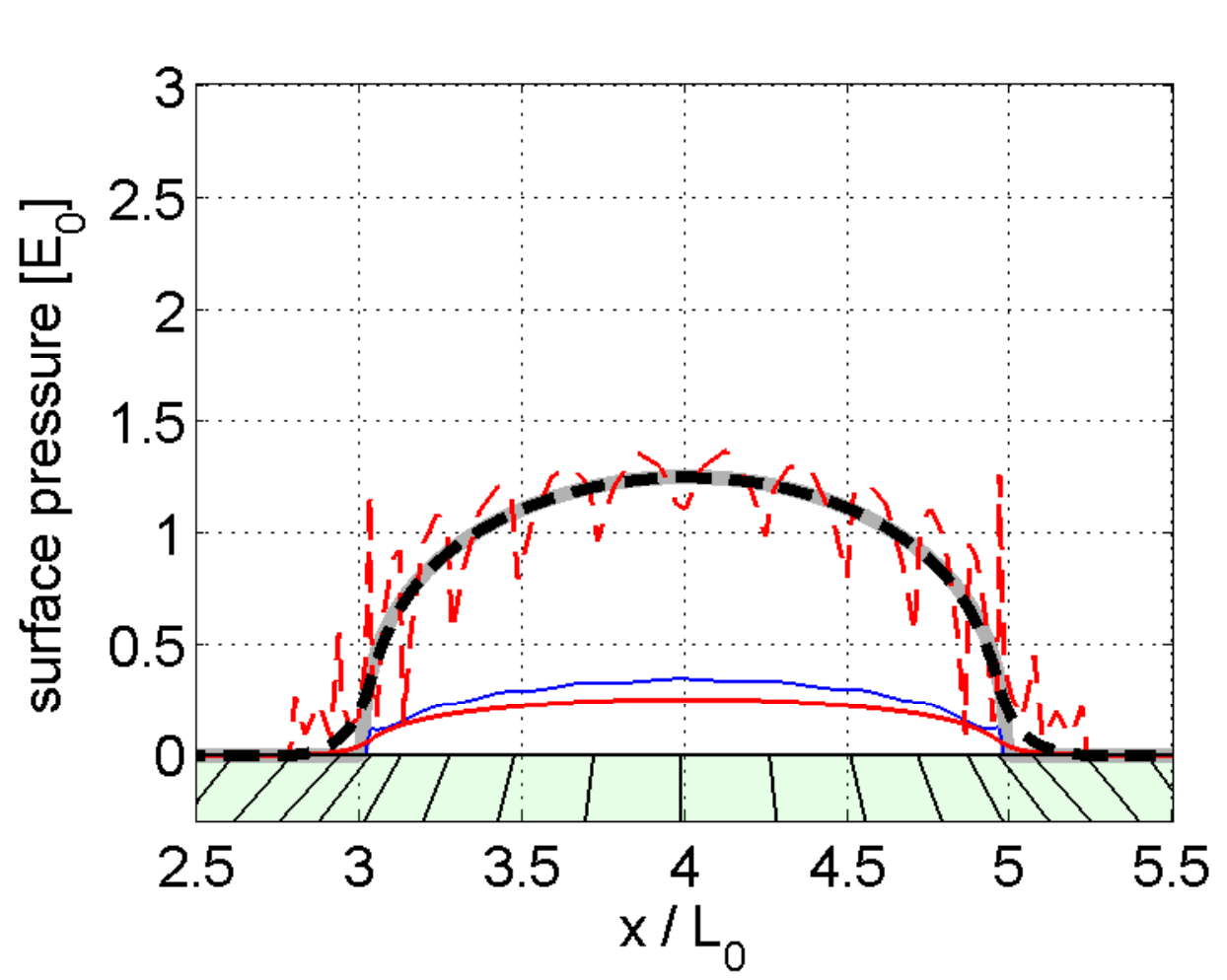}} 
\put(2.8,5){\includegraphics[width=52mm]{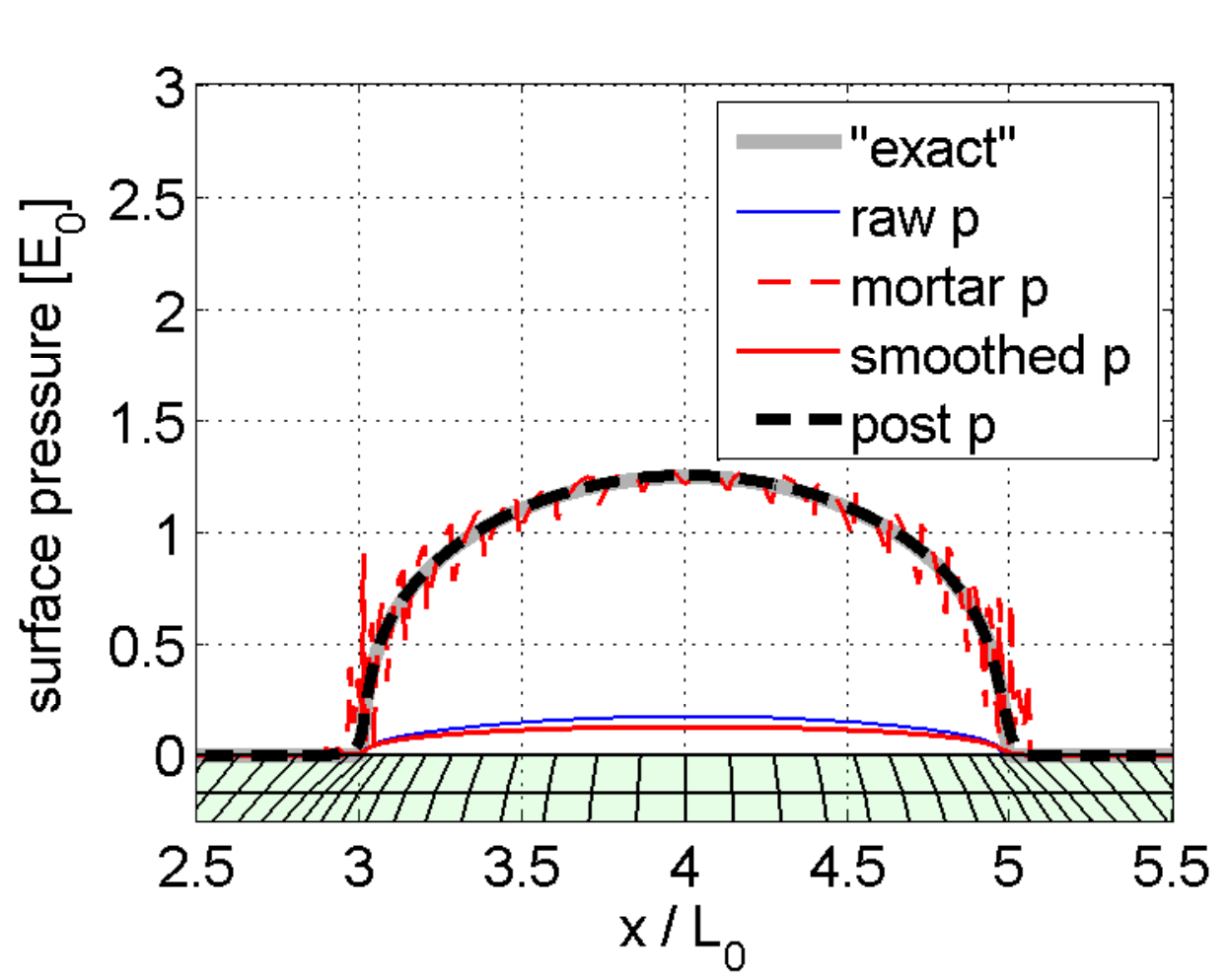}} 
\put(-8.1,0){\includegraphics[width=52mm]{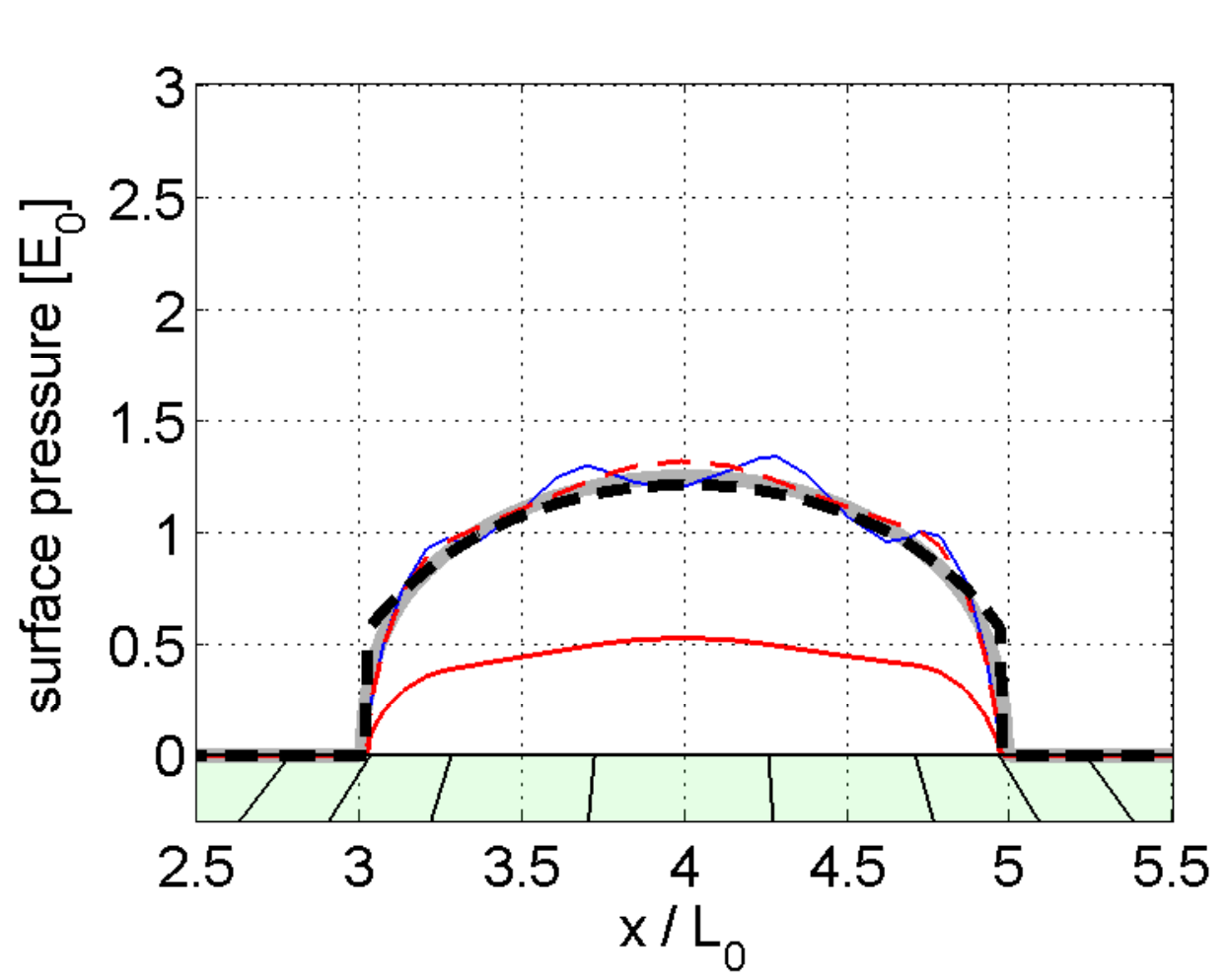}}  
\put(-2.6,0){\includegraphics[width=52mm]{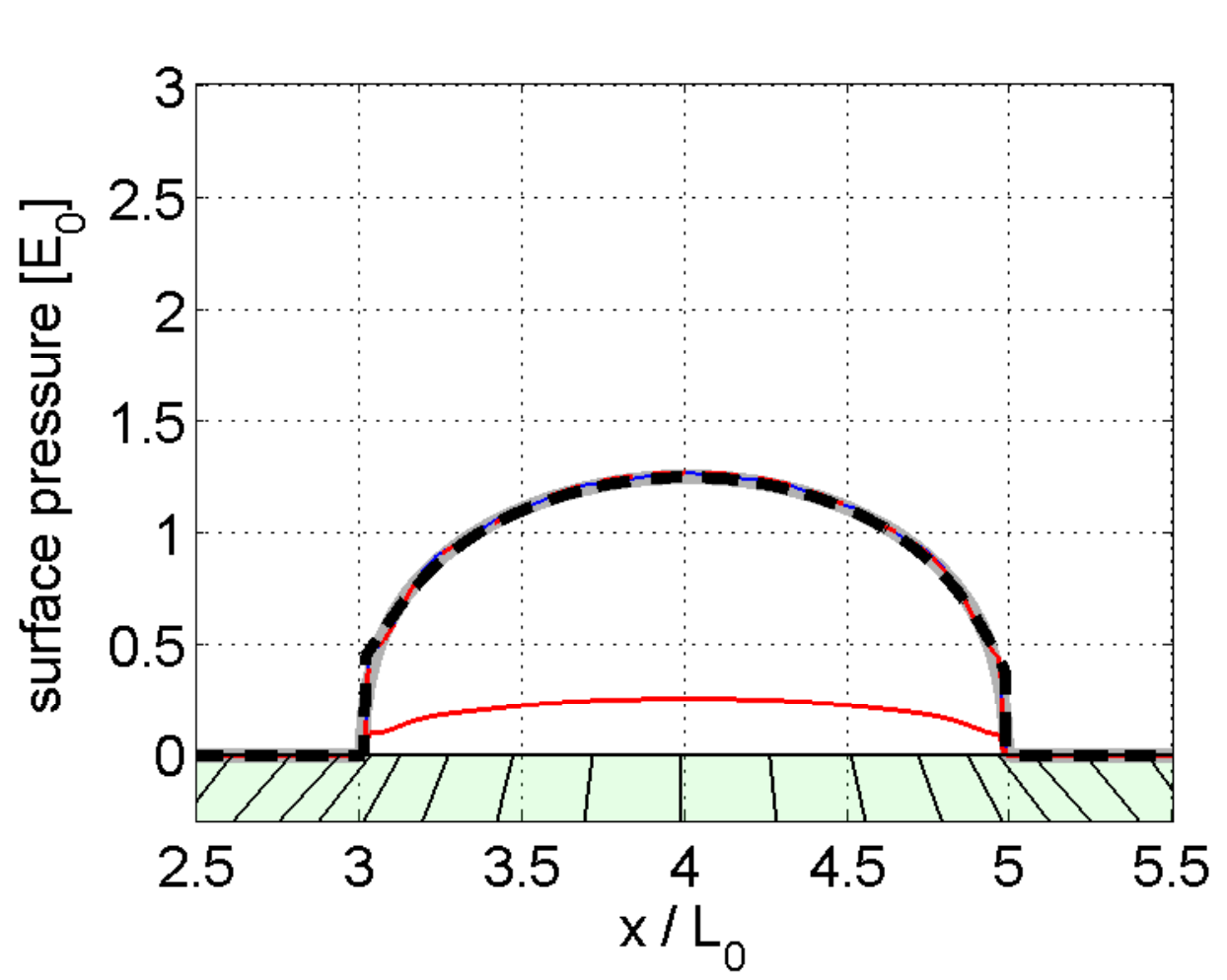}} 
\put(2.8,0){\includegraphics[width=52mm]{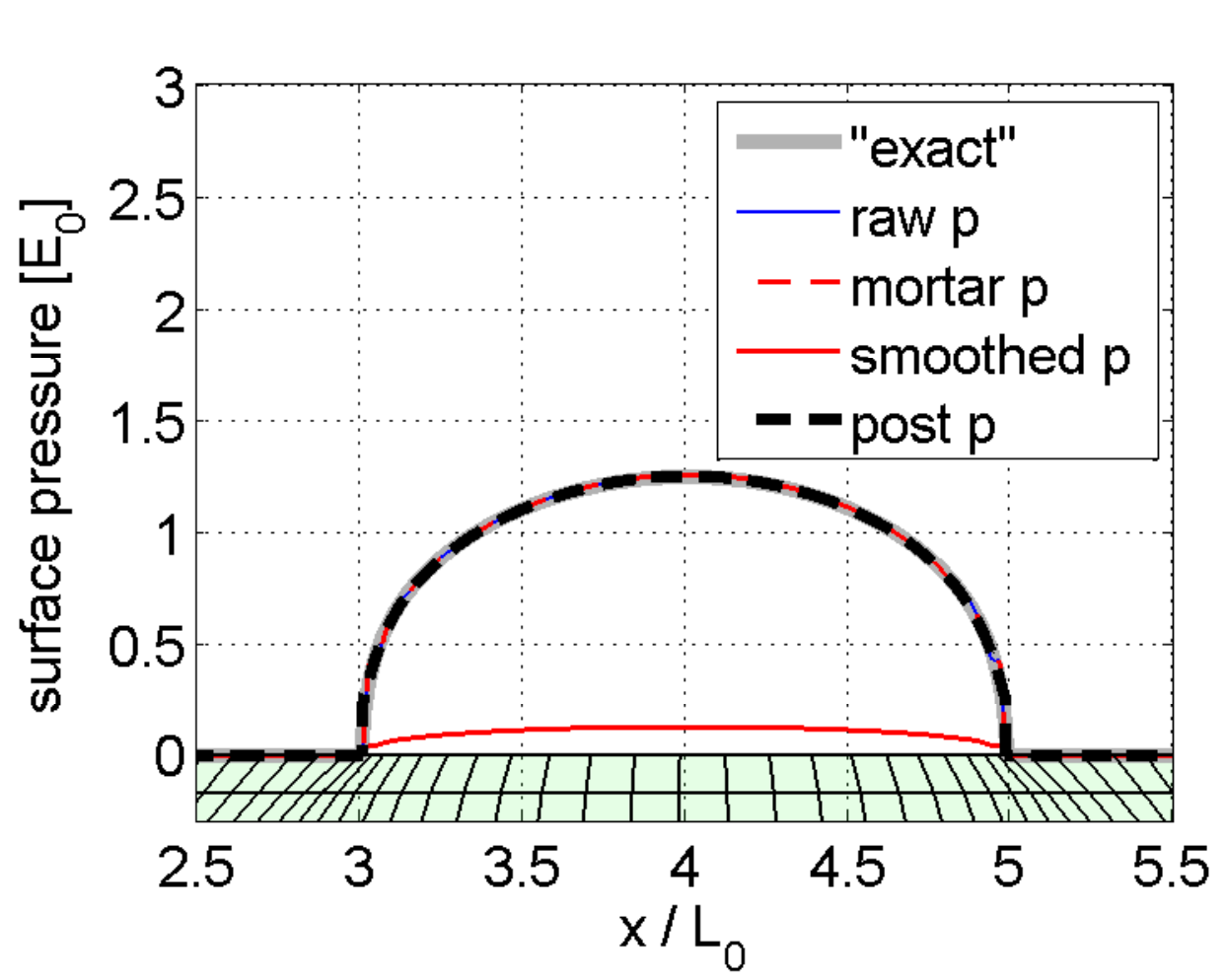}} 
\put(-7.6,14.8){a. }
\put(-7.6,9.8){b.}
\put(-7.6,4.8){c.}
\put(-7.6,-0.2){d.}
\end{picture}
\caption{2D indentation: contact pressure plotted on the flattened current surface  for $\bar{u} = -2\,L_0$ for different mesh refinement (left to right), considering various contact formulations: a.~GPTS with RBQ,  b.~Standard mortar with M-LmLS, c.~Standard mortar with M-LcLS, and d.~Extended mortar with M-LmLS.  Note that both the raw and the smoothed pressures are not physically meaningful in b, c and d. Only the mortar pressure and its post procession are.}
\label{f:sconpress}
\end{center}
\end{figure} 

The first example examines two-dimensional contact between a rigid cylinder and a deformable slab. The computational results are compared for three contact formulations: GPTS--RBQ, standard mortar, and extended mortar. In particular, for standard mortar, the M-LmLS and M-LcLS shape functions are used, while for extended mortar M-LmLS shape function is considered. \\[1.5mm]
The initial mesh and boundary conditions are shown in Fig.~\ref{f:indencyl}a. Here, the bulk is discretized by linear elements, while the contact surface is enriched by quadratic NURBS elements \citep{nece}.
Fig.~\ref{f:indencyl}b shows the deformed configuration colored by the first stress invariant $I_1=\tr\bsig$.
Note that in this example, the full-pass algorithm is used. But no segmentation is required even for standard mortar, since one of the bodies is rigid and thus no master mortar shape functions are appearing in integration. However, the contact pressure has a weak physical discontinuity at the contact boundary. While this kind of discontinuity is embedded in GPTS and extended mortar, it is not accounted for in standard mortar (see Sec.~\ref{s:stdmortar}). \\[1.5mm]
The comparison of the load--displacement curve (Fig.~\ref{f:indencyl}c-d) shows that extended mortar provides comparable accuracy as GPTS, while standard mortar,  especially with the M-LmLS shape function, gives much less accurate results (see Fig.~\ref{f:indencyl}c).   \\[1.5mm]
Fig.~\ref{f:cylpen} shows the penetration at the contact interface for various contact formulations.  As observed in Fig.~\ref{f:cylpen}c, standard mortar with M-LmLS results in an irregular penetration around the contact boundary.  This implies that the weak physical discontinuity at the contact boundary induces a significant error in the standard mortar formulation with standard (M-LmLS*) or weighted standard (M-LmLS) shape functions. \\[1.5mm]
 In Fig.~\ref{f:sconpress}, we examine different measures of the contact pressure as the mesh is refined.  The post-processed mortar contact pressure is obtained according to post-processing scheme of \citet{sauer-ece2}.  Three mesh refinement levels, $\mrm:=2,4$ and $8$, are plotted. Here, the exact contact pressure is obtained with a very fine mesh (mesh level $\mrm=128$). As seen, all the considered contact formulations converge with mesh refinement. However, in contrast to the standard mortar method, the extended mortar formulation can capture the discontinuities at the contact boundary much better with coarse meshes.
\subsection{2D ironing}\label{sec:iron2D}
\begin{figure}[!htp]
\begin{center} \unitlength1cm
\begin{picture}(0,8.5)

\put(-9.7,4.13){\includegraphics[width=1.2\textwidth]{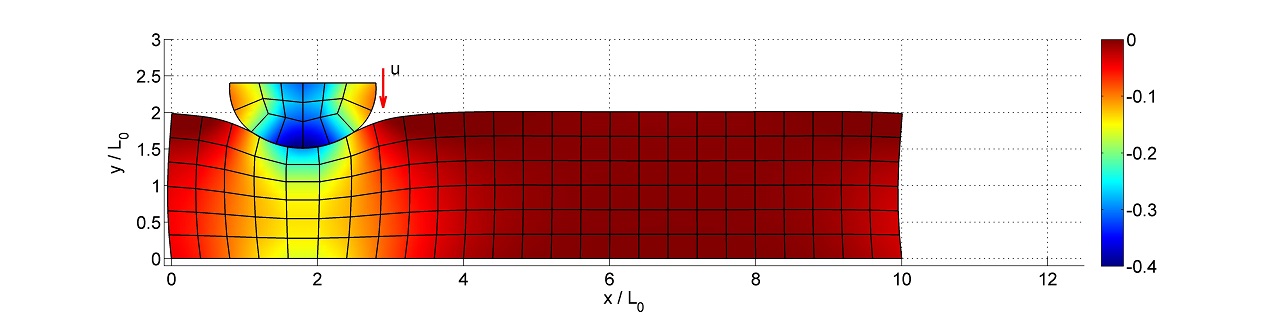}}
\put(-9.7,0.0){\includegraphics[width=1.2\textwidth]{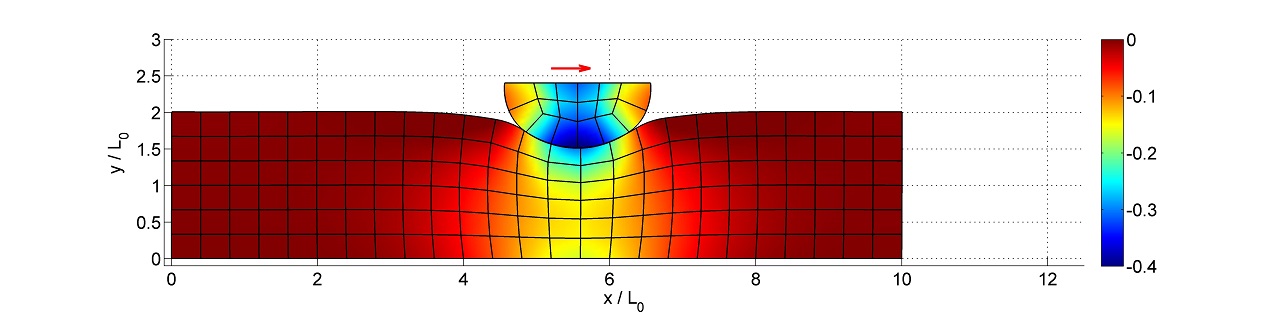}}

\put(-6.9,4.3){a. }
\put(-6.9,0.2){b.}
\end{picture}
\caption{2D ironing: a half-cylinder with radius $R_0=1$ is pressed downward until $\bar{u}=-0.6\,L_0$ (a.) and then moved horizontally (b.). The color shows the first stress invariant $I_1$.}
\label{f:iron2D_cf}
\end{center}
\end{figure}

\begin{figure}[!h]
\begin{center} \unitlength1cm
\begin{picture}(0,13.5)

\put(-8.0,6.8){\includegraphics[width=75mm]{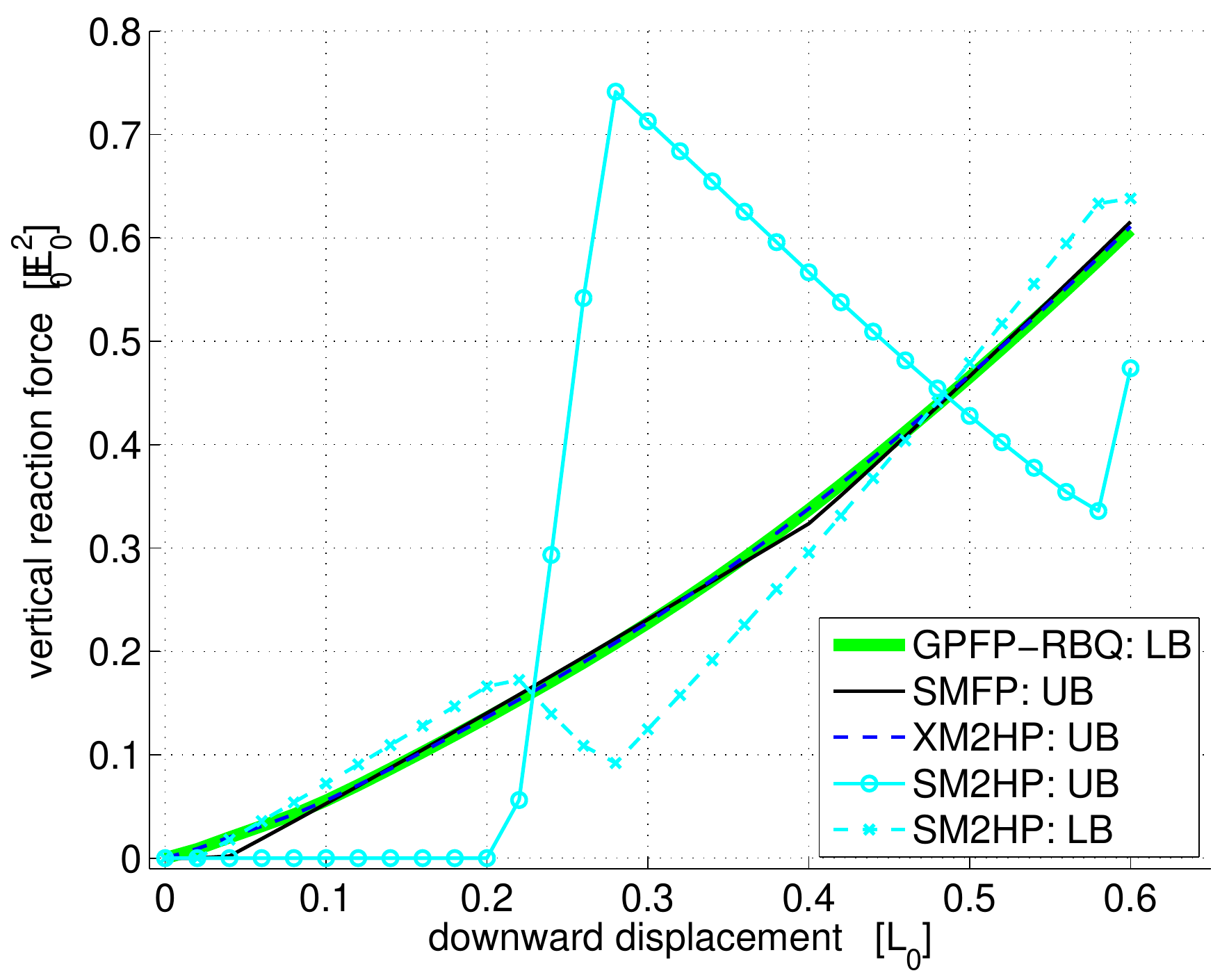}}
\put(-0.2,6.8){\includegraphics[width=80mm]{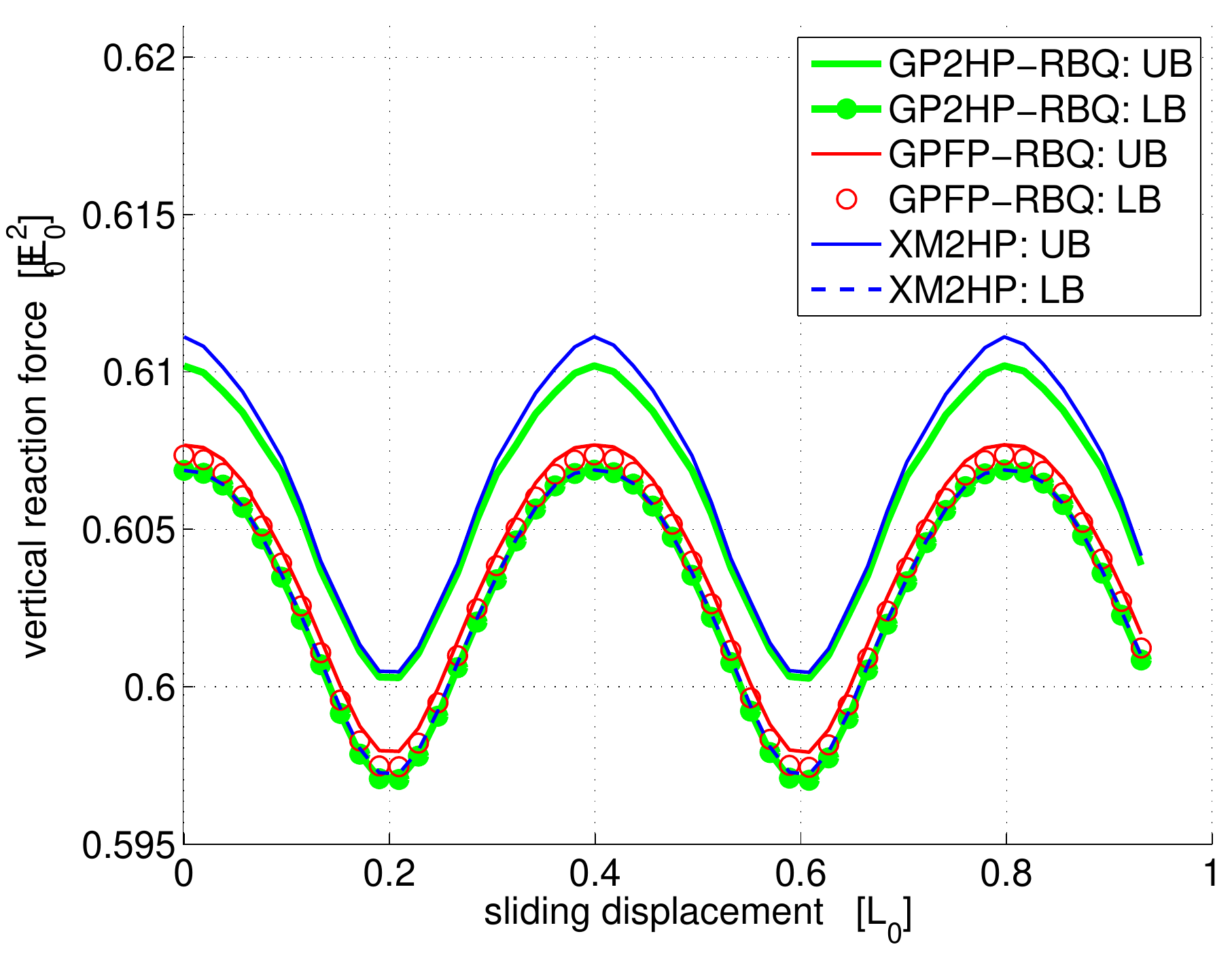}}

\put(-7.9,0){\includegraphics[width=75mm]{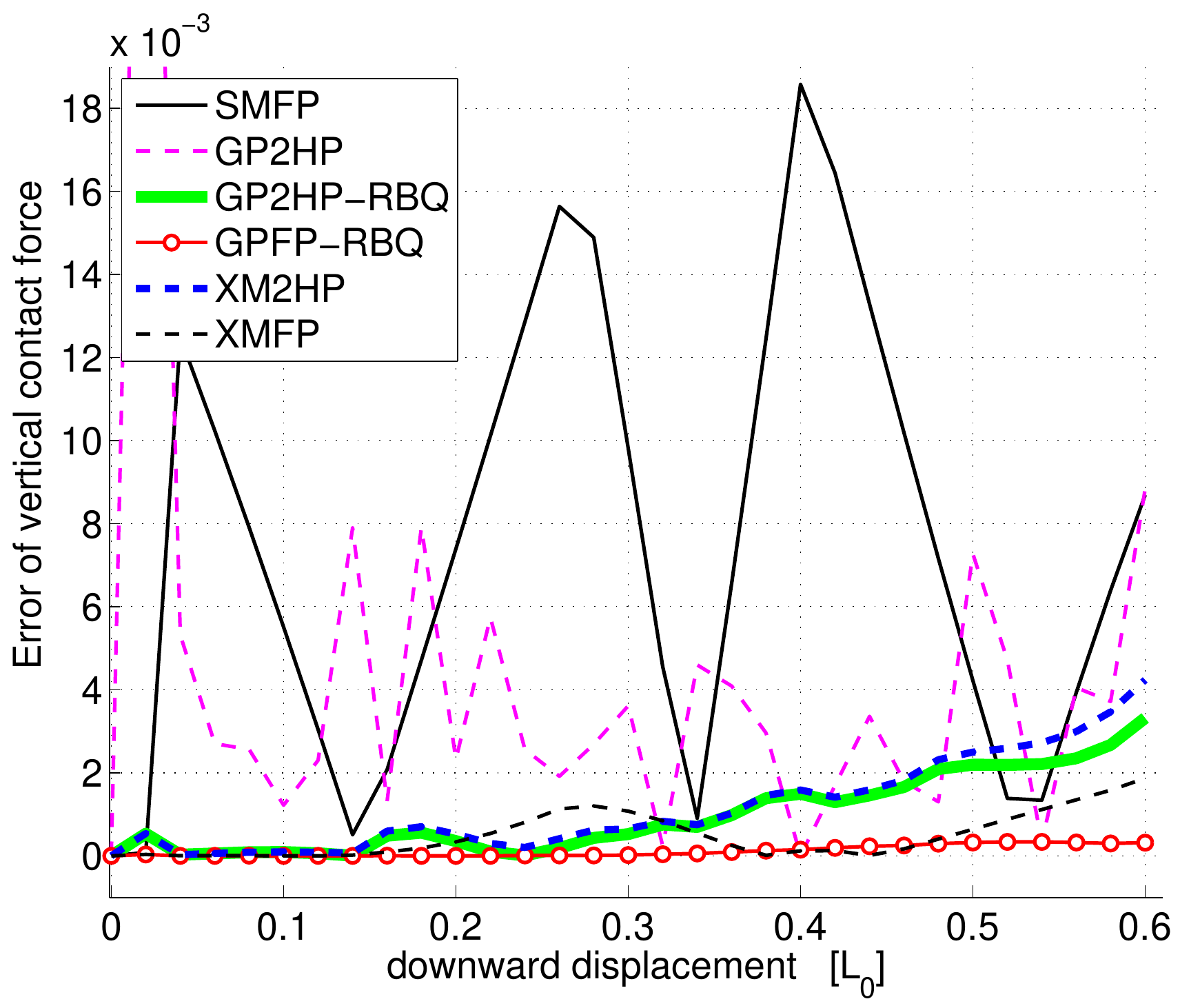}}
\put(0.3,0){\includegraphics[width=75mm]{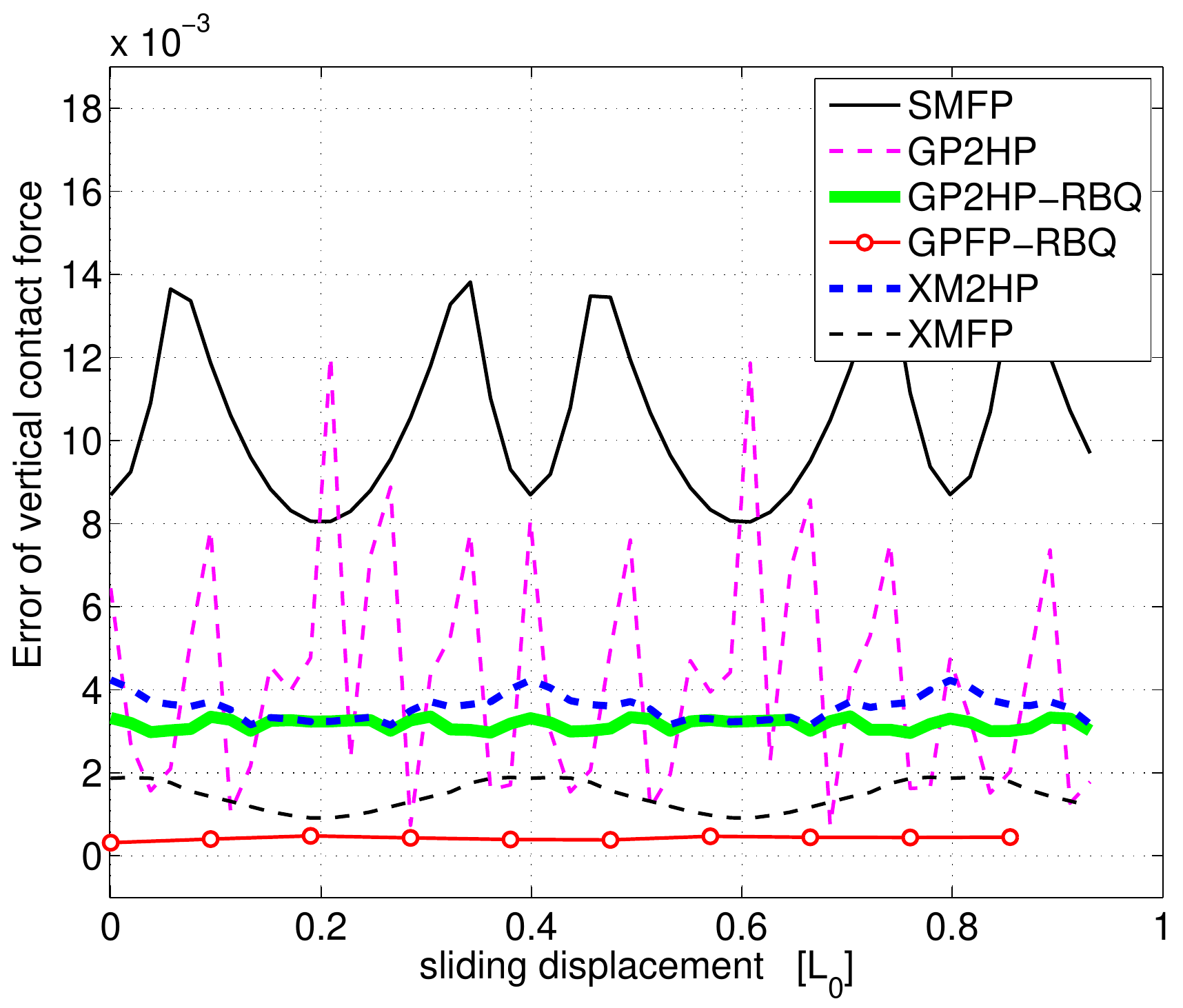}}

\put(-6.9,6.8){a. }
\put(1,6.8){b.}

\put(-6.9,0){c. }
\put(1,0){d.}

\end{picture}
\vspace{-0.3cm}
\caption{2D ironing: Load--displacement curves for a.~pressing phase and b.~sliding phase; c. and d. show the bias error, defined by $|R_\mathrm{UB} - R_\mathrm{LB}|$, for pressing and sliding, respectively. Here, $R$ denotes the reaction force either on the upper (UB) or lower (LB) body.}
\label{f:iron2D_load}
\end{center}
\end{figure}

Next, a 2D ironing problem with two deformable bodies is considered. The problem setup is shown in Fig.~\ref{f:iron2D_cf}. A periodic boundary condition is applied to the left and the right sides of the slab. The bulk of the bodies is discretized by linear finite elements, while its contact surface is enriched by quadratic NURBS elements \citep{nece}.
This problem has been considered in \cite{spbc} for assessing the GP2HP formulation.  Here, we will examine both standard mortar and extended mortar in comparison with GPTS-RBQ (considering both the full-pass and the two-half-pass approach). In the example, the shape functions M-LmLS and M-LmLS* are used for standard mortar and extended mortar, respectively. For the full-pass mortar method, we use many Gauss points ($n_{\mrg\mrp} = 20$) to avoid segmentation. For the two-half-pass mortar method, only $7$ Gauss points are used. \\[1.5mm]
In this example we observe that all the considered contact formulations can complete the simulation  without any convergence problems. 
Plotting the load-displacement curve (see Fig.~\ref{f:iron2D_load}) reveals that SM2HP yields highly inaccurate results  (Fig.~\ref{f:iron2D_load}a). XM2HP on the other hand provides an accuracy even higher than GP2HP (Fig.~\ref{f:iron2D_load}c,d).
Furthermore, by using RQB for GPFP, the accuracy in terms of the bias error is reduced significantly as is shown in Fig.~\ref{f:iron2D_load}b,d. and Tab.~\ref{tab:ironEvaEr}. For GP2HP, RBQ also yields smoother results. In comparison with standard mortar, the accuracy of extended mortar is also improved for both the full-pass and two-half-pass approach. These results are shown in Fig.~\ref{f:iron2D_load}c-d.
%
\begin{table}[!htp]
\begin{center}
\def\arraystretch{1.2}\tabcolsep=7pt
\begin{tabular}{|l|l|l|l|l|l| }
\hline 
GPFP-RBQ & GP2HP& GP2HP-RBQ &   SMFP&    XMFP & XM2HP  \\  
\hline   
 $4.23\times10^{-04}$ & $4.59\times 10^{-03}$ &$3.18\times 10^{-03}$ & $1.02\times 10^{-02}$ & $1.46\times 10^{-03}$ & $3.61\times 10^{-03}$\\ 
\hline
\end{tabular} 
\end{center}
\label{tab:ironEvaEr}   
\vspace{-3mm}
\caption{2D ironing:  the average error per sliding period obtained from Fig.~\ref{f:iron2D_load}d for each contact method.} 
\end{table}


\subsection{3D ironing}
\begin{figure}[!h]
\begin{center} \unitlength1cm
\setlength{\belowcaptionskip}{-0.7cm}
\begin{picture}(0,12.9)


\put(-1,0){\includegraphics[width=0.62\textwidth]{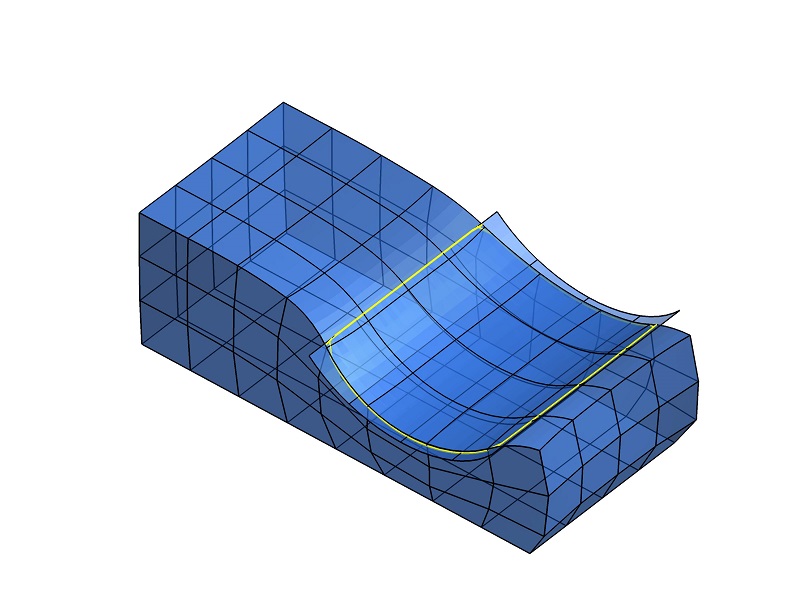}}

\put(-9.2,0){\includegraphics[width=0.62\textwidth]{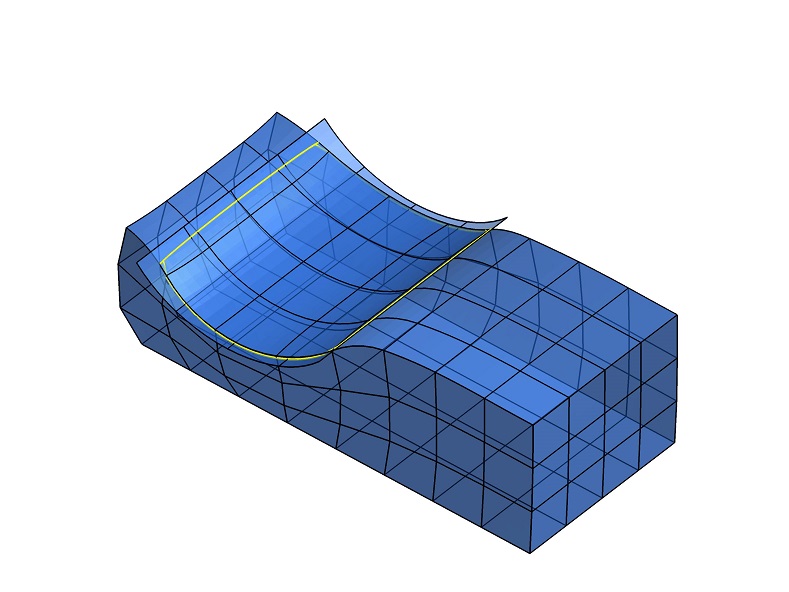}}
\put(-9.5,6.1){\includegraphics[width=0.62\textwidth]{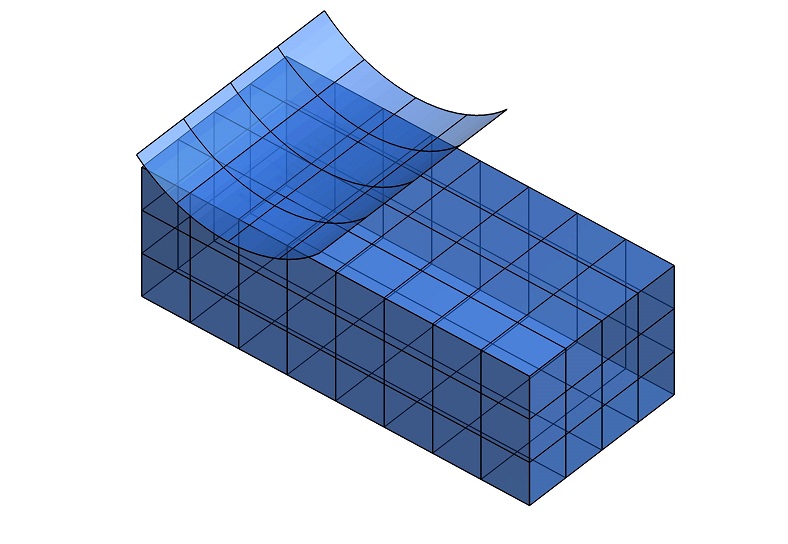}}

\put(-0.1,6.4){\includegraphics[width=0.5\textwidth]{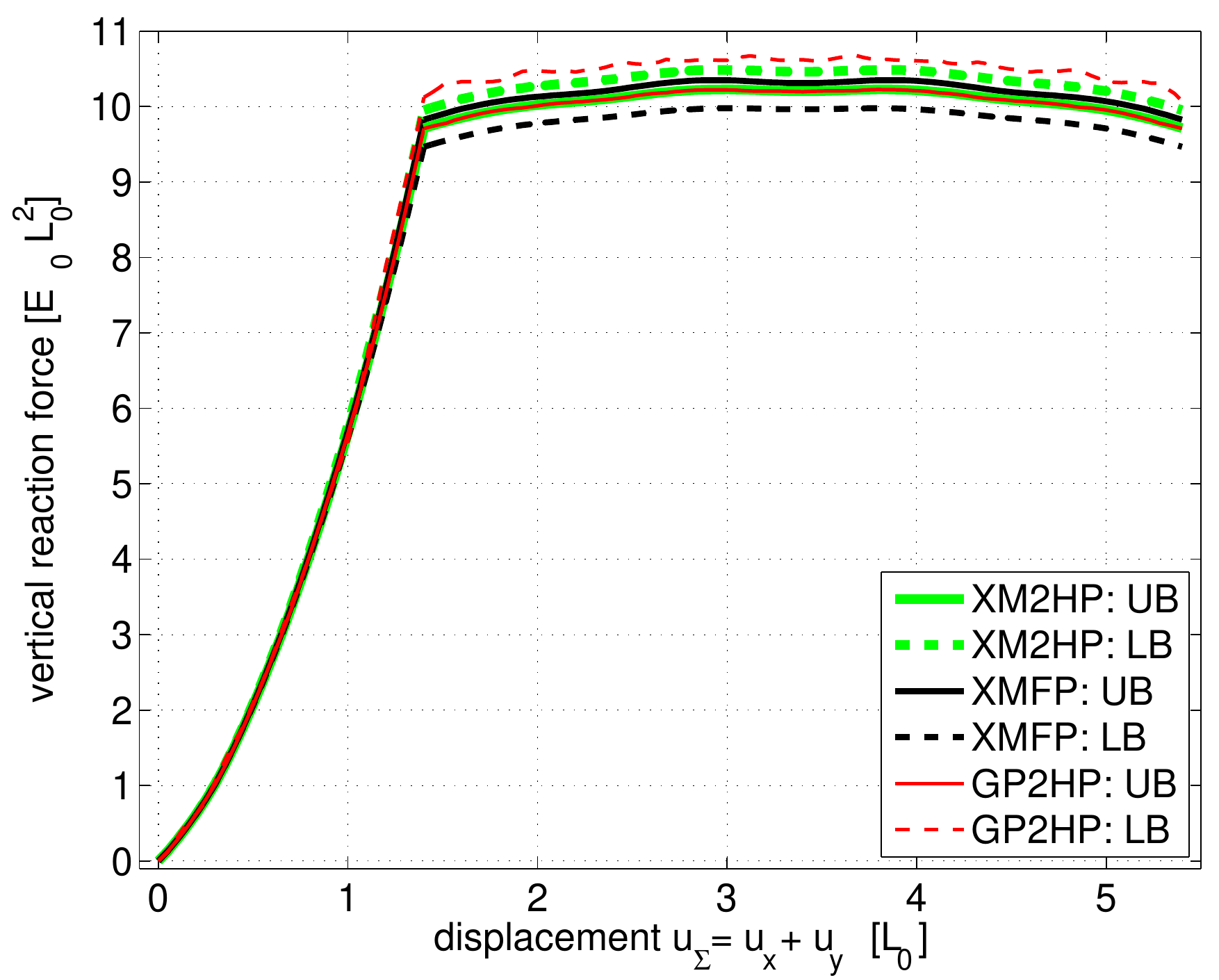}}
\put(3,9.2){\includegraphics[width=0.22\textwidth]{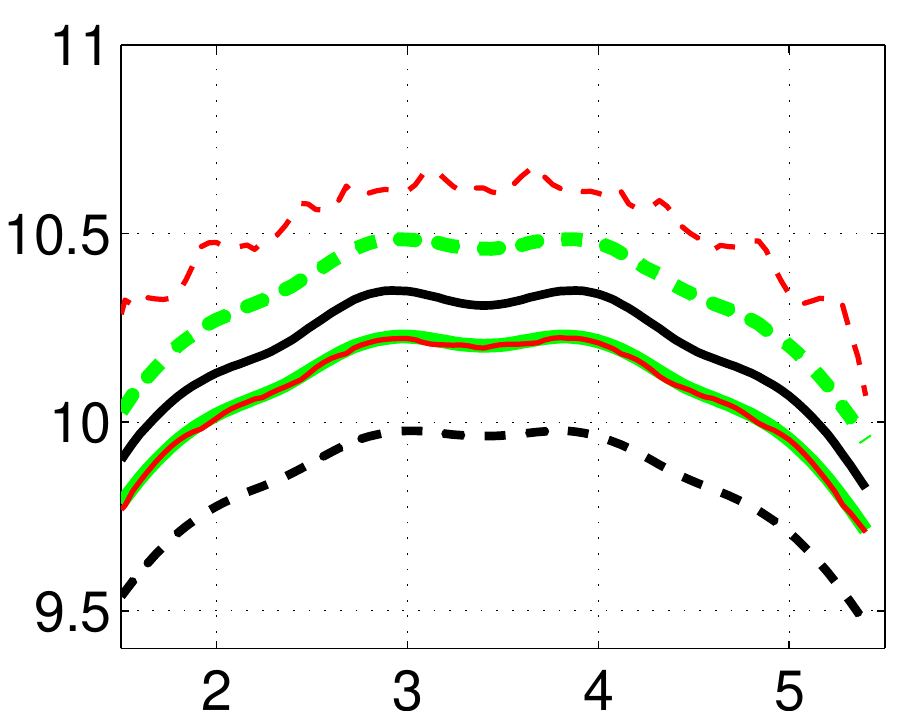}}

\put(-7.9,8){a. $u_\Sigma = 0$ }
\put(-7.9,1.8){b. $u_\Sigma = 1.4$ }
\put(0.4,1.8){c. $u_\Sigma = 4.4$ }
\put(0.4,6.2){d.}
\end{picture}
\caption{3D ironing: a.~initial configuration.  b.~and c.~Deformed configurations at the end of the pressing and the sliding phase for XM2HP. Here, the yellow curve shows the detected contact boundary  based on the RBQ technique \citep{rbq}. d.~Load-displacement curve for GP2HP, XMFP, and XM2HP.} 
\label{f:iron3D}
\end{center}
\end{figure}

3D ironing contact  between a ring  and slab is considered. In this example, extended mortar (full-pass/two-half-pass) is compared with GP2HP. The setup is depicted in Fig.~\ref{f:iron3D}a. The geometry and the material parameters are adopted from the example presented by \cite{puso04a}. The ring has $E=1000\,E_0$, $\nu=0.3$, radius $R=3\, L_0$, thickness $T=0.2\,L_0$, and width $W=5.2\,L_0$, and the slab has $E=1\,E_0$,  $\nu=0.3$ and dimension $9\,L_0\times4\,L_0\times3\,L_0$. The ring is discretized by quadratic NURBS shell elements \citep{solidshell}. The bulk of the slab is discretized by 3D linear finite elements, while its contact surface is enriched by quadratic NURBS elements \citep{nece}.\\[1.5mm]
The ring is pressed downward by $u_y=-1.4\,L_0$, and then moved in $X$-direction. $20\times20$ quadrature points per element are used for the quadrature of the contact terms. The snapshots of the deformed configurations at the end of each phase are shown in Fig.~\ref{f:iron3D}b-c.\\[1.5mm]
As seen in Fig.~\ref{f:iron3D}d, extended mortar gives comparable accuracy as GP2HP. Both extended mortar and GPTS can complete the simulation.
For a more irregular overlapping of the meshes during sliding, we consider a twisting example in the following.

\subsection{Twisted blocks}
\begin{figure}[!htp]
\begin{center} \unitlength1cm
\begin{picture}(0,12.9)

\put(-8.5,7.6){\includegraphics[height=55mm]{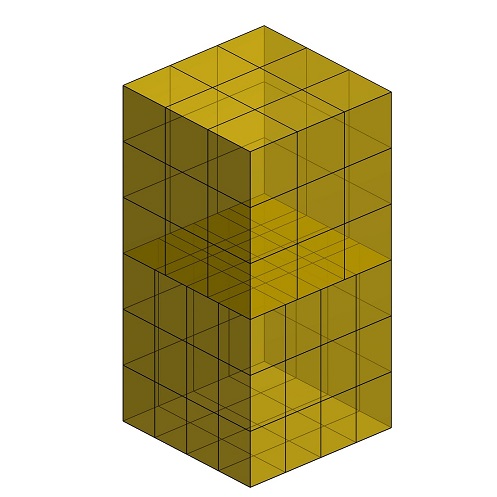}}

\put(-3.5,7.6){\includegraphics[height=55mm]{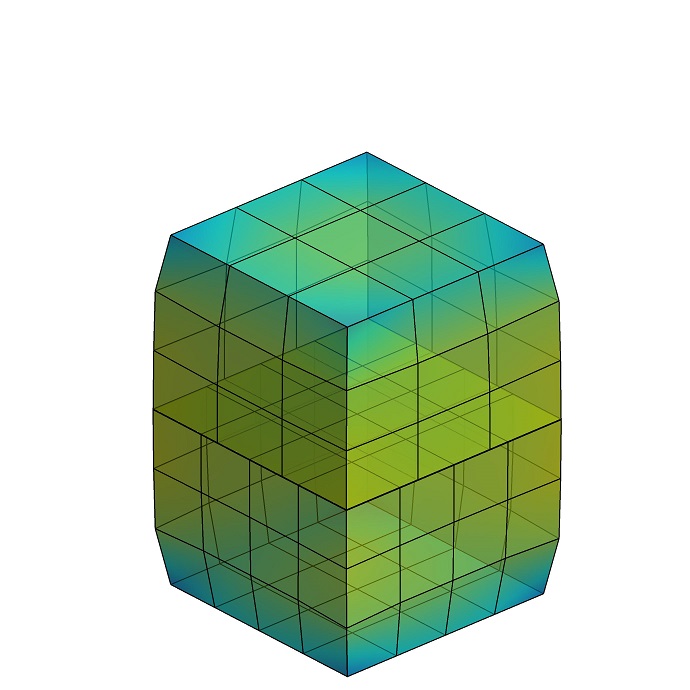}}
\put(1.5,7.6){\includegraphics[height=55mm]{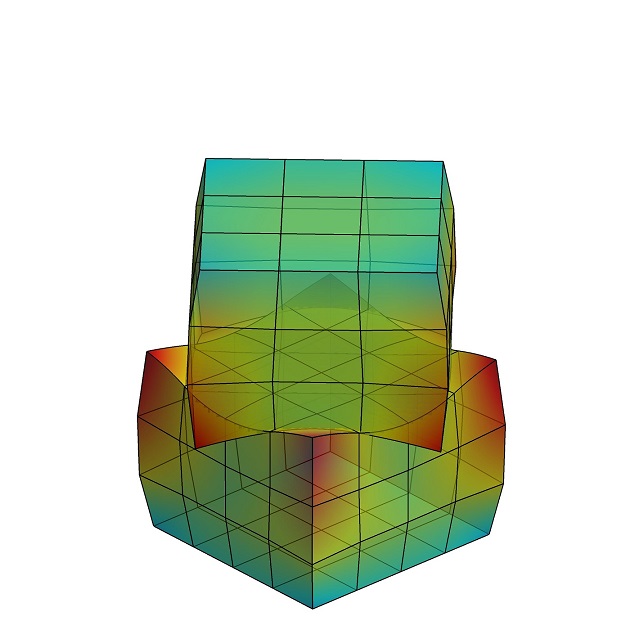}}

\put(6.55 ,7.6){\includegraphics[height=55mm]{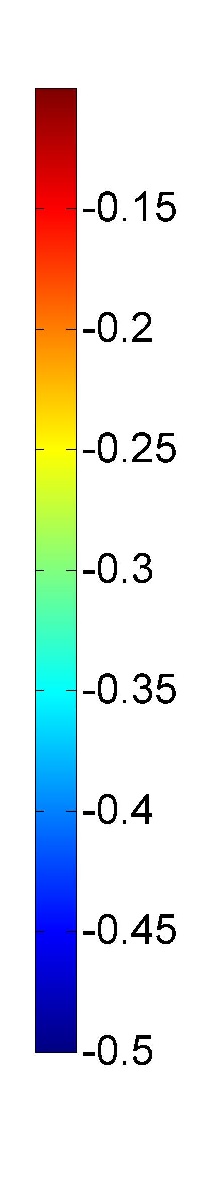}}

\put(-7.6,0.0){\includegraphics[height=70mm]{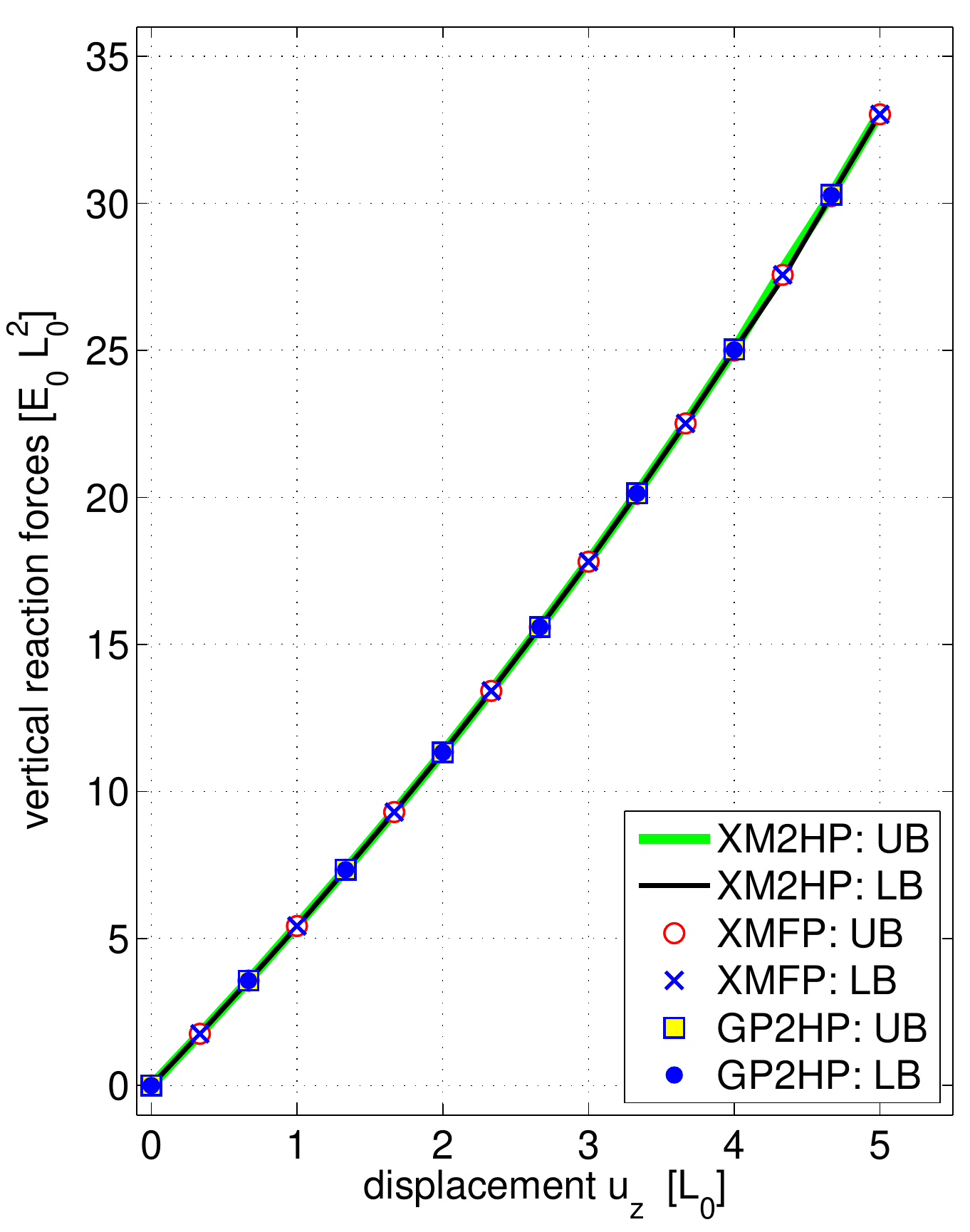}}
\put(-1.2,0.1){\includegraphics[height=70mm]{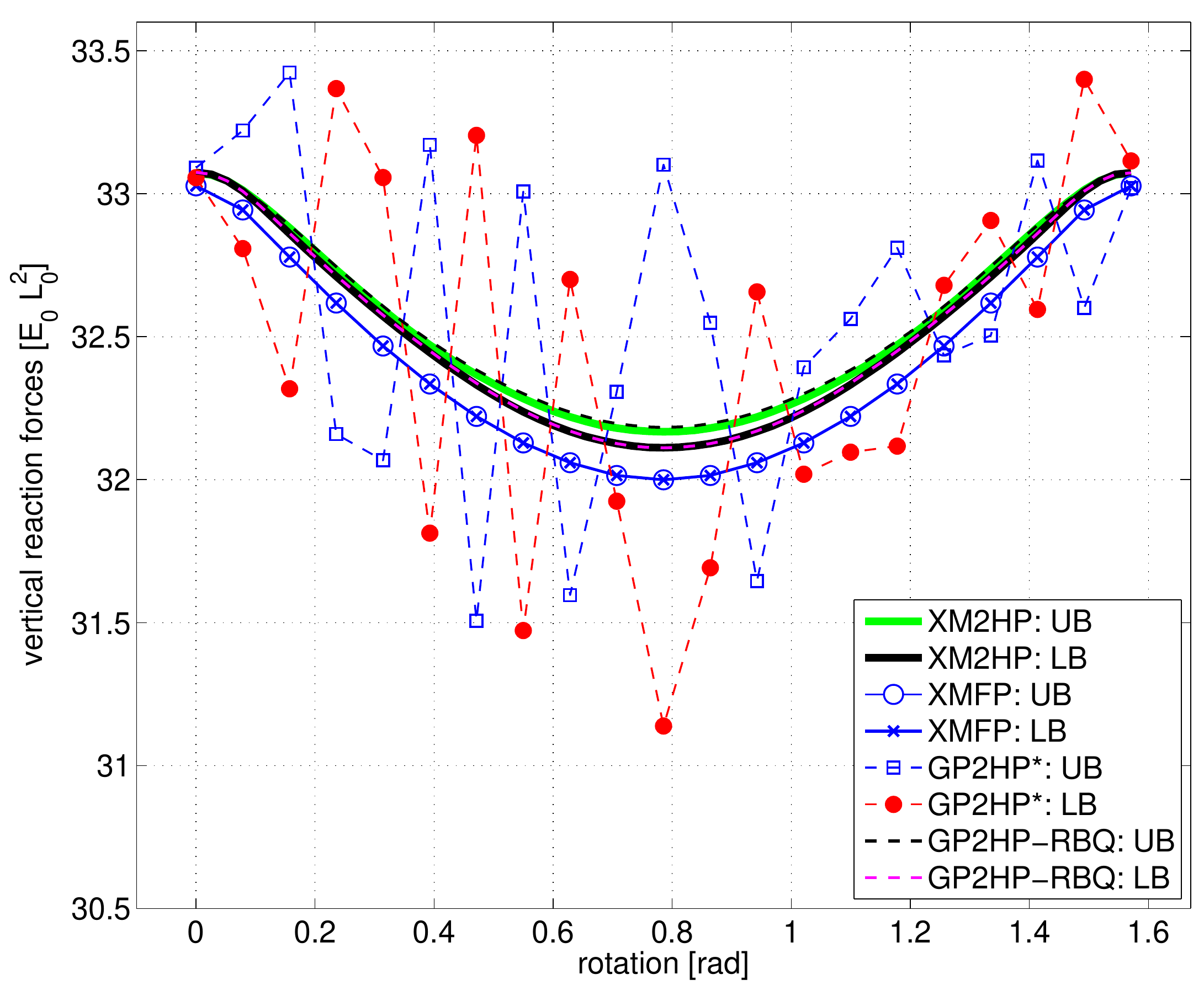}}

\put(-7.0,7.6){a.}
\put(-2.0,7.6){b.}
\put(3.0,7.6){c.}
\put(-7.0,0){d.}
\put(-0.5,0){e.}

\end{picture}
\caption{Twisting blocks: a.~initial configuration, and deformed configurations at the end of b.~the pressing phase and c.~the twisting phase (at $\phi=\pi/4$). Here, the color represents the vertical stress. d-e.~Load-displacement curves for various contact formulations. } 
\label{f:rot3D_load}
\end{center}

\begin{center} \unitlength1cm
\begin{picture}(0,7.5)
\put(-7.7,0){\includegraphics[height=70mm]{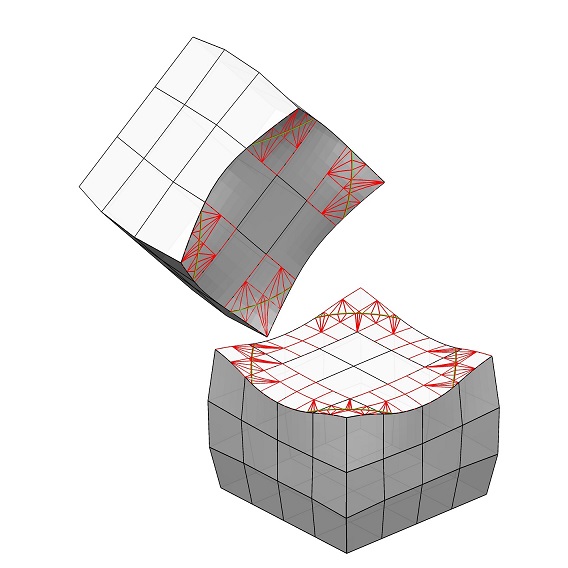}}
\put(-1,0){\includegraphics[height=70mm]{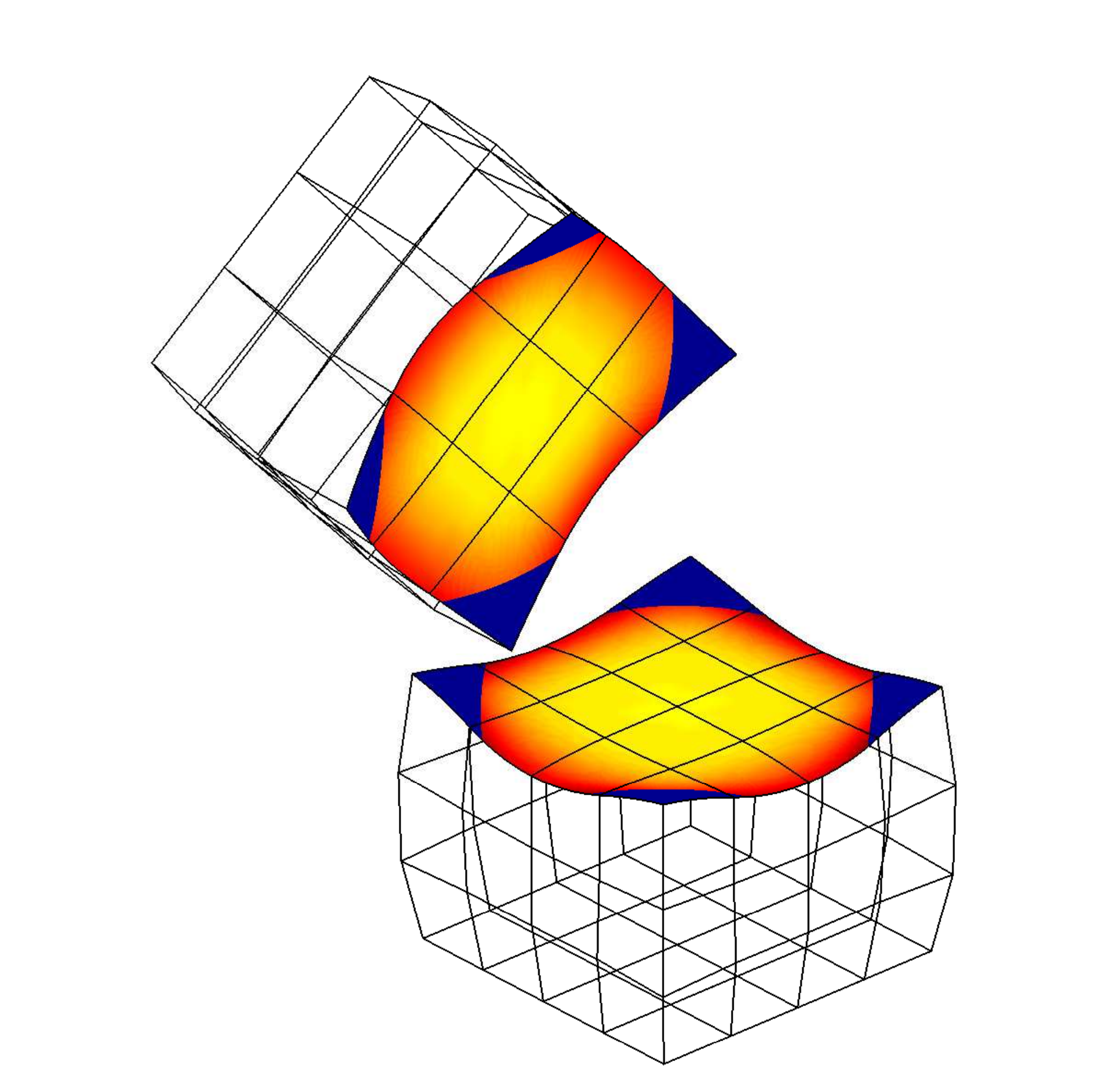}}
\put(6.2,0.0){\includegraphics[height=65mm]{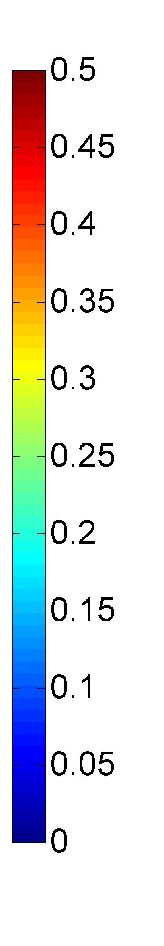}}
\end{picture}
\caption{Twisting block: partitioning of the refined boundary quadrature (left) for the twisting phase at $\phi=\pi/4$, and post-processed contact pressure (right).}
\label{f:rot3D_segm}
\end{center}
\end{figure}

In this example, the contact of two 3D blocks with dimensions $10L_0\times10L_0\times 10 L_0$ is simulated. The two blocks are first  pressed together and then twisted  against each other by prescribing $u_z=-5\,L_0$ and $\phi=\pi/2$, respectively, at the upper boundary of the upper block (see Figs.~\ref{f:rot3D_load}a-c). The load-displacement curves are plotted in Figs.~\ref{f:rot3D_load}d-e for extended mortar in comparison with GP2HP. In particular, both standard GP2HP and GP2HP-RBQ are examined. The penalty parameter $\epsilon=10E_0/L_0$ and $5\times5$ quadrature points per element are considered in the contact simulations.\\[1.5mm]
During the pressing phase, the computational results are almost identical for all considered contact formulations. This is due to the fact that no edge contact appears in this case.
During the twisting phase, however, GP2HP (and GPFP) fails to converge.  For continuation, the freezing of the active-set strategy is applied here (denoted as GP2HP*), but the results tend to strongly oscillate as is seen in Fig.~\ref{f:rot3D_load}e. This is caused by the sudden contact changes at the quadrature points. In contrast to this, both GP2HP-RBQ and extended mortar (either full-pass or two-half-pass) give smooth results (see Fig.~\ref{f:rot3D_load}e). XM2HP is shown to have comparable accuracy as GP2HP-RBQ.   Furthermore, in this simulation, XMFP is observed to have the least bias error in the vertical contact force among all considered contact formulations, even though we have not used segmentation at the overlapping element boundaries. This is seen in Fig.~\ref{f:rot3D_load}e. These results demonstrate the effectiveness of RBQ applied in the XMFP formulation. \\[1.5mm]
Furthermore, Fig.~\ref{f:rot3D_segm}a shows the contact boundary  captured explicitly by the RBQ partitioning. With this, apart from the benefit of improving the accuracy of the interpolation and the numerical quadrature at contact boundaries, an accurate post-processing for the contact pressure can also be carried out \citep{rbq}. This is illustrated in Fig.~\ref{f:rot3D_segm}b.
%
\section{Conclusions}\label{s:conclude}
A summary of the existing and proposed mortar contact formulations  that have been discussed in this paper is presented in Tab.~\ref{tab:compare}. 
%
\begin{table}[!htp]
\begin{small}
\begin{center}
\def\arraystretch{1.5}\tabcolsep=3pt
\resizebox{\textwidth}{!}{%
\begin{tabular}{|l|c|c|c|c|}
\hline 
 ~~~~~~~~~~~~~~Items &SMFP & SM2HP & XMFP &  XM2HP\\ 
\hline 
Interpolation of \textit{strong} physical discontinuities&yes & yes  & yes  & yes\\ 
And how?&consistency & consistency  & XFEM  & XFEM\\ 
\hline 
Accurate quadrature of \textit{strong} physical discontinuities&yes & yes  & yes  & yes\\ 
And how?&segmentation & segmentation & RBQ  & RBQ\\
\hline 
Interpolation of \textit{weak} physical discontinuities&no & no  & yes  & yes\\ 
And how?& -  &  -  & XFEM  & XFEM\\ 
\hline 
Accurate quadrature of \textit{weak} physical discontinuities&no & no  & yes  & yes\\ 
And how?& - & - & RBQ  & RBQ\\
\hline 
Master/slave choice& biased  & unbiased & biased & unbiased \\ 
\hline
Satisfaction of (local) momentum conservation& intrinsically  &as $h \rightarrow 0$ & intrinsically & as $h \rightarrow 0$  \\ 
\hline
\end{tabular}}

\caption{\textcolor{black}{A comparison between the standard and the extended mortar contact formulation (both with either full-pass and two-half-pass). $h$ stands for the element size.}}
\end{center} 
\label{tab:compare}  
\end{small}
 \end{table}   
In order to accurately capture  \textit{physical} discontinuities in the contact pressure, we propose here an \textit{extended mortar method} that is based on isogeometric analysis and XFEM enrichment. Both \textit{weak} and \textit{strong} physical discontinuities are captured.  For the accurate quadrature at the contact boundary, the \textit{refined boundary quadrature} (RBQ) technique \citep{rbq} is applied. Unlike the consistency treatment at the contact boundary of \citet{Cichosz2011}, extended mortar does not encounter ill-conditioning as long as the contact domain is properly refined.  \\[1.5mm]
Furthermore, unbiased mortar contact formulations are obtained by applying the \textit{two-half-pass}  approach \citep{spbc}. The computational results with these formulations are independent of the choice of the slave and master surfaces. At the same time, the mortar coupling term vanishes with the two-half-pass approach. An expensive segmentation is thus not needed. The numerical examples show that SM2HP gives highly inaccurate results, while XM2HP is as robust as SMFP, and as accurate as classical GPTS.\\[1.5mm]
Additionally, the contact patch test is examined for various contact formulations. 
\textcolor{black}{The testing results show that the two-half-pass version of the classical penalty mortar formulation (SM2HP) does not pass the patch test. Meanwhile,} the full-pass extended mortar with RBQ can pass the patch test within quadrature error. RBQ can also help GPTS full-pass to pass the patch test within quadrature error. Especially, the patch test is passed for the two-half-pass extended mortar without using segmentation and RBQ.\\[1.5mm]
%
An important extension of this work is to the use of the Lagrange multiplier method. This would require to choose a suitable level-set function. One possible choice for this can be found in \cite{rbq}.  However, how this choice affects the accuracy and the efficiency should be investigated in future research. The proposed mortar method can also be extended to frictional contact following the approaches e.g.~of \citet{puso04b} and \citet{Temizer2012f}. Another interesting extension is a consistent linearization of the RBQ procedure in order to improve the convergence of Newton iterations. In this paper, we have used an active-set strategy that avoids the linearization of the terms related to the contact boundaries. Nevertheless, it is possible to fully linearize the RBQ contributions, so that all kinds of non-linearities can be accounted for in the tangent matrices.   


\bigskip

\appendix

\section{Consistent linearization of contact formulations}\label{sap:contact}
This appendix provides a consistent linearization of various quantities for different contact formulations including classical GPTS, mortar methods, and extended isogeometric mortar method. From Eq.~\eqref{e:vircontactW}, in general we have
\eqb{l}
\Delta\delta\Pi_\mrc = \ds\sum_{e\in\sS_\mrs}\, \int_{\Gamma^e_0}  (p^*\, \Delta\delta{g}_{\mrn}  +  \Delta p^*\, \delta{g}_{\mrn}) \,\dif A_e~,
\label{e:LinvircontactW}
\eqe
where $ \Delta\delta{g}_{\mrn}$  is the usual term and can be found e.g. in \cite{spbf}. The term  $\Delta p^*$, however, depends on the contact formulation and the choice of the  mortar shape function. 

\subsection{Linearization of $g_\mrn$}
The linearization of raw gap~\eqref{e:rawgn} is analogous to variation~\eqref{e:vargn}, i.e.
\eqb{l}
\Delta g_\mrn = \bn_\mrp\cdot\big(\Delta\bx - \Delta\bx_\mrp \big)~,
\label{e:lingn}
\eqe
where $\Delta\bx = \mN_e\,\Delta\mx_e$ denotes the increment of $\bx$ due to a nodal increment $\Delta\mx_e$. Note that $\bx_\mrp$ depends on both $\mx_{\hat{e}}$ and $\mx_e$, i.e. $\bx_\mrp\big(\mx_{\hat{e}},~ \bxi_\mrp(\bx,\mx_{\hat{e}})\big)$, but variation~\eqref{e:vargn} and linearization~\eqref{e:lingn} are still exact due to $\bn_\mrp\cdot\delta\bn_\mrp= 0$ and $\bn_\mrp\cdot\ba_\alpha^\mrp = 0$. 
\subsection{Linearization of $\delta g_\mrn$}
For variation~\eqref{e:vargn}, its linearization reads
\eqb{l}
\Delta\delta g_\mrn := \Delta\bn_\mrp\cdot\big(\delta\bx - \delta\bx_\mrp \big)~.
\label{e:linvargn}
\eqe
Since $\bn_\mrp = \bn(\bxi_\mrp)$  is a normalized vector and computed at projection point~$\bx_\mrp$, it is a complicated function \citep{spbc}, 
\eqb{l}
\bn_\mrp = \bn_\mrp\big( \bx,~ \bxi_\mrp(\bx,\mx_{\hat{e}}),~ g_\mrn(\bx,\bx_\mrp)\big)~. 
\eqe
The linearization of $\bn_\mrp$ can be found in \citet{spbc} as
\eqb{l}
\Delta\bn_\mrp = \mP_\mrs\,\Delta\mx_e +  \mP_\mrm\,\Delta\mx_{\hat{e}}~,
\label{e:linnp}
\eqe
where
\eqb{llll}
\mP_\mrs \dis \ds\pa{\bn_\mrp}{\mx^\mrs_e} = \quad\ds\frac{1}{g_\mrn}\,(\bI - \bn_\mrp\otimes\bn_\mrp - c_\mrp^{\alpha\beta}\,\ba_{\alpha}^\mrp\otimes\ba_{\beta}^\mrp)\,\mN_e~,\\[3mm]
\mP_\mrm \dis \ds\pa{\bn_\mrp}{\mx_{\hat{e}}} = -\ds\frac{1}{g_\mrn}\,(\bI - \bn_\mrp\otimes\bn_\mrp - c_\mrp^{\alpha\beta}\,\ba_{\alpha}^\mrp\otimes\ba_{\beta}^\mrp)\,\mN_\mrp - c_\mrp^{\alpha\beta}\,\ba_{\alpha}^\mrp\otimes\bn_\mrp\,\mN_{\mrp,\beta}~.
\eqe
with $c_\mrp^{\alpha\beta}:=[a^\mrp_{\alpha\beta} - g_\mrn\,b^\mrp_{\alpha\beta}]^{-1}$. Inserting expression~\eqref{e:linnp} into Eq.~\eqref{e:linvargn} we find
\eqb{lll}
\Delta \delta g_\mrn \is  \delta\mx_e\cdot\,(\mN_e^\mrT\,\mP_\mrs\,\Delta\mx_e + \mN_e^\mrT\,\mP_\mrm\,\Delta\mx_{\hat{e}})~\\[2mm]
 \mi \delta\mx_{\hat{e}} \cdot\, (\mN^\mrT_\mrp\,\mP_\mrs\,\Delta\mx_e + \mN^\mrT_\mrp\,\mP_\mrm\,\Delta\mx_{\hat{e}})~.
\label{e:lingnN}
\eqe
As seen, $\Delta \delta g_\mrn$ can be computed in elementwise manner.
\subsection{Linearization of $p^*$ for GPTS}
For GPTS, one has $p^*:=\chi\, p = \chi\, J\,\bar{\epsilon}_\mrn g_\mrn = \chi\, \epsilon_\mrn\,g_\mrn$ (see Tab.~\ref{t:vircontactW}). Since an active set strategy is used, $\chi$ is considered to be fixed. We thus have $\Delta p^*= \chi\,\Delta p$. Furthermore, if  $\epsilon_\mrn$ is considered fixed, the linearization of $p$ reads
\eqb{l}
\Delta p\textcolor{black}{|_{\epsilon_{\mrn} \, \mathrm{fixed}}} = \epsilon_\mrn\,\bn_\mrp\cdot\big(\Delta\bx - \Delta\bx_\mrp \big) =  \epsilon_\mrn\,\bn_\mrp\cdot\big(\mN_e\,\Delta\mx_e - \mN_\mrp\,\Delta\mx_{\hat{e}} \big)~,
\label{e:linp1}
\eqe
due to Eq.~\eqref{e:lingn}. However, if $\bar{\epsilon}_\mrn$ is considered fixed, we need the linearization of $J$, which can be found in e.g.~\citet{membrane} as
\eqb{l}
\Delta J = J\,\ba^\alpha\cdot\Delta \ba_\alpha =  \textcolor{black}{J\,\ba^\alpha\cdot}\,\mN_{e,\alpha}\, \Delta\mx_e~,
\eqe
due to Eq.~\eqref{e:tangentsl}. In this case, we find
\eqb{l}
\Delta p\textcolor{black}{|_{\bar{\epsilon}_{\mrn} \, \mathrm{fixed}}} = \epsilon_\mrn\,\bn_\mrp\cdot\big(\mN_e\,\Delta\mx_e - \mN_\mrp\,\Delta\mx_{\hat{e}} \big) + \epsilon_\mrn\,\ba^\alpha\cdot\mN_{e,\alpha}\, \Delta\mx_e ~,
\label{e:linp2}
\eqe
with $\epsilon_\mrn = J\,\bar{\epsilon}_\mrn$. 

\subsection{Linearization of $p^*$ and $\tilde{\mpp}$}
As seen from Eq.~\eqref{e:unifiedP}, in general, the  linearization of $\tilde{\mpp}$ for the mortar formulations requires the linearization of $\mG$, 
$\mPhi$,  
and $p$. Nevertheless, in our case, $\mG$ and $\mPhi$ 
are fixed since the active-set strategy is employed and the mortar shape functions $M_A$ can be constructed such that it is independent from element distortion.
 Therefore, the linearization of $\tilde\mpp$ simplifies to
\eqb{l}
\Delta\tilde{\mpp}= \mG\, \ds\int_{\Gamma^\mrs_0} {\mPhi}^\mrT_\mrs \,\Delta p\,\dif A_\mrs ~.
\label{e:LinP}
\eqe
Inserting Eq.~\eqref{e:linp1} into Eq.~\eqref{e:LinP} for $\epsilon_\mrn$ fixed, gives
\eqb{l}
\Delta\tilde{\mpp}= \mG\, \ds\int_{\Gamma^\mrs_0} \Big( {\mPhi}^\mrT_\mrs \, \epsilon_\mrn\,\bn_\mrp^\mrT\,\mN_e\,\Delta\mx_e - {\mPhi}^\mrT_\mrs \, \epsilon_\mrn\,\bn_\mrp^\mrT\,\mN_\mrp\,\Delta\mx_{\hat{e}} \Big)\,\dif A_\mrs ~.
\label{e:LinP1}
\eqe
Here, the integral implies here the assembly into a global matrix. Thus, we can rewrite  Eq.~\eqref{e:LinP1} into
\eqb{l}
\Delta\tilde{\mpp}= \mG\,(\hat{\mM}_{\mrs\mrs}\,\Delta\mx^\mrs - \hat{\mM}_{\mrs\mrm}\,\Delta\mx^\mrm) ~,
\label{e:LinP2}
\eqe
where 
\eqb{l}
\hat{\mM}_{\mrs\mrs}:=  \ds\int_{\Gamma^\mrs_0} {\mPhi}^\mrT_\mrs \, \epsilon_\mrn\,\bn_\mrp^\mrT\,\mN_e\,\dif A_\mrs~,\quad \hat{\mM}_{\mrs\mrm}:=  \ds\int_{\Gamma^\mrs_0} {\mPhi}^\mrT_\mrs \, \epsilon_\mrn\,\bn_\mrp^\mrT\,\mN_\mrp\,\dif A_\mrs~.
\eqe
Furthermore, for the linearization of $p^*$ (see Eq.~\eqref{e:unifiedpdual}), we have
\eqb{l}
\Delta p^* = \ds\sum_{A=1}^{n_e}\,\Phi_A\,\Delta\tilde{p}_A = \mPhi_\mrs\,\Delta \tilde{\mpp}^e~,
\label{e:linunifiedpdual}
\eqe 
since $\mPhi$ is fixed. Note that $\Delta\tilde{\mpp}^e$ is the elemental array extracted from global array~$\Delta\tilde{\mpp}$.
\subsection{Linearization of $\Delta\delta\Pi_{\mrc}$: the first term}
By inserting Eq.~\eqref{e:lingnN} into the first term of Eq.~\eqref{e:LinvircontactW}, we have 
\eqb{lll}
\Delta\delta\Pi_{\mrc}^1 \is \delta\mx\cdot\,\mk^1\,\Delta\mx~,
\eqe
where $\mx:=[\mx^\mrs;~\mx^\mrm]$ contain positions of all nodes on the two contact surfaces, and $\mk^1$ is defined by
\eqb{lll}
\mk^1:= \left[\begin{array}{cc}
\mk^1_{\mrs\mrs} & \mk^1_{\mrs\mrm} \\ 
\mk^1_{\mrm\mrs} & \mk^1_{\mrm\mrm}
\end{array} \right]~,
\label{e:link1}
\eqe
and
\eqb{lll}
\mk^1_{\mrs\mrs} \dis   \ds\int_{\Gamma^\mrs_0} \mN_e^\mrT\,p^*\,\mP_\mrs\,\dif A_\mrs~,\quad  \mk^1_{\mrs\mrm}:= \ds\int_{\Gamma^\mrs_0} \mN_e^\mrT\,p^*\,\mP_\mrm\,\dif A_\mrs~,\\[4mm]
\mk^1_{\mrm\mrs} \dis   \ds\int_{\Gamma^\mrs_0} \mN_\mrp^\mrT\, p^*\, \mP_\mrs\,\dif A_\mrs~,\quad  \mk^1_{\mrm\mrm}:= \ds\int_{\Gamma^\mrs_0} \mN_\mrp^\mrT\,p^*\,\mP_\mrm\,\dif A_\mrs~.
\eqe
\subsection{Linearization of $\Delta\delta\Pi_{\mrc}$: the second term}
For the second term in  Eq.~\eqref{e:LinvircontactW}, we consider GPTS and mortar formulation separately.
\subsubsection{For GPTS}
 Inserting Eqs.~\eqref{e:vargn} and \eqref{e:linp1} (considering $\epsilon_\mrn$ fixed) into the second term of Eq.~\eqref{e:LinvircontactW}, leads to
\eqb{lll}
\Delta\delta\Pi_{\mrc}^2 = \delta\mx\cdot\,\mk^2\,\Delta\mx~,
\label{e:stiffterm2}
\eqe
where  $\mk^2$ is the tangent matrix associated with the second term of Eq.~\eqref{e:LinvircontactW}, given by
\eqb{lll}
\mk^2:= (\chi\,\epsilon_\mrn)\,\left[\begin{array}{cc}
\mk^2_{\mrs\mrs} & \mk^2_{\mrs\mrm} \\ 
\mk^2_{\mrm\mrs} & \mk^2_{\mrm\mrm}
\end{array} \right]~,
\label{e:link2}
\eqe
with
\eqb{lll}
\mk^2_{\mrs\mrs} \dis   \ds\int_{\Gamma^\mrs_0} \mN_e^\mrT\,\bn_\mrp\otimes\bn_\mrp\,\mN_e\,\dif A_\mrs~,\quad  \mk^2_{\mrs\mrm}:= \ds\int_{\Gamma^\mrs_0} \mN_e^\mrT\,\bn_\mrp\otimes\bn_\mrp\,\mN_\mrp\,\dif A_\mrs~,\\[4mm]
\mk^2_{\mrm\mrs} \dis   \ds\int_{\Gamma^\mrs_0} \mN_\mrp^\mrT\, \bn_\mrp\otimes\bn_\mrp\, \mN_e\,\dif A_\mrs~,\quad  \mk^2_{\mrm\mrm}:= \ds\int_{\Gamma^\mrs_0} \mN_\mrp^\mrT\,\bn_\mrp\otimes\bn_\mrp\,\mN_\mrp\,\dif A_\mrs~.
\eqe
\subsubsection{For mortar formulations}
 Inserting Eqs.~\eqref{e:vargn} and \eqref{e:linunifiedpdual} into  the second term of Eq.~\eqref{e:LinvircontactW}, gives
 \eqb{lll}
\Delta\delta\Pi_{\mrc}^2 = \left(\ds \delta\mx^\mrs\cdot\, \int_{\Gamma^\mrs_0} \mN_e^\mrT\,\bn_\mrp\,\mPhi_\mrs \dif A_\mrs - \delta\mx^\mrm\cdot\, \int_{\Gamma^\mrs_0} \mN_\mrp^\mrT\,\bn_\mrp\,\mPhi_\mrs \dif A_\mrs\right)\,\Delta\tilde{\mpp}
\eqe
The integrals here imply the assembly into a global matrix. Thus, we can rewrite  this expression as
\eqb{l}
\Delta\delta\Pi_{\mrc}^2= \Big(\delta\mx^\mrs\cdot\mM^\mrT_{\mrs\mrs}- \delta\mx^\mrm\cdot\mM^\mrT_{\mrs\mrm}\Big)\,\Delta\tilde{\mpp}~,
\label{e:LinTerm2}
\eqe
where 
\eqb{l}
\mM_{\mrs\mrs}:=  \ds\int_{\Gamma^\mrs_0}\mPhi^\mrT_\mrs \, \epsilon_\mrn\,\bn_\mrp^\mrT\,\mN\,\dif A_\mrs~,\quad \mM_{\mrs\mrm}:=  \ds\int_{\Gamma^\mrs_0} \mPhi^\mrT_\mrs \, \epsilon_\mrn\,\bn_\mrp^\mrT\,\mN_\mrp\,\dif A_\mrs~.
\eqe
Furthermore, by inserting Eq.~\eqref{e:LinP2} into Eq.~\eqref{e:LinTerm2}, we finally get the same form as Eq.~\eqref{e:stiffterm2}, where $\mk^2$ is now given by
\eqb{lll}
\mk^2:= \left[\begin{array}{cc}
\mk^2_{\mrs\mrs} & \mk^2_{\mrs\mrm} \\ 
\mk^2_{\mrm\mrs} & \mk^2_{\mrm\mrm}
\end{array} \right]~,
\label{e:link3}
\eqe
and
\eqb{lll}
\mk^2_{\mrs\mrs} \dis \mM^\mrT_{\mrs\mrs}\,\mG\,\hat{\mM}_{\mrs\mrs} ~,\quad  \mk^2_{\mrs\mrm}:= -\mM^\mrT_{\mrs\mrs}\,\mG\,\hat{\mM}_{\mrs\mrm}~,\\[4mm]
\mk^2_{\mrm\mrs} \dis   - \mM^\mrT_{\mrs\mrm}\,\mG\,\hat{\mM}_{\mrs\mrs}~,\quad  \mk^2_{\mrm\mrm}:= \mM^\mrT_{\mrs\mrm}\,\mG\,\hat{\mM}_{\mrs\mrm}~.
\eqe
Here $\mM_{\mrs\mrs}$, $\mM_{\mrs\mrm}$,  $\hat{\mM}_{\mrs\mrs}$, $\hat{\mM}_{\mrs\mrm}$, and $\mG$ are all global matrices.
\subsubsection{For two-half-pass algorithm}
For two-half-pass algorithm, since the master surface is considered to be variationally fixed, all associated stiffness matrices, i.e.~$\mk_{\mrm\mrm}$ and $\mk_{\mrm\mrs}$, in Eqs.~\eqref{e:link1}, \eqref{e:link2}, \eqref{e:link3}  are disregarded.

\vspace{1cm}
{\Large{\bf Acknowledgements}}

The authors are grateful to the German Research Foundation (DFG)
for supporting this research under grants  GSC 111 and SA1822/8-1. 

\bigskip

\bibliographystyle{apalike}
\bibliography{Bibliography,sauerduong}
\end{document}

%% file: neco_updated.tex
\newcommand{\bitm}{\begin{itemize}}
\newcommand{\eitm}{\end{itemize}}
\newcommand{\bnumr}{\begin{enumerate}}
\newcommand{\enumr}{\end{enumerate}}

\newcommand{\mcalC}{\mathcal{C}}

\newcommand{\mrT}{\mathrm{T}}

\newcommand{\mrin}{\mathrm{in}}

\newcommand {\eqb}[1]{\begin{equation}\begin{array}{#1}}
\newcommand {\eqe}{\end{array}\end{equation}}

\newcommand {\esb}[1]{\begin{equation*}\begin{array}{#1}}
\newcommand {\ese}{\end{array}\end{equation*}}
\newcommand {\ds}{\displaystyle}

\newcommand {\pa}[2]{\frac{\partial{#1}}{\partial{#2}}}

\newcommand {\back}{\! \! \!}
\newcommand {\is}{\back &=& \back}
\newcommand {\dis}{\back &:=& \back}

\newcommand {\mi}{\back &-& \back}

\newcommand {\norm}[1]{\|#1\|}
\newcommand {\tr}{\mathrm{tr}\,}
\newcommand {\sign}{\mathrm{sign}\,}

\newcommand {\dif}{\mathrm{d}}

\newcommand {\II}{{I\kern-.3em I}}
\newcommand {\III}{{I\kern-.3em I\kern-.3em I}}

\newcommand {\mrc}{\mathrm{c}}

\newcommand {\mrg}{\mathrm{g}}

\newcommand {\mrm}{\mathrm{m}}
\newcommand {\mrn}{\mathrm{n}}
\newcommand {\mro}{\mathrm{o}}
\newcommand {\mrp}{\mathrm{p}}

\newcommand {\mrs}{\mathrm{s}}

\newcommand {\mrx}{\mathrm{x}}

\newcommand {\mf}{\mathbf{f}}
\newcommand {\mg}{\mathbf{g}}

\newcommand {\mk}{\mathbf{k}}

\newcommand {\mn}{\mathbf{n}}

\newcommand {\mpp}{\mathbf{p}}

\newcommand {\mx}{\mathbf{x}}

\newcommand {\ba}{\boldsymbol{a}}

\newcommand {\bn}{\boldsymbol{n}}

\newcommand {\bx}{\boldsymbol{x}}

\newcommand {\bxi}{\mbox{\boldmath$\xi$}}

\newcommand {\mA}{\mathbf{A}}

\newcommand {\mG}{\mathbf{G}}

\newcommand {\mM}{\mathbf{M}}
\newcommand {\mN}{\mathbf{N}}

\newcommand {\mP}{\mathbf{P}}

\newcommand {\mW}{\mathbf{W}}

\newcommand {\mPhi}{\mathbf{\Phi}}

\newcommand {\bA}{\boldsymbol{A}}

\newcommand {\bI}{\boldsymbol{I}}

\newcommand {\bT}{\boldsymbol{T}}

\newcommand {\bsig}{\mbox{\boldmath$\sigma$}}

\newcommand {\IR}{{\rm\kern.24em
   \vrule width.02em height1.53ex depth-.05ex
   \kern-.3em R}}
\newcommand {\ic}{{\rm\kern.20em
   \vrule width.02em height1.0ex depth-.05ex
   \kern-.22em c}}
\newcommand {\ia}{{\rm\kern.20em
   \vrule width.02em height1.05ex depth-.0ex
   \kern-.25em a}}
\newcommand {\IC}{{\rm\kern.24em
   \vrule width.02em height1.4ex depth-.05ex
   \kern-.26em C}}
\newcommand {\ID}{{\rm\kern.34em
   \vrule width.02em height1.5ex depth-.05ex
   \kern-.36em D}}
\newcommand {\IS}{{\rm\kern.24em
   \vrule width.02em height1.6ex depth.05ex
   \kern-.26em S}}
\newcommand {\IT}{{\rm\kern.50em
   \vrule width.02em height1.55ex depth-.05ex
   \kern-.52em T}}

\newcommand {\IE}{{\rm\kern.24em
   \vrule width.02em height1.55ex depth-.05ex
   \kern-.33em E}}
\newcommand {\IEa}{{\rm\kern.24em
   \vrule width.02em height1.55ex depth-.05ex
   \kern-.33em E}^{1}_{ijkl}}
\newcommand {\IEb}{{\rm\kern.24em
   \vrule width.02em height1.55ex depth-.05ex
   \kern-.33em E}^{2}_{ijkl}}

\newcommand {\sE}{\mathcal{E}}

\newcommand {\sP}{\mathcal{P}}

\newcommand {\sS}{\mathcal{S}}

\newcommand {\Ass}[2]{\kern 0.9ex \vrule width0.45em height0.2ex depth0ex \kern -2.1ex \bigwedge_{#1}^{#2}}
\newcommand {\ASS}[2]{\kern 1.45ex \vrule width0.5em height0.2ex depth0ex \kern -2.65ex \bigwedge_{#1}^{#2}}

%% file: MortarXFEM_R2.bbl
\begin{thebibliography}{}

\bibitem[Bartels et~al., 1996]{Richard96}
Bartels, R., Beatty, J., and Barsky, B. (1996).
\newblock {\em An Introduction to Splines for Use in Computer Graphics and
  Geometric Modeling}.
\newblock Morgan Kaufmann.

\bibitem[Brivadis et~al., 2015]{Buffa15}
Brivadis, E., Buffa, A., Wohlmuth, B., and Wunderlich, L. (2015).
\newblock Isogeometric mortar methods.
\newblock {\em Comput. Methods Appl. Mech. Engrg.}, {\bf 284}(Supplement C):292
  -- 319.

\bibitem[Cichosz and Bischoff, 2011]{Cichosz2011}
Cichosz, T. and Bischoff, M. (2011).
\newblock Consistent treatment of boundaries with mortar contact formulations
  using dual {L}agrange multipliers.
\newblock {\em Comput. Methods Appl. Mech. Engrg.}, {\bf{200}}:1317--1332.

\bibitem[Corbett and Sauer, 2014]{nece}
Corbett, C.~J. and Sauer, R.~A. (2014).
\newblock {NURBS}-enriched contact finite elements.
\newblock {\em Comput. Methods Appl. Mech. Engrg.}, {\bf 275}:55--75.

\bibitem[{De~Lorenzis} et~al., 2014]{Laura2014}
{De~Lorenzis}, L., , Wriggers, P., and Hughes, T. J.~R. (2014).
\newblock Isogeometric contact: {A} review.
\newblock {\em GAMM Mitteilungen}, {\bf {37}}:85--123.

\bibitem[{De~Lorenzis} et~al., 2012]{Lorenzis2012}
{De~Lorenzis}, L., Wriggers, P., and Zavarise, G. (2012).
\newblock A mortar formulation for {3D} large deformation contact using
  {NURBS}-based isogeometric analysis and the augmented {L}agrangian method.
\newblock {\em Comput. Mech.}, {\bf{49}}:1--20.

\bibitem[De~Luycker et~al., 2011]{Luycker11}
De~Luycker, E., Benson, D.~J., Belytschko, T., Bazilevs, Y., and Hsu, M.~C.
  (2011).
\newblock X-fem in isogeometric analysis for linear fracture mechanics.
\newblock {\em Int. J. Numer. Meth. Engrg.}, {\bf{87}}:541--565.

\bibitem[Dittmann et~al., 2014]{Dittmann14}
Dittmann, M., Franke, M., Temizer, I., and Hesch, C. (2014).
\newblock Isogeometric analysis and thermomechanical mortar contact problems.
\newblock {\em Comp. Meth. Appl. Mech. Engrg.}, {\bf{274}}:192--212.

\bibitem[Duong, 2017]{duong-phd}
Duong, T.~X. (2017).
\newblock {\em Efficient contact computations based on isogeometric
  discretization, mortar methods and refined boundary quadrature}.
\newblock PhD thesis, RWTH Aachen.

\bibitem[Duong et~al., 2017a]{solidshell}
Duong, T.~X., Roohbakhshan, F., and Sauer, R.~A. (2017a).
\newblock A new rotation-free isogeometric thin shell formulation and a
  corresponding continuity constraint for patch boundaries.
\newblock {\em Comput. Methods Appl. Mech. Engrg.}, {\bf 316}:13--28.

\bibitem[Duong and Sauer, 2015]{rbq}
Duong, T.~X. and Sauer, R.~A. (2015).
\newblock An accurate quadrature technique for the contact boundary in {3D}
  finite element computations.
\newblock {\em Comput. Mech.}, {\bf 55}(1):145--166.

\bibitem[Duong et~al., 2017b]{duong17a}
Duong, X.~T., Lorenzis, L.~D., and Sauer, R.~A. (2017b).
\newblock On the shape functions for the contact pressure in mortar methods.
\newblock In von Scheven, M., Keip, M.-A., and Karajan, N., editors, {\em
  Proceedings of the 7th GACM Colloquium on Computational Mechanics}, pages
  130--133.

\bibitem[Fries and Belytschko, 2006]{Fries2006}
Fries, T.-P. and Belytschko, T. (2006).
\newblock The intrinsic {XFEM}: a method for arbitrary discontinuities without
  additional unknowns.
\newblock {\em Int. J. Numer. Meth. Engrg.}, {\bf 68}:1358--1385.

\bibitem[Graveleau et~al., 2015]{Graveleau2015}
Graveleau, M., Chevaugeon, N., and Mo{\"e}s, N. (2015).
\newblock The inequality level-set approach to handle contact: membrane case.
\newblock {\em Adv Model Simul Eng Sci.}, 2(1):16--31.

\bibitem[Hartmann and Ramm, 2008]{Hartmann2008}
Hartmann, S. and Ramm, E. (2008).
\newblock A mortar based contact formulation for non-linear dynamics using dual
  {L}agrange multipliers.
\newblock {\em Finite Elem. Anal. Des.}, {\bf{44}}:245--258.

\bibitem[Hesch and Betsch, 2008]{Hesch2008}
Hesch, C. and Betsch, P. (2008).
\newblock A mortar method for energy-momentum conserving schemes in
  frictionless dynamic contact problems.
\newblock {\em Int. J. Numer. Meth. Engrg.}, {\bf{77}}:1468--1500.

\bibitem[Hesch and Betsch, 2011]{hesch2011}
Hesch, C. and Betsch, P. (2011).
\newblock Transient three-dimensional contact problems: {M}ortar method.
  {M}ixed methods and conserving integration.
\newblock {\em Comput. Mech.}, {\bf{48}}:461--475.

\bibitem[Hughes et~al., 2005]{hughes05}
Hughes, T. J.~R., Cottrell, J.~A., and Bazilevs, Y. (2005).
\newblock Isogeometric analysis: {CAD}, finite elements, {NURBS}, exact
  geometry and mesh refinement.
\newblock {\em Comp. Meth. Appl. Mech. Engrg.}, {\bf{194}}:4135--4195.

\bibitem[Kim and Youn, 2012]{Kim12}
Kim, J.-Y. and Youn, S.-K. (2012).
\newblock Isogeometric contact analysis using mortar method.
\newblock {\em Int. J. Numer. Meth. Engrg.}, {\bf 89}(12):1559--1581.

\bibitem[McDevitt and Laursen, 2000]{McDevitt2000}
McDevitt, T.~W. and Laursen, T.~A. (2000).
\newblock A mortar-finite element formulation for frictional contact problems.
\newblock {\em Int. J. Numer. Meth. Engrg.}, {\bf{48}}:1525--1547.

\bibitem[Mo{\"e}s et~al., 1999]{Moes99}
Mo{\"e}s, N., Dolbow, J., and Belytschko, T. (1999).
\newblock A finite element method for crack growth without remeshing.
\newblock {\em Int. J. Numer. Meth. Engrg.}, {\bf{46}}:131--150.

\bibitem[Ogden, 1987]{ogden}
Ogden, R.~W. (1987).
\newblock {\em Non-Linear Elastic Deformations}.
\newblock Dover Edition, Mineola.

\bibitem[Popp et~al., 2013]{Apop2013}
Popp, A., A.~Seitz, M.~G., and Wall, W. (2013).
\newblock Improved robustness and consistency of {3D} contact algorithms based
  on a dual mortar approach.
\newblock {\em Comput. Methods Appl. Mech. Engrg.}, {\bf{264}}:67--80.

\bibitem[Popp et~al., 2012]{Apop2012}
Popp, A., Wohlmuth, B.~I., Gee, M.~W., and Wall, W.~A. (2012).
\newblock Dual quadratic mortar finite element methods for {3D} finite
  deformation contact.
\newblock {\em SIAM J. Sci. Comput.}, {\bf{34}}:B421--B446.

\bibitem[Puso and Laursen, 2004a]{puso04a}
Puso, M.~A. and Laursen, T.~A. (2004a).
\newblock A mortar segment-to-segment contact method for large deformation
  solid mechanics.
\newblock {\em Comput. Methods Appl. Mech. Engrg.}, {\bf 193}:601--629.

\bibitem[Puso and Laursen, 2004b]{puso04b}
Puso, M.~A. and Laursen, T.~A. (2004b).
\newblock A mortar segment-to-segment frictional contact method for large
  deformations.
\newblock {\em Comput. Methods Appl. Mech. Engrg.}, {\bf 193}:4891--4913.

\bibitem[Sauer, 2013]{sauer-ece2}
Sauer, R.~A. (2013).
\newblock Local finite element enrichment strategies for {2D} contact
  computations and a corresponding postprocessing scheme.
\newblock {\em Comput. Mech.}, {\bf 52}(2):301--319.

\bibitem[Sauer and {De~Lorenzis}, 2013]{spbc}
Sauer, R.~A. and {De~Lorenzis}, L. (2013).
\newblock A computational contact formulation based on surface potentials.
\newblock {\em Comput. Methods Appl. Mech. Engrg.}, {\bf 253}:369--395.

\bibitem[Sauer and {De~Lorenzis}, 2015]{spbf}
Sauer, R.~A. and {De~Lorenzis}, L. (2015).
\newblock An unbiased computational contact formulation for {3D} friction.
\newblock {\em Int. J. Numer. Meth. Engrg.}, {\bf{101}}:251--280.

\bibitem[Sauer et~al., 2014]{membrane}
Sauer, R.~A., Duong, T.~X., and Corbett, C.~J. (2014).
\newblock A computational formulation for constrained solid and liquid
  membranes considering isogeometric finite elements.
\newblock {\em Comput. Methods Appl. Mech. Engrg.}, {\bf 271}:48--68.

\bibitem[Seitz et~al., 2016]{Apop2016}
Seitz, A., Farah, P., Kremheller, J., Wohlmuth, B.~I., Wall, W.~A., and Popp,
  A. (2016).
\newblock Isogeometric dual mortar methods for computational contact mechanics.
\newblock {\em Comput. Methods Appl. Mech. Engrg.}, {\bf{301}}:259--280.

\bibitem[Temizer, 2012]{temizer2012}
Temizer, I. (2012).
\newblock A mixed formulation of mortar-based frictionless contact.
\newblock {\em Comput. Methods Appl. Mech. Engrg.}, {{\bf{223-224}}}:173--185.

\bibitem[Temizer et~al., 2011]{temizer2011}
Temizer, I., Wriggers, P., and Hughes, T. (2011).
\newblock Contact treatment in isogeometric analysis with {NURBS}.
\newblock {\em Comput. Methods Appl. Mech. Engrg.}, {\bf{200}}:1100--1112.

\bibitem[Temizer et~al., 2012]{Temizer2012f}
Temizer, I., Wriggers, P., and Hughes, T. J.~R. (2012).
\newblock Three-dimensional mortar-based frictional contact treatment in
  isogeometric analysis with {NURBS}.
\newblock {\em Comput. Methods Appl. Mech. Engrg.}, {\bf{209-212}}:115--128.

\bibitem[Wilking and Bischoff, 2017]{Bischoff17}
Wilking, C. and Bischoff, M. (2017).
\newblock Alternative integration algorithms for three-dimensional mortar
  contact.
\newblock {\em Comput. Mech.}, {\bf{59}}:203--218.

\bibitem[Wriggers, 2006]{wriggers-contact}
Wriggers, P. (2006).
\newblock {\em Computational Contact Mechanics}.
\newblock Springer, 2$^{\text{nd}}$ edition.

\bibitem[Yang et~al., 2005]{yang05}
Yang, B., Laursen, T.~A., and Meng, X. (2005).
\newblock Two dimensional mortar contact methods for large deformation
  frictional sliding.
\newblock {\em Int. J. Numer. Meth. Engng}, {\bf 62}:1183--1225.

\end{thebibliography}
